\documentclass[useAMS, usenatbib, a4paper]{mnras}
\usepackage{graphicx}
\usepackage{dblfloatfix}
\usepackage{microtype}
\usepackage{xcolor}
\usepackage{fixltx2e}
\usepackage{booktabs}
\usepackage{siunitx}
\usepackage{color}
\usepackage{enumerate}
\usepackage{pdflscape}
\usepackage{rotating}
\usepackage{nicefrac}

\usepackage{dsfont}
\usepackage{bbm}

\usepackage[T1]{fontenc} 
\usepackage[utf8]{inputenc}

\usepackage{newtxtext}
\usepackage[varvw,smallerops]{newtxmath}

\usepackage{hyperref}
\usepackage{xr-hyper}

\hypersetup{colorlinks=True, linkcolor=blue!50!black, citecolor=black,
  urlcolor=blue!50!black}

\usepackage{etoolbox}
\robustify\bfseries
\robustify\itshape

\usepackage{bm}

\usepackage{aastex-compat}

\newcommand\hii{\ion{H}{ii}}

\title[Bow shock shapes]{True versus apparent shapes of bow shocks}

\newcommand\AddressCRyA{Instituto de Radioastronom\'{\i}a y Astrof\'{\i}sica,
  Universidad Nacional Aut\'onoma de M\'exico, Apartado Postal 3-72,
  58090 Morelia, Michoac\'an, M\'exico}
\author[Tarango-Yong \& Henney]{
  Jorge A. Tarango-Yong\thanks{E-mail: j.tarango@irya.unam.mx}
  \& William J. Henney\thanks{E-mail: w.henney@irya.unam.mx}\\
  \AddressCRyA
}

\date{Accepted XXX. Received YYY; in original form ZZZ}

\pubyear{2017}

\DeclareMathOperator{\sgn}{sgn}
\DeclareMathOperator{\Sin}{\mathcal{S}}
\DeclareMathOperator{\Cos}{\mathcal{C}}

\DeclareMathOperator{\GammaFunc}{\Gamma}
\DeclareMathOperator{\Depart}{\Delta}
\newcommand\ecc{\ensuremath{e}}
\newcommand\w{\ensuremath{\mathrm{w}}}
\newcommand\C{\ensuremath{\mathrm{c}}}
\providecommand{\abs}[1]{\lvert#1\rvert}
\providecommand{\Abs}[1]{\left\lvert#1\right\rvert}
\newenvironment{Vector}{\left(\begin{array}{c}}{\end{array}\right)}
\newcommand\uvec[1]{\bm{\hat{#1}}}
\newcommand\T{_{\mathrm{\scriptscriptstyle T}}}

\begin{document}
\label{firstpage}
\pagerange{\pageref{firstpage}--\pageref{lastpage}}
\maketitle
\begin{abstract}
  Astrophysical bow shocks are a common result of the interaction
  between two supersonic plasma flows, such as winds or jets from
  stars or active galaxies, or streams due to the relative motion
  between a star and the interstellar medium.
  For cylindrically symmetric bow shocks, we develop a general theory
  for the effects of inclination angle on the apparent shape. We
  propose a new two-dimensional classification scheme for bow shapes,
  which is based on dimensionless geometric ratios that can be
  estimated from observational images.  The two ratios are related to
  the flatness of the bow's apex, which we term \textit{planitude} and
  the openness of its wings, which we term \textit{alatude}.  We
  calculate the expected distribution in the planitude--alatude plane
  for a variety of simple geometrical and physical models: quadrics of
  revolution, wilkinoids, cantoids, and ancantoids.  We further test
  our methods against numerical magnetohydrodynamical simulations of
  stellar bow shocks and find that the apparent planitude and alatude
  measured from infrared dust continuum maps serve as accurate
  diagnostics of the shape of the contact discontinuity, which can be
  used to discriminate between different physical models.  We present
  an algorithm that can determine the planitude and alatude from observed
  bow shock emission maps with a precision of 10 to 20\%. 
\end{abstract}



\section{Introduction}
\label{sec:intro}


The archetypal bow shock is formed when a solid body moves
supersonically through a compressible fluid.  Terrestrial examples
include the atmospheric re-entry of a space capsule, or the sonic boom
produced by a supersonic jet \citep{van-Dyke:1982a}.  In astrophysics
the term bow shock is employed more widely, to refer to many different
types of curved shocks that have approximate cylindrical symmetry.
Instead of a solid body, astrophysical examples usually involve the
interaction of \emph{two} supersonic flows, such as the situation of a
stellar wind emitted by a star that moves supersonically through the
interstellar medium \citep{van-Buren:1988a, Kobulnicky:2010a,
  van-Marle:2011a, Mackey:2012b, Mackey:2015a}.  In such cases, two
shocks are generally produced, one in each flow.  Sometimes,
especially in heliospheric studies \citep{Zank:1999a, Scherer:2014a},
the term ``bow shock'' is reserved for the shock in the ambient
medium, with the other being called the ``wind shock'' or
``termination shock''.  However, in other contexts such as colliding
wind binaries \citep{Stevens:1992a, Gayley:2009a} such a distinction
is not so useful.  


\begin{figure}
  \centering
  \bigskip
  \includegraphics[width=0.9\linewidth]{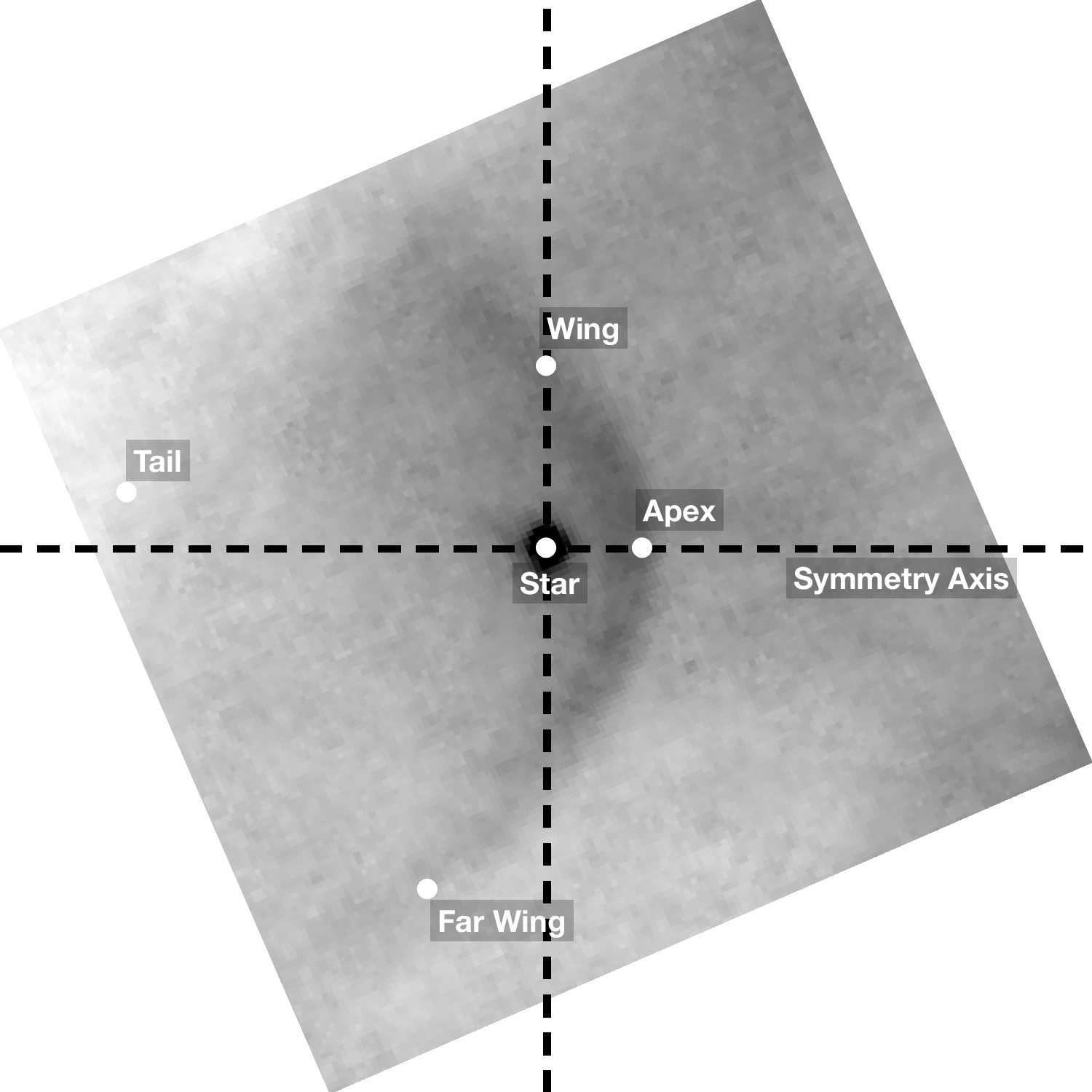}
  \caption{Descriptive terminology for a stellar bow shock.  The apex
    is the closest approach of the bow to the star, while the wings
    are the parts of the bow that curve back past the star.}
  \label{fig:bow-terminology}
\end{figure}
A further class of astrophysical bow shock is driven by highly
collimated, supersonic jets of material, such as the Herbig Haro
objects \citep{Schwartz:1978b, Hartigan:1987a} that are powered by
jets from young stars or protostars.  Additional examples are seen in
planetary nebulae \citep{Phillips:2010a, Meaburn:2013a}, active
galaxies \citep{Wilson:1987a}, and in galaxy clusters
\citep{Markevitch:2002a}.  In the jet-driven case, the term ``working
surface'' is often applied to the entire structure comprising the two
shocks plus the shocked gas in between them, separated by a
\textit{contact discontinuity}.  The working surface may be due to the
interaction of the jet with a relatively quiescent medium, or may be an
``internal working surface'' within the jet that is due to
supersonic temporal variations in the flow velocity
\citep{Raga:1990a}.

In empirical studies the relationship between these theoretical
constructs and the observed emission structures is not always clear.
In such cases the term ``bow shock'' is often used in a more general
sense to refer to the entire arc of emission.  In this paper, we will
concentrate on \textit{stellar bow shocks}, in which the position of
the star can serve as a useful reference point for describing the bow
shape.  The empirical terminology that we will employ is illustrated
in Figure~\ref{fig:bow-terminology}.  The \textit{apex} is the point
of closest approach of the bow to the star, which lies on the
approximate symmetry axis, and the region around the apex is sometimes
referred to as the \textit{head} of the bow.  The \textit{wings} are
the swept-back sides of the bow, which lie in a direction from the
star that is orthogonal to the axis, with the \textit{far wings} being
the wing region farthest from the apex. Finally, the \textit{tail} is
the region near the axis but in the opposite direction from the apex.

\begin{figure}
  \includegraphics[width=\linewidth]{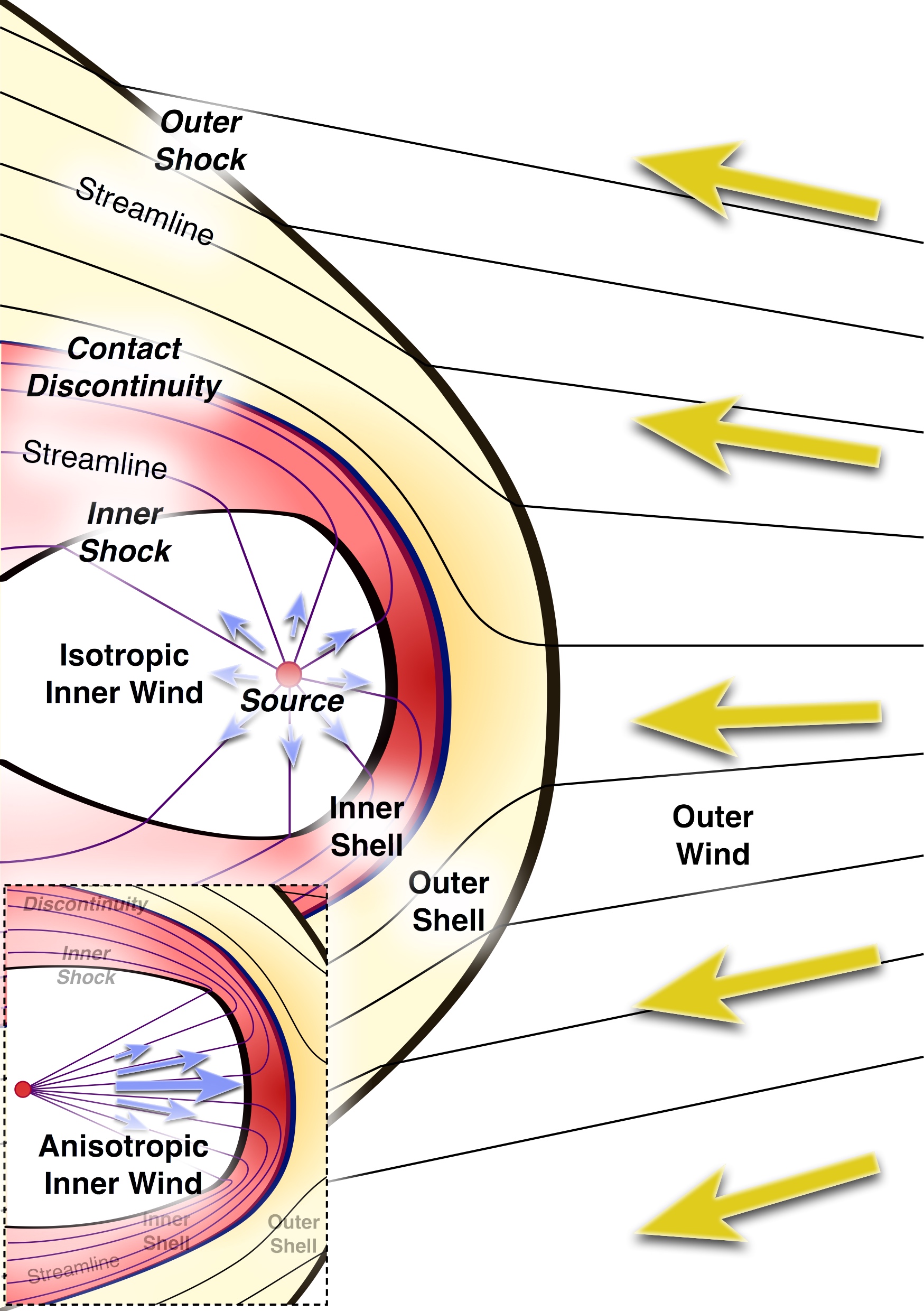}
  \caption{Quasi-stationary bow shock structure formed by the
    interaction of two supersonic winds.  Lower-left inset box shows
    the case where the inner wind is anisotropic. The streamlines
    (thin lines) are drawn to be qualitatively realistic: they are
    straight in regions of hypersonic flow, but curved in subsonic
    regions, responding to pressure gradients in the shocked
    shells. Streamline slopes are discontinuous across oblique
    shocks.}
\label{fig:2-winds}
\end{figure}
\newcommand\Mach{\ensuremath{\mathcal{M}}} Figure~\ref{fig:2-winds}
shows an idealized schematic of how a double bow-shock shell is formed
from the interaction of two supersonic streams: an \textit{inner wind}
and an \textit{outer wind}, with the inner wind being the weaker of
the two (in terms of momentum), so that the shell curves back around
the inner source.  The outer wind may be from another star, or may be
a larger scale flow of the interstellar medium, such as the
\textit{champagne flow} produced by the expansion of an \hii{} region
away from a molecular cloud \citep{Tenorio-Tagle:1979a, Shu:2002a,
  Medina:2014a}.  Alternatively, it may be due to the supersonic
motion of the inner source through a relatively static medium, in
which case the outer wind will not be divergent as shown in the figure
but rather plane-parallel.  The thickness of the shocked shells at the
apex depends on the Mach number, \Mach{}, of the flows and the
efficiency of the post-shock cooling.  For sufficiently strong
cooling, the post-shock cooling zone thickness is negligible and the
shock can be considered isothermal.  In this case, the shell thickness
is of order \(\Mach^{-2}\) times the source-apex separation
\citep{Henney:2002a}, which can become very small for high Mach
numbers.  The shell thickness will tend to increase towards the wings,
due to the increasing shock obliqueness, which reduces the
perpendicular Mach number.

\begin{figure}
  \centering
  \includegraphics[width=\linewidth]{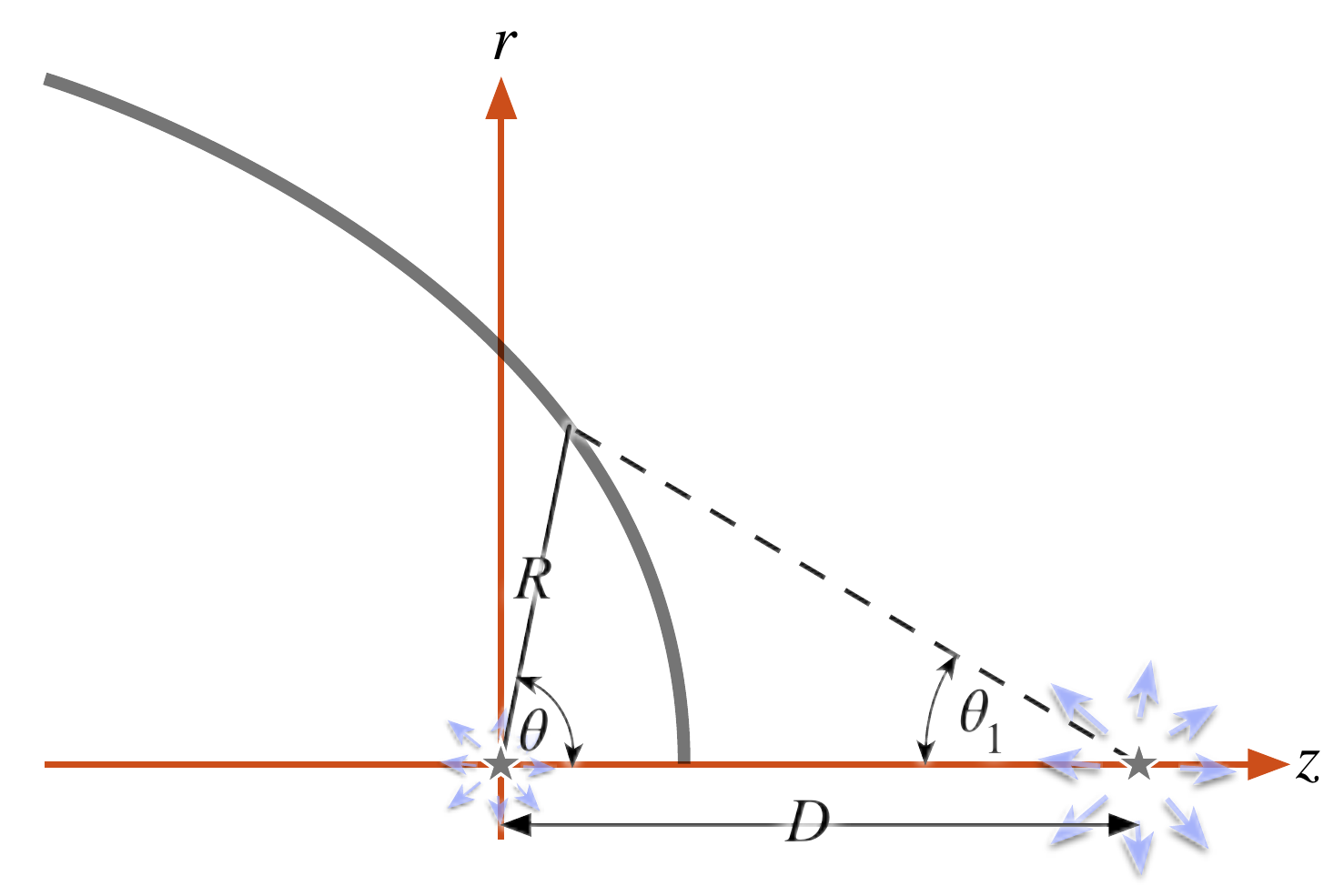}
  \caption[]{Schematic diagram of cylindrically symmetric two-wind
    interaction problem in the thin-shell limit, following
    \citet{Canto:1996}.}
  \label{fig:crw-schema}
\end{figure}
In the extreme thin-shell limit, the entire bow structure can be
treated as a surface.  The bow radius measured from the inner source
(star) is \(R(\theta, \phi)\), where \(\theta\) is the polar angle, measured from
the star-apex axis, and \(\phi\) is the azimuthal angle, measured around
that axis.  Assuming cylindrical symmetry about the axis, this reduces
to \(R(\theta)\), which is illustrated in Figure~\ref{fig:crw-schema},
following \citet{Canto:1996}.  The separation between the two sources
is \(D\) and the complementary angle, as measured at the position of
the outer source, is \(\theta_1\).  The minimum value of \(R(\theta)\) is the
stagnation radius, \(R_0\), which occurs at the apex (\(\theta = 0\)).  In
a steady state, ram-pressure balance on the axis implies that
\begin{equation}
  \label{eq:stagnation-radius}
  \frac{R_0} {D} = \frac{\beta^{1/2}} {1 + \beta^{1/2}} ,
\end{equation}
where \(\beta\) is the momentum ratio between the two winds.  If the winds
are isotropic, with inner wind mass-loss rate \(\dot{M}_{\w}\) and
terminal velocity \(V_{\w}\), while the outer wind has corresponding
values \(\dot{M}_{\w1}\) and \(V_{\w1}\), then the momentum ratio is
\begin{equation}
  \label{eq:beta-definition}
  \beta = \frac{\dot{M}_{\w} V_{\w}} {\dot{M}_{\w1} V_{\w1}} .
\end{equation}
The case where the outer wind is a parallel stream
\citep{Wilkin:1996a} corresponds to the limit \(\beta \to 0\), in which case
\(D\) is no longer a meaningful parameter.


The paper is organized as follows.
In \S~\ref{sec:plan-alat-bow} we outline the geometric parameters that
are necessary for describing bow shapes and introduce two
dimensionless ratios: planitude and alatude.
In \S~\ref{sec:projection} we derive general results for the
projection of bow shapes on to the plane of the sky.
In \S~\ref{sec:conic} we apply the results to the simplest possible
class of geometric bow models: the quadrics of revolution, which
comprise spheroids, paraboloids, and hyperboloids, each of which
occupies a distinct region of the planitude--alatude plane.
In \S~\ref{sec:crw-scenario} we consider thin-shell hydrodynamic
models for the parallel-stream case (wilkinoids) and wind-wind case
(cantoids), including extension to an anisotropic inner wind
(ancantoids).  We calculate the location of the models in the
planitude--alatude plane as a function of the inclination of the bow
shock axis to the plane of the sky.
In \S~\ref{sec:more-realistic-bow} we test our methods against the
results of more realistic numerical simulations of bow shocks,
including the derivation of the shape parameters from maps of infrared
dust emission.
In \S~\ref{sec:obs} we apply our methods to example observations of
proplyd bow shocks in the Orion Nebula, paying close attention to the
systematic uncertainties that arise when our algorithms are applied to
real data.
In \S~\ref{sec:conc} we summarise our results and outline how
following papers will apply these ideas to a more extensive set of
observations, models and numerical simulations.


\section{Planitude and alatude of bow shapes}
\label{sec:plan-alat-bow}

\begin{figure}
  \centering
  \includegraphics[width=\linewidth]{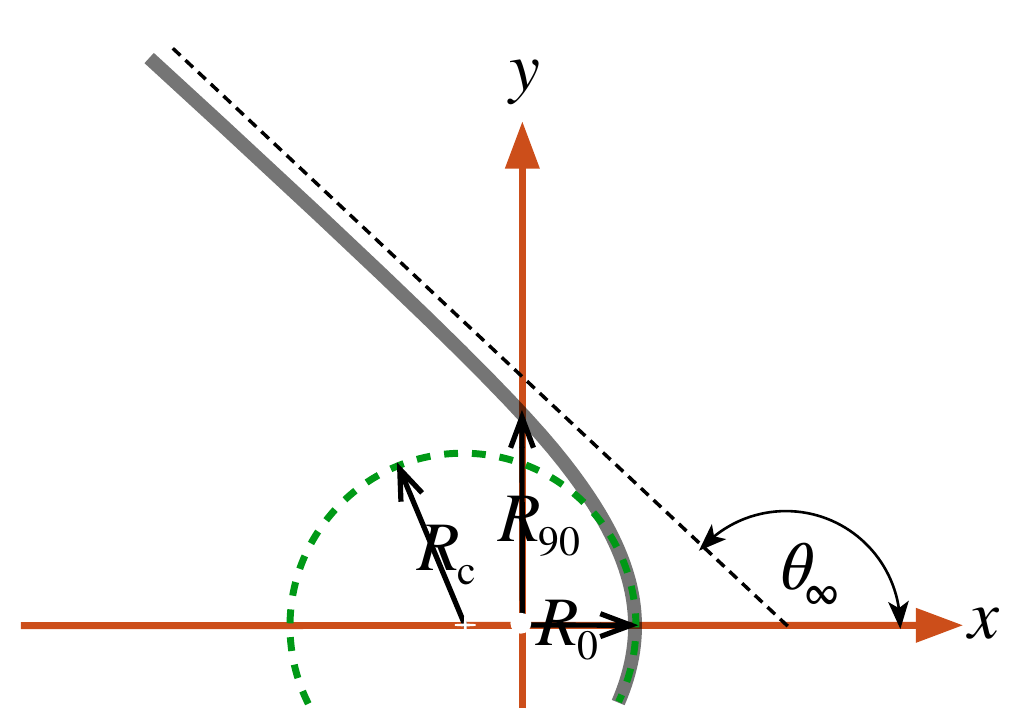}
  \caption[]{Parameters for characterizing a bow shape.  Bow radius
    from the star, measured parallel (\(R_0\)) and perpendicular
    (\(R_{90}\)) to the symmetry axis, together with radius of
    curvature (\(R_{\C}\)) at apex and asymptotic opening angle
    (\(\theta_\infty\)) of the far wings. }
  \label{fig:characteristic-radii}
\end{figure}

The stagnation radius \(R_0\) describes the linear scale of the bow
shock, but in order to characterize its shape more parameters are
required.  To efficiently capture the diversity of bow shapes, we
propose the parameters shown in Figure~\ref{fig:characteristic-radii}.
The perpendicular radius \(R_{90}\) is the value of \(R(\theta)\) at
\(\theta = 90^\circ\), whereas \(R_{\C}\) is the radius of curvature of the bow
at the apex (\(\theta = 0\)).  For a cylindrically symmetric bow, we show
in Appendix~\ref{sec:radius-curvature} that this is given by
\begin{equation}
  \label{eq:radius-curvature}
  R_{\C} 
  = \frac{R_0^2}{R_0 - R_{\theta\theta,0}} \ , 
\end{equation}
where \(R_{\theta\theta,0}\) is \(d^2 \!R / d\theta^2\) evaluated at \(\theta = 0\).

A fourth parameter is the asymptotic opening angle of the far wings,
\(\theta_\infty\), which is useful in the case that the wings are asymptotically
conical.  However, in many bow shocks the wings tend towards the
asymptotic angle only slowly, making \(\theta_\infty\) difficult to measure,
especially since the emission from the far wings is often weak at
best.  In contrast, the three radii, \(R_0\), \(R_{90}\), and
\(R_{\C}\). are straightforward to determine from observations.  One
simple method to estimate the radius of curvature is to make use of
the Taylor expansion\footnote{%
  This method assumes both that \(R(\theta)\) is even (true for a
  cylindrically symmetric bow) and that the orientation of the axis is
  already known.  Generalization to cases where these assumptions do
  not hold is discussed in Appendix~\ref{app:rcurv-empirical}.} %
of \(R(\theta)\) about the apex (with \(\theta\) in radians):
\begin{equation}
  \label{eq:taylor-R-theta}
  R(\theta) = R_0 + \frac12 R_{\theta\theta,0} \,\theta^2 + \mathcal{O}(\theta^4) \ ,
\end{equation}
so that fitting a polynomial in \(\theta^2\) to \(R(\theta)\) for
\(|\theta| < \Delta\theta \) yields \(R_0\) and \(R_{\theta\theta,0}\) from the first two
coefficients, and hence \(R_{\C}\) from
equation~\eqref{eq:radius-curvature}.  Experience has shown that
\(\Delta\theta = 30^\circ\) and three terms in the polynomial are good choices,
where the third term is used only as a monitor (if the co-efficient of
\(\theta^4\) is not small compared with \(R_0\), then it may indicate a
problem with the fit).

Since we have three radii, we can construct two independent
dimensionless parameters:
\begin{align}
  \label{eq:planitude}
  \text{Planitude} \quad \Pi & \equiv  \frac{R_{\C}} {R_0} \\
  \label{eq:alatude}
  \text{Alatude} \quad \Lambda & \equiv  \frac{R_{90}} {R_0}
\end{align}
and these will be the principal shape parameters that we will use in
the remainder of the paper.  The \textit{planitude}, \(\Pi\), is a
measure of the flatness of the head of the bow around the apex, while
the \textit{alatude}, \(\Lambda\), is a measure of the openness of the bow
wings.  Although ``planitude'' can be found in English dictionaries,
``alatude'' is a new word that we introduce here, derived from the
latin \textit{ala} for ``wing''.

Several previous studies have discussed the relation between
\(R_{90}\) and \(R_0\) as a diagnostic of bow shape (for example
\citealp{Robberto:2005a, Cox:2012a, Meyer:2016a}), but as far as we
know, we are the first to include \(R_{\C}\).  \citet{Robberto:2005a}
\S~4.2 use the ratios \(R_0/D\) and \(R_{90}/D\) in analyzing proplyd
bow shapes in the Trapezium cluster in the center of the Orion Nebula
\citep{Hayward:1994a, Garcia-Arredondo:2001a, Smith:2005a}.  In that
case, the source of the outer wind is known, and so \(D\) is
well-determined (at least, in projection), but for many bow shocks
\(D\) is not known, and is not even defined for the moving-star or
parallel-stream case. \citet{Cox:2012a} \S~4.1 compare the observed
shapes of bow shocks around cool giant stars with an analytic model,
and use \(A\) and \(B\) for the projected values of \(R_0\) and
\(R_{90}\), respectively (see next section for discussion of
projection effects).  \citet{Meyer:2016a} \S~3.2 analyze the
distribution of \(R_0 / R_{90}\) (the reciprocal of our \(\Lambda\)) for
hydrodynamic simulations of bow shocks around runaway OB stars.







\section{Projection onto the plane of the sky}
\label{sec:projection}

In this section we calculate the apparent shape on the plane of the
sky of the limb-brightened border of a shock or shell that is
idealized as an arbitrary cylindrically symmetric surface.


\subsection{Frames of reference}
\label{sec:ref-frames}

Consider body-frame cartesian coordinates $(x,y,z)$, where \(x\) is
the symmetry axis, and spherical polar coordinates
\((R, \theta, \phi)\), where \(\theta\) is the polar angle and
\(\phi\) the azimuthal angle.  Since the surface is cylindrically
symmetric, it is can be specified as $R = R(\theta)$, so that
cartesian coordinates on the surface are:
\begin{equation}
  \bm{r} \equiv
  \begin{Vector}
    x \\ y \\ z
  \end{Vector} 
  = 
  \begin{Vector}
    R(\theta)\, \cos\theta \\
    R(\theta)\, \sin\theta\cos\phi \\
    R(\theta)\, \sin\theta\sin\phi
  \end{Vector}.
  \label{eq:body-frame}
\end{equation} 
Suppose that the viewing direction makes an angle \(i\) with the
\(z\)~axis, so that we can define observer-frame coordinates
\((x', y', z')\), which are found by rotating the body-frame
coordinates about the \(y\)~axis.  The same vector, \(\bm{r}\), expressed in the observer frame is then 
\begin{equation}
  \bm{r} = 
  \begin{Vector}
    x' \\ y' \\ z'
  \end{Vector}
  = \mathbfss{A}_y(i)\,
  \begin{Vector}
    x \\ y \\ z
  \end{Vector}
  =
  \begin{Vector}
    x\cos i - z\sin i\\
    y \\
    z\cos i + x\sin i
  \end{Vector} \ ,
  \label{eq:Trans}
\end{equation}
where the rotation matrix \(\mathbfss{A}_y(i)\) is given in
Appendix~\ref{sec:plane-sky-projection}.  The inclination angle \(i\)
is defined so that \(i = \ang{0}\) when the surface is viewed
perpendicular to its axis (\textit{side on}) and \(i = \pm\ang{90}\)
when it is viewed along its axis (\textit{end on}), with positive
\(i\) when the apex points towards the observer.

\begin{figure}
  \centering
  \includegraphics[width=\linewidth]{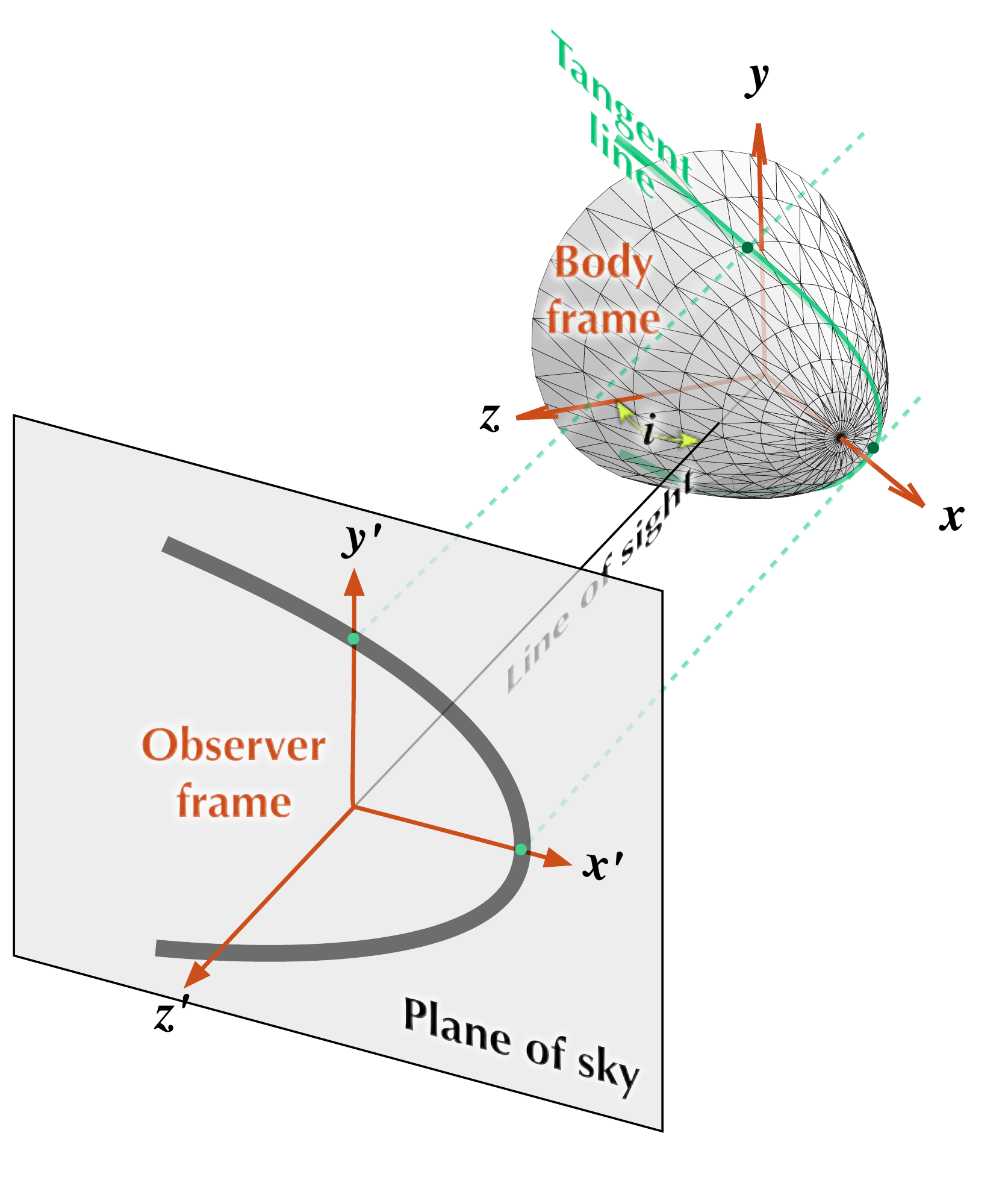}
  \caption{Relationship between body frame (unprimed coordinates) and
    observer frame (primed coordinates).  Note that the plane of the
    sky is a projective plane, not a geometric plane in Euclidean
    3-space, see discussion in text.}
  \label{fig:projection-pos}
\end{figure}

The relationship between the two frames is illustrated in
Figure~\ref{fig:projection-pos}.  All quantities in the observer's
frame are denoted by attaching a prime to the equivalent quantity in
the body frame. There are two ways of interpreting the primed
coordinates. On the one hand, the 3-vector \((x', y', z')\) specifies
a point in Euclidean space, \(\mathds{R}^3\), but an alternative
interpretation is to take the 2-vector \((x', y')\) as specifying a
point in a \textit{projective space}, \(\mathds{P}^2\) (see, for
example, \S~15.6 of \citealp{Penrose:2004a}). Each ``point'' in
\(\mathds{P}^2\) is equivalent to a line in \(\mathds{R}^3\),
specifically: a line of sight that passes through the observer. Thus,
\((x', y')\) gives the celestial coordinates on the \textit{plane of
  the sky}, with \(x'\) being the projected symmetry axis of the
surface.  We assume that the observer is located at a very large
distance, relative to the size of the bow, so that all lines of sight
are effectively parallel to the \(z'\) axis, with the observer at
\(z' = -\infty\).  But, from the point of view of the plane of the sky, the
\(z'\) coordinate is strictly irrelevant since it is a projective
plane, and not a Euclidean plane.  In the following, we will switch
between the \(\mathds{R}^3\) and \(\mathds{P}^2\) interpretations as
convenient, resolving ambiguity where necessary via the adjectives
``Euclidean'' for \(\mathds{R}^3\) and ``plane-of-sky'' or
``projected'' for \(\mathds{P}^2\).

\subsection{Unit vectors normal and tangential to the surface}
\label{sec:unit-vectors-normal}
\begin{figure}
  \centering
  \includegraphics[width=\linewidth]{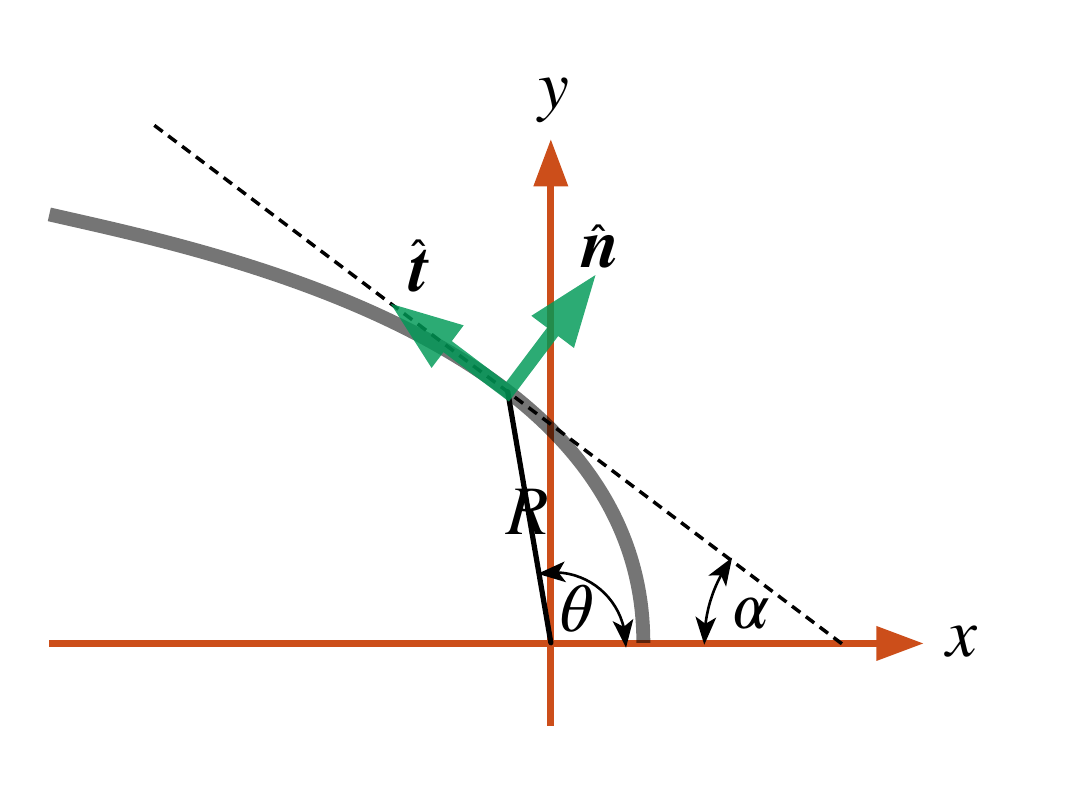}
  \caption{Unit vectors in the body frame that are normal and
    tangential to the surface \(R(\theta)\) in a plane of constant
    azimuth, \(\phi\).}
  \label{fig:unitvec}
\end{figure}
We define unit vectors \(\uvec{n}, \uvec{t}\), such that \(\uvec{n}\) is normal to the surface, while \(\uvec{t}\) is tangent to the surface in a plane of constant \(\phi\). For \(\phi = 0\) the surface lies in the \(xy\) plane and it is straightforward to show (Fig.~\ref{fig:unitvec}) that in this case the unit vectors are given by 
\begin{equation}
  \label{eq:uvec-phi0}
  \uvec{t}_0 =
  \begin{Vector}
    -\cos\alpha \\ \sin\alpha \\ 0
  \end{Vector}
  \quad \mathrm{and} \quad
  \uvec{n}_0 =
  \begin{Vector}
    \sin\alpha \\ \cos\alpha \\ 0
  \end{Vector}, 
\end{equation}
where \(\alpha\) is the \textit{slope angle}, given by
\begin{equation}
  \label{eq:alpha}
  \tan\alpha = -\left.\frac{dy}{dx}\right\vert_{R(\theta)} 
  = \frac{1 + \omega \tan\theta}{\tan\theta - \omega}
\end{equation}
and \(\omega\) is a dimensionless \textit{local growth factor}:
\begin{equation}
  \label{eq:omega}
  \omega(\theta) = \frac{1}{R} \frac{dR}{d\theta} . 
\end{equation}
For general \(\phi \ne 0\), we find \(\uvec{n}\) and \(\uvec{t}\) by
rotating equations~(\ref{eq:uvec-phi0}) around the \(x\)-axis with the
matrix \(\mathbfss{A}_x(\phi)\) (eq.~[\ref{eq:Rotation-matrix-x}]):
\begin{align}
  \label{eq:uvec-n-phi}
  \uvec{n} &= \mathbfss{A}_x(\phi) \, \uvec{n}_0 =
  \begin{Vector}
    \sin\alpha \\ \cos\alpha\cos\phi \\ \cos\alpha\sin\phi
  \end{Vector}  \\
  \label{eq:uvec-t-phi}
  \uvec{t} &= \mathbfss{A}_x(\phi) \, \uvec{t}_0 =
  \begin{Vector}
    -\cos\alpha \\ \sin\alpha\cos\phi \\ \sin\alpha\sin\phi
  \end{Vector} \ .
\end{align}

\subsection{Tangent line}
\label{sec:tangent-line}

 


The boundary on the plane of the sky of the projected surface is the
locus of those lines of sight that graze the surface tangentially.
This corresponds to a curved line on the surface itself, which we
denote the \textit{tangent line}, and which is defined by the
condition  
\begin{equation}
  \label{eq:tangent-line-condition}
  \uvec{n} \cdot \uvec{z}' = 0.
\end{equation}
We denote by \(\phi\T\) that value of $\phi$ that satisfies this relation
for a given inclination, \(i\), and polar angle, \(\theta\).  From
equations~(\ref{eq:uvec-n-phi}, \ref{eq:tangent-line-condition},
\ref{eq:zunit-prime}, \ref{eq:alpha}) this is
\begin{equation}
\sin\phi\T = -\tan i\tan\alpha = \tan i \frac{1+\omega\tan\theta}{\omega-\tan\theta} \ .
\label{eq:tanphi}
\end{equation}
From equations~(\ref{eq:body-frame}, \ref{eq:Trans}) it follows that
the observer-frame coordinates of the tangent line are given by
\begin{equation}
\left(\begin{array}{c}
x'\T \\ y'\T \\ z'\T
\end{array}\right)= R(\theta)\left(\begin{array}{c}
\cos\theta\cos i - \sin\theta\sin\phi\T \sin i \\
\sin\theta(1-\sin^2\phi\T)^{1/2} \\
\cos\theta\sin i +\sin\theta\sin\phi\T\cos i
\end{array}\right).
\label{eq:tangential}
\end{equation} 
Note that, in general, \(z'\T\) is not a linear function of \(x'\T\)
and \(y'\T\), so that the tangent line \((x'\T, y'\T, z'\T)\) is not a
plane curve in 3-dimensional Euclidean space, \(\mathds{R}^3\).
However, for the projected shape \((x'\T, y'\T)\) of the tangent line
on the plane of the sky, \(\mathds{P}^2\), the value of \(z'\T\) does
not matter (see above).  The projected shape can also be described in
polar form as \(R'(\theta')\), where
\begin{equation}
  \label{eq:R-prime-theta-prime}
  R' = (x'^2\T+ y'^2\T)^{1/2} 
  \quad \text{and} \quad
  \tan\theta' = y'\T / x'\T.
\end{equation}

Equation~(\ref{eq:tanphi}) will not have a solution for
arbitrary values of $\theta$ and $i$, but only when
$|\tan i\tan\alpha|<1$. In particular, if $i\neq 0$, then the tangent line
only exists for \(\theta > \theta_{0}\) where \(\theta_{0}\) is the value of
\(\theta\) on the tangent line's projected symmetry axis
(\(\theta' = 0\)).  From equations~(\ref{eq:tangential},
\ref{eq:R-prime-theta-prime}) it follows that \(\sin^2\phi\T = 1\) at
\(\theta = \theta_0\), which yields the implicit equation
\begin{align}
\tan\theta_{0} = \frac{|\tan i| + \omega(\theta_{0})}{1-\omega(\theta_{0}) |\tan i|} . 
\label{eq:thetapar}
\end{align}
In addition, if the surface is sufficiently ``open''
(\(\alpha \ge \alpha_{\mathrm{min}} > 0\) for all \(\theta\)), then for those
inclinations with
\(\vert i\vert > (\ang{90} - \alpha_{\mathrm{min}}) \) the tangent line does not exist
for any value of \(\theta\).  In other words, when the viewing angle is
sufficiently close to face-on, the projected surface has no ``edge''
and will no longer look like a bow shock to the observer.

After completing this work, it was brought to our attention that the
principal results of this section had already been derived in
Appendix~B of the PhD thesis \citet{Wilkin:1997a}.  For instance,
Wilkin's equation~(8) is equivalent (apart from differences in
notation) to our equation~\eqref{eq:tanphi}.

\subsection{Characteristic radii on the plane of the sky}

In order to compare the shell shape given by $R(\theta)$ with observations,
it is convenient to define the following apparent radii in the
observer frame: $R'_{0}$ and $R'_{90}$. These are projected distances
of the shell tangent line from the origin. The first is measured in
the direction of the symmetry axis, and the second in a perpendicular
direction. More concretely $R'_{0} = x'\T(y'\T=0)$ and
$R'_{90} = y'\T(x'\T=0)$. From equations (\ref{eq:tanphi}) and
(\ref{eq:tangential}) we find that:
\begin{align}
R'_{0} = R(\theta_{0})\cos(\theta_{0} + i) \label{eq:Rpar} 
\end{align}
Where $\theta_{0}$ is the solution of equation (\ref{eq:thetapar}), and
\begin{align}
  \label{eq:R90prime}
R'_{90} = R(\theta_{90})\sin\theta_{90}\left(1-\sin^2\bigl(\phi\T(\theta_{90})\bigr)\right)^{1/2}
\end{align}
Where $\theta_{90}$ is the solution of the implicit equation:
\begin{align}
  \label{eq:th90}
\cot\theta_{90} = \frac{1-\left(1+\omega(\theta_{90})^2\sin^22i\right)^{1/2}}{2\omega(\theta_{90})\cos^2 i}
\end{align}
The projected alatude (see \S~\ref{sec:plan-alat-bow}) is then given
by \(\Lambda' = R_{90}' / R_0'\).

Similarly, the projected planitude is \(\Pi' = R_{\C}' / R_0'\), where
\(R_{\C}'\) is found by applying the equivalent of equation
\eqref{eq:radius-curvature} for primed quantities:
\begin{equation}
  \label{eq:projected-radius-curvature}
  R_{\C}' 
  = \frac{(R_0')^2}{R_0' - R_{\theta'\theta',0}'} \ .
\end{equation}

\subsection{Line-of-sight velocities on the tangent line}
\label{sec:line-sight-veloc}
Motions in a thin shocked shell will be predominantly tangential to
the shell surface. In addition, for the particular case of wind-wind
bowshocks, the flow in each azimuthal slice can be shown to be
independent \citep{Wilkin:2000a}, which implies that the shell
velocity is parallel to \(\uvec{t}\). The projected line-of-sight
shell velocity is therefore
  \begin{equation}
    \label{eq:vlos}
    v_{\mathrm{los}} = (\uvec{t}' \cdot -\uvec{z}') \, v_{\parallel}(\theta) = \frac{v_{\parallel}(\theta) (1+\omega^2)^{1/2} \sin i }{\sin\theta - \omega\cos\theta} ,
  \end{equation}
  where \( v_{\parallel}(\theta)\) is the gas velocity along the shell and the standard sign convention has been adopted such that velocities away from the observer are deemed positive.


\newcommand{\Q}{\ensuremath{\mathcal{Q}}}
\newcommand{\thetaQ}{\ensuremath{\theta_{\scriptscriptstyle \Q}}}

\begin{figure*}
  \setlength\tabcolsep{2em}
  \begin{tabular}{ll}
    (a) & (b) \\
    \includegraphics[height=0.45\linewidth]{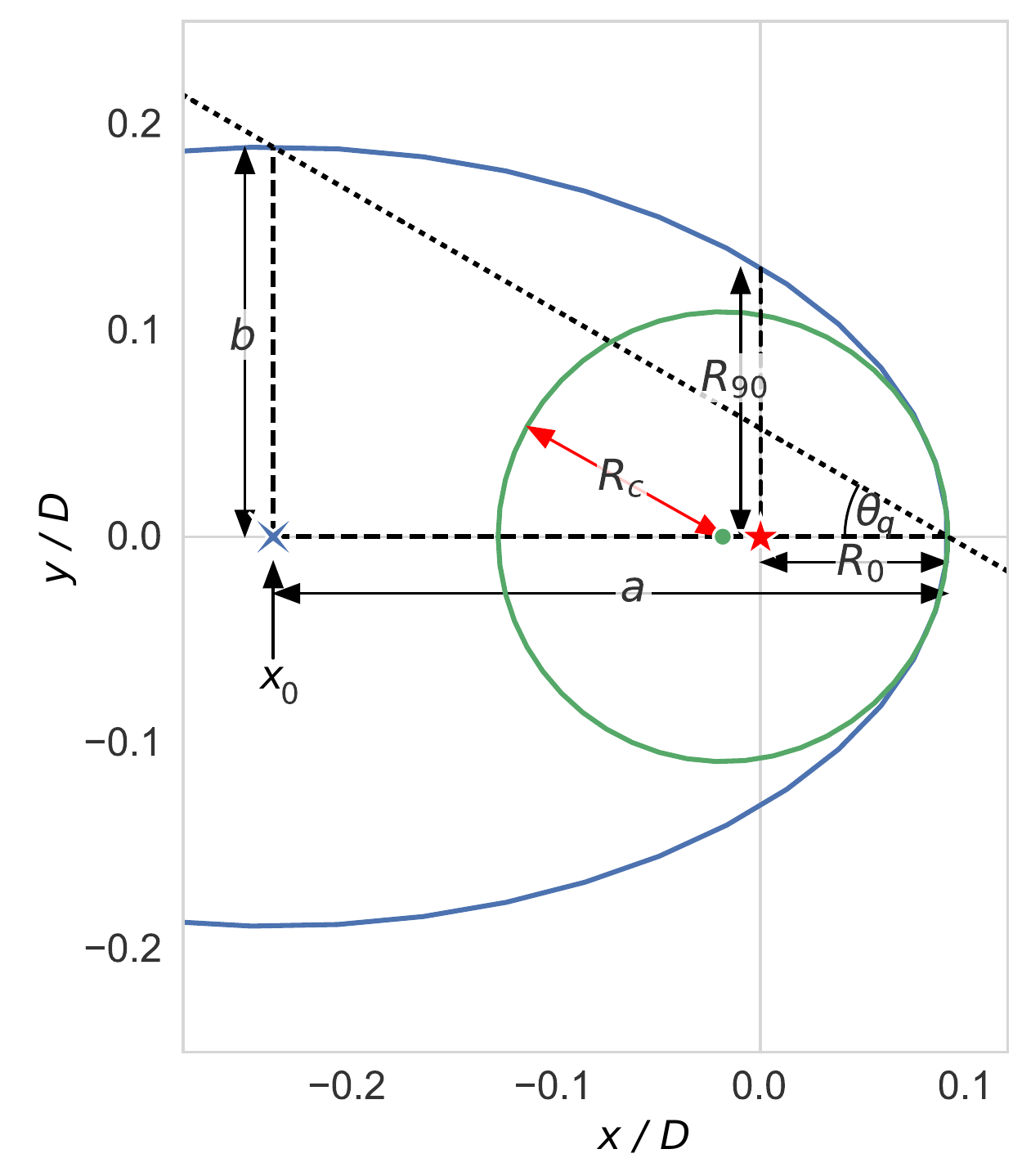}
        & \includegraphics[height=0.45\linewidth]{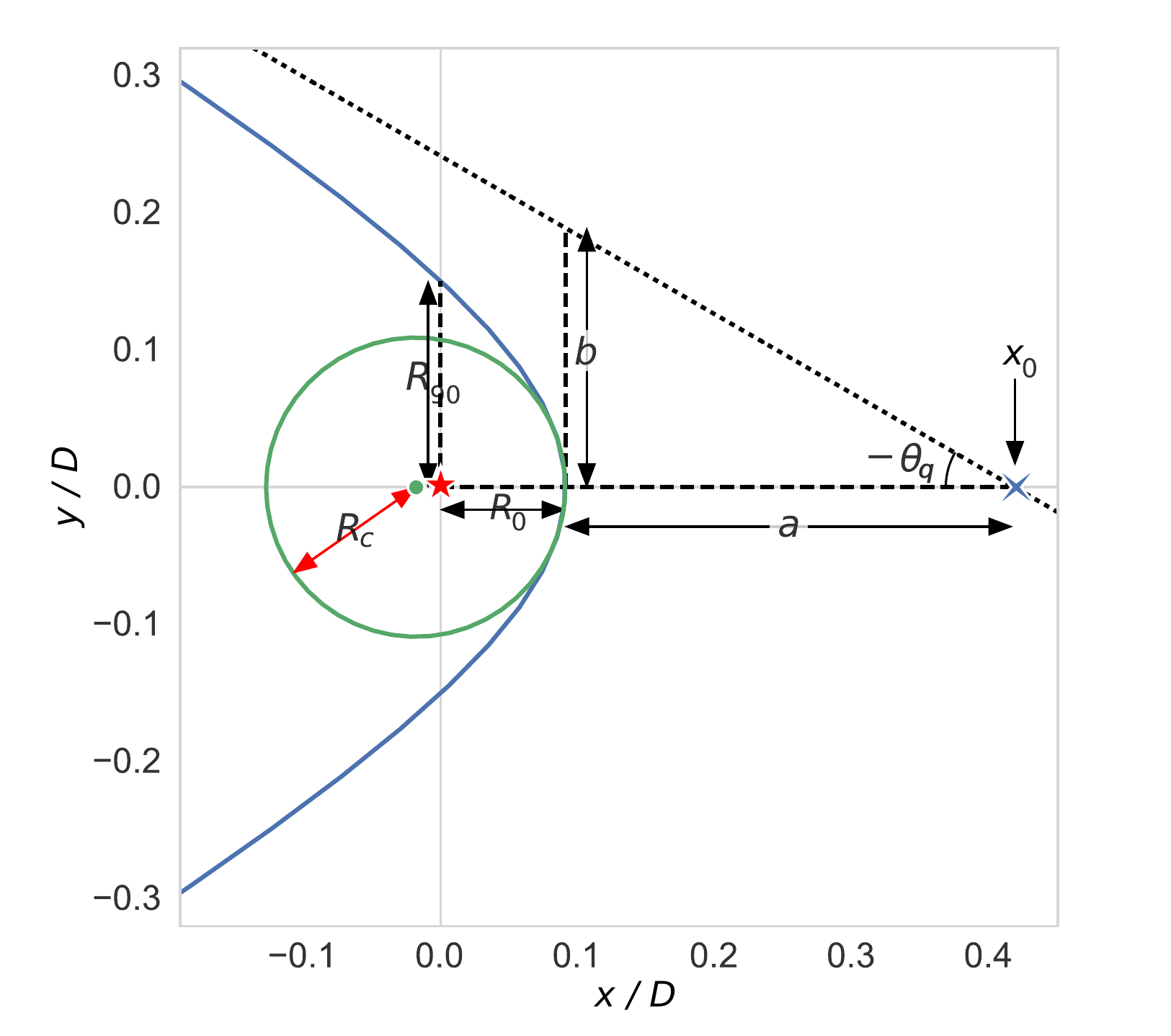}
  \end{tabular}
  \caption[]{Example off-center conic sections that can form quadrics of
    revolution: (a) ellipse, (b) hyperbola.  The relationship is shown
    between the conic section parameters \(a\), \(b\), \(\thetaQ\) and the
    bowshock characteristic radii \(R_0\), \(R_{90}\), \(R_c\), as
    defined in Fig.~\ref{fig:characteristic-radii}. The origin (center
    of the weaker flow) is indicated by a red star, the center of
    curvature of the apex of the bow shock is indicated by a green
    dot, and the geometric center of the conic section is indicated by
    a blue cross, which is offset by \(x_0\) from the origin.  Note
    that \(R_0\), \(R_{90}\), \(R_c\), \(a\), and \(b\) are all
    \emph{lengths} and are always positive, whereas \(x_0\) is a
    \emph{displacement} and may be positive or negative.}
  \label{fig:conics}
\end{figure*}
\begin{figure}
  \includegraphics[width=\linewidth]{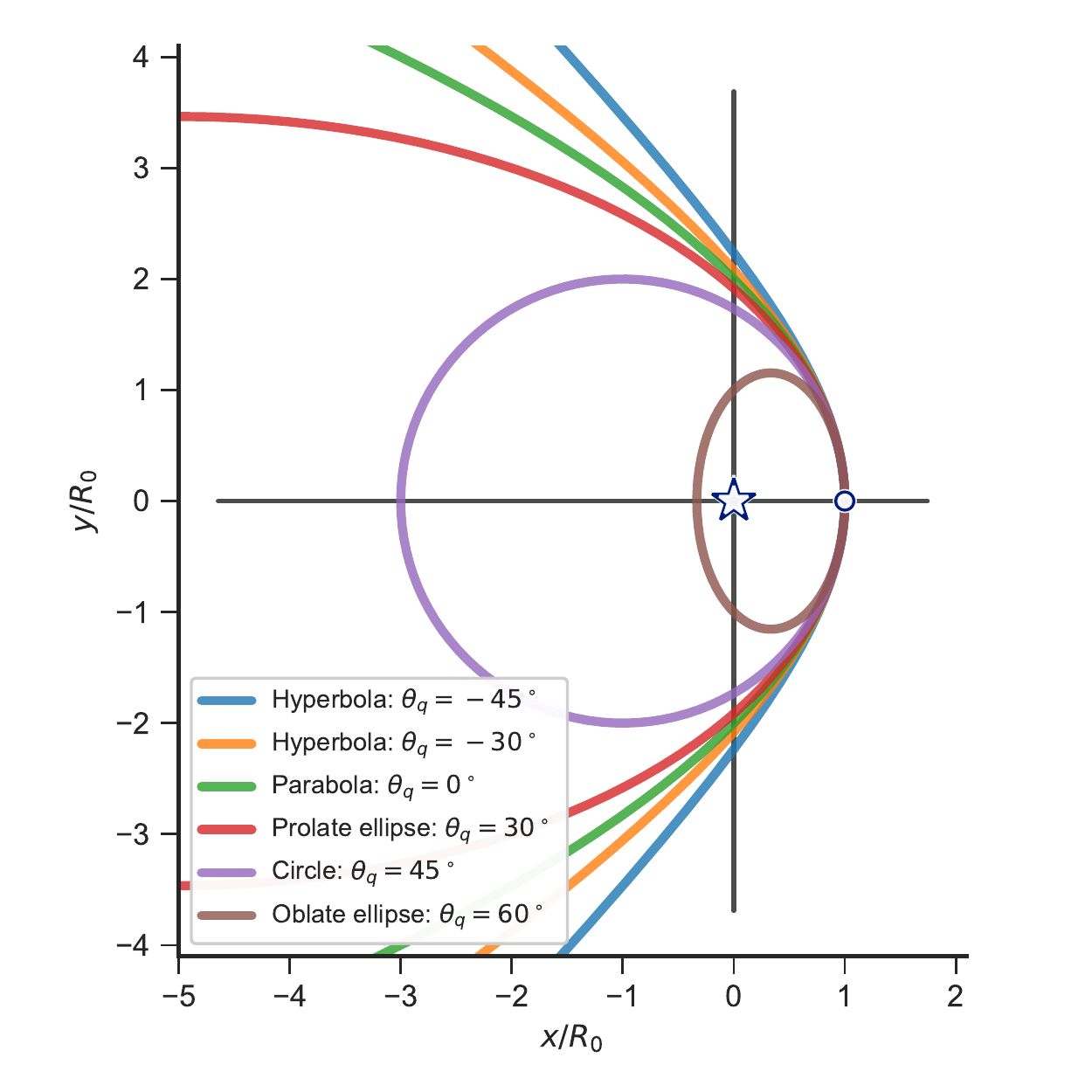}
  \caption[]{Example of a family of conic sections, all with the same
    planitude (flatness at apex, marked by white dot):
    \(\Pi = R_c/R_0 = 2 \). The quadric angle, \(\thetaQ\), varies over
    the family (see text), with lower values of \(\thetaQ\) giving
    larger values of the alatude, \(\Lambda = R_{90}/R_0 \), meaning more
    open wings.  Different values of \(\Pi\) can be achieved for the
    exact same quadrics by sliding them along the \(x\)-axis, which
    will also change the axis scales since these are normalized by
    \(R_0\).}
  \label{fig:conics-family}
\end{figure}

\section{Quadrics of revolution}
\label{sec:conic}

For an arbitrary surface of revolution, application of
equations~(\ref{eq:tanphi}, \ref{eq:tangential}) to determine the
projected shape of the tangent line is not straightforward and in
general requires numerical techniques.  However, analytical results
can be found for the important class of surfaces known as
\textit{quadrics of revolution} \citep{Goldman:1983a, Gfrerrer:2009a},
which are formed by rotating a conic section plane curve about its
symmetry axis.  Examples are the sphere, spheroids (oblate and
prolate), and right circular paraboloids and hyperboloids.\footnote{We
  consider only the case of a single sheet of a 2-sheet hyperboloid or
  paraboloid, since these are the versions that resemble a bow,
  whereas the 1-sheet versions resemble the waist of an hourglass.}
We ignore the degenerate cases of cylinders, cones, and pairs of
parallel planes.  While mathematically simple, these quadrics are
sufficiently flexible that they can provide a useful approximation to
more complex bow shock shapes.



The shape of the quadric curves in the \(xy\) plane (\(\phi = 0\)) are
shown in Figure~\ref{fig:conics}(a) and (b) for the ellipse and
hyperbola case, respectively.  The conic section itself is fully
described by two lengths, \(a\) and \(b\), which are the
semi-axes.\footnote{Note that we do not require that \(a > b\), so
  either \(a\) or \(b\) may be the semi-major axis.}  However, the
curve can be translated along the \(x\) axis to an arbitrary point
with respect to the star, so that the apex distance \(R_0\) has no
necessary relation to \(a\) or \(b\) and therefore the star/bow
combination requires \emph{three} independent lengths for its
specification.  The displacement \(x_0\) from the star to the
``center'' of the conic section is
\begin{equation}
  \label{eq:conic-x0}
  x_0 = R_0 - \sigma a
  \quad \text{with} \quad
  \sigma = \begin{cases}
    +1 & \text{ellipse}\\
    -1 & \text{hyperbola}
  \end{cases} \ .
\end{equation}
For hyperbolas the center is ``outside'' of the bow and \(x_0\) is
always positive, whereas for ellipses the center is ``inside'' the bow
and \(x_0\) is usually negative, except when \(a < R_0\) (see
Figure~\ref{fig:conics}).

A general parametric form\footnote{%
  The special case of the parabola needs to be treated differently,
  see Appendix~\ref{app:parabola}.} %
for the \(xy\) coordinates of the quadrics (in the \(\phi = 0\) plane,
and with the star at the origin) as a function of \(t = [0, \pi]\) is
then
\begin{gather}
  \label{eq:par-xy}
  \begin{aligned}
    x &= x_0 + \sigma a \Cos(t) \\ 
    y &= b\Sin(t) 
  \end{aligned}
\end{gather}
where
\begin{equation}
  \label{eq:sin-sinh-etc}
  \Sin{}, \Cos = \begin{cases}
    \sin{}, \cos & \text{ellipse}\\
    \sinh{}, \cosh & \text{hyperbola}
  \end{cases}
\end{equation}
Except for the circle case (\(\sigma = +1\), \(a = b\)), the parametric
variable \(t\) is not actually an angle in physical space.  Instead,
the polar form of the bow shape \(R(\theta)\) must be found by substituting
equations~\eqref{eq:par-xy} into \(\theta = \tan^{-1} y/x\) and
\(R = (x^2 + y^2)^{1/2}\).

The type of quadric surface can be characterized by the
\textit{quadric parameter}:
\begin{equation}
  \label{eq:Tq}
  \Q \equiv \sigma \frac{b^2} {a^2} \ , 
\end{equation}
where \(\Q < 0\) corresponds to open surfaces (hyperboloids) and
\(\Q > 0\) corresponds to closed surfaces (oblate spheroids with
\(\Q > 1\) and prolate spheroids with \(\Q < 1\)).  Special cases are
the sphere (\(\Q = 1\)) and the paraboloid (\(\Q = 0\)).
Alternatively, one can define a \textit{quadric angle}:
\begin{equation}
  \label{eq:thetaQ}
  \thetaQ = \sigma \tan^{-1} (b/a) \ ,
\end{equation}
which is marked in Figure~\ref{fig:conics}.  In the case of
hyperboloids, the asymptotic opening angle of the wings
(\S~\ref{sec:plan-alat-bow} and Fig.~\ref{fig:characteristic-radii})
is \(\theta_\infty = \pi + \thetaQ\) (note that \(\thetaQ < 0\) in this case), and
the minimum slope angle is \(\alpha_{\mathrm{min}} = \abs{\thetaQ}\), see
discussion following equation~\eqref{eq:thetapar}.

\begin{figure}
  \includegraphics[width=\linewidth]{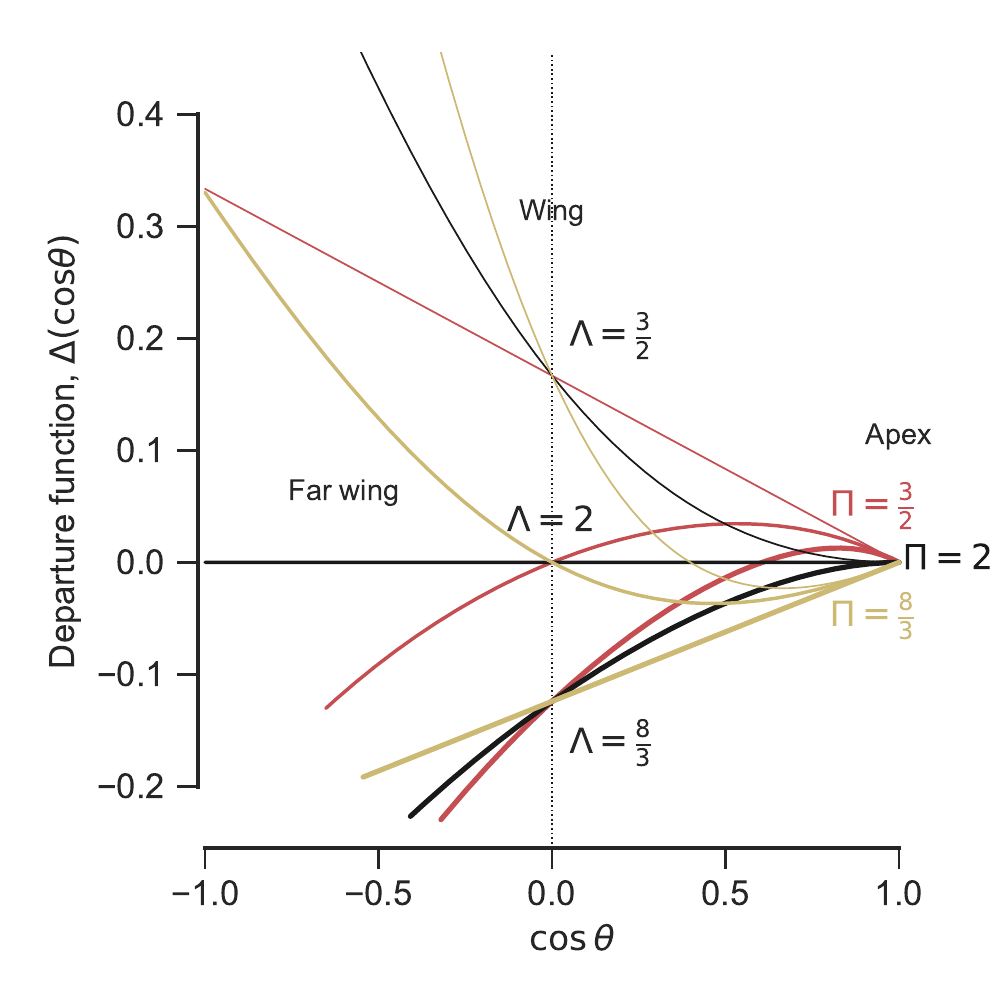}
  \caption[]{Parabolic departure function, \(\Depart(\cos\theta)\), for
    conic sections with different planitude and alatude, centered on
    that of the confocal parabola, \((\Pi, \Lambda) = (2, 2)\), which has
    \(\Depart(\cos\theta) = 0\).  Planitude (shown by different line
    colors) determines the slope of \(\Depart\) at the apex,
    \(\cos\theta = 1\).  Alatude (shown by different line widths)
    determines the value of \(\Depart\) at \(\cos\theta = 0\).  All conics
    with \(\Pi = \Lambda\) have departure functions that are straight lines.}
  \label{fig:conic-departure}
\end{figure}

\begin{figure*}
  \centering
  \setkeys{Gin}{width=0.48\linewidth}
  \begin{tabular}{@{}ll@{}}
    (a) & (b) \\
    \includegraphics{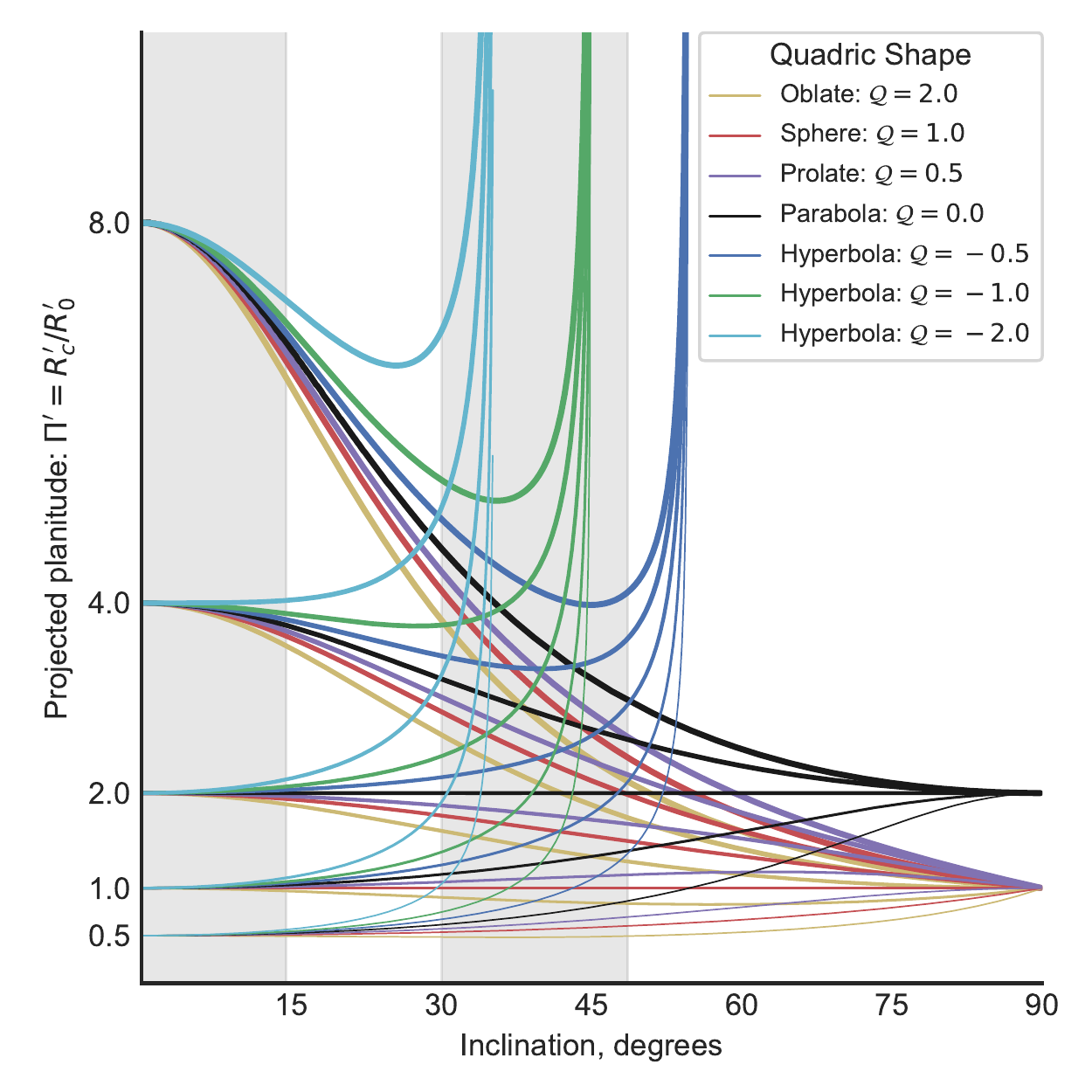}
        & \includegraphics{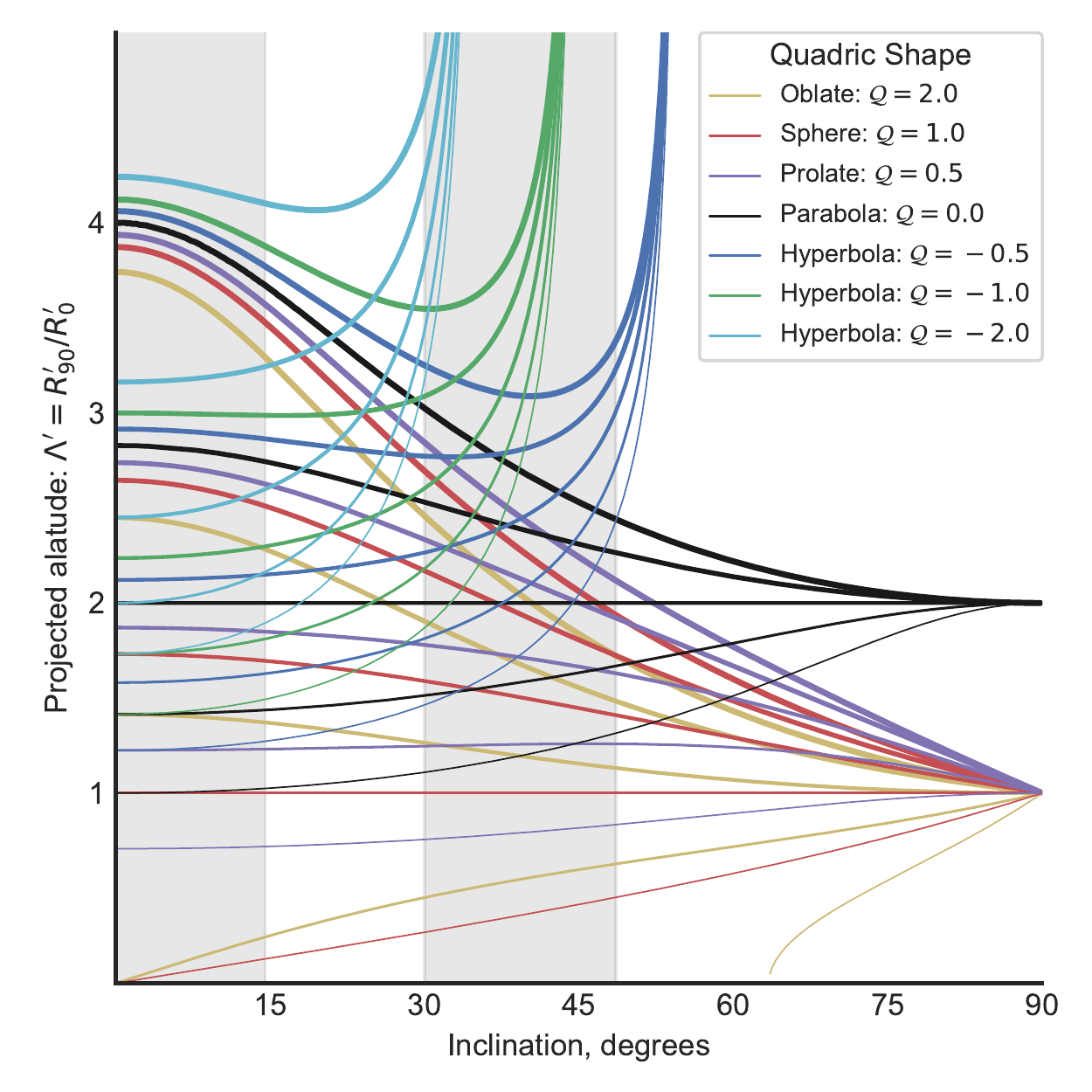}
  \end{tabular}
  \caption[]{Effects of projection on quadrics of revolution:
    variation with inclination, \(\abs{i}\), of bow size and shape.
    Different line colors correspond to varying quadric parameter,
    \(\Q\), (see key), while variation in line width corresponds to
    variation in the ``true'' planitude, \(\Pi\), or apex radius of
    curvature. Vertical gray rectangles show quartiles of
    \(\abs{\sin i}\), which will be equally populated for an isotropic
    distribution of orientations. (a)~Projected planitude:
    \(\Pi'\). (b)~Projected alatude, \(\Lambda'\).}
  \label{fig:quadric-projection}
  \bigskip
  \bigskip
  \begin{tabular}{@{}ll@{}}
    (a) & (b) \\
    \includegraphics{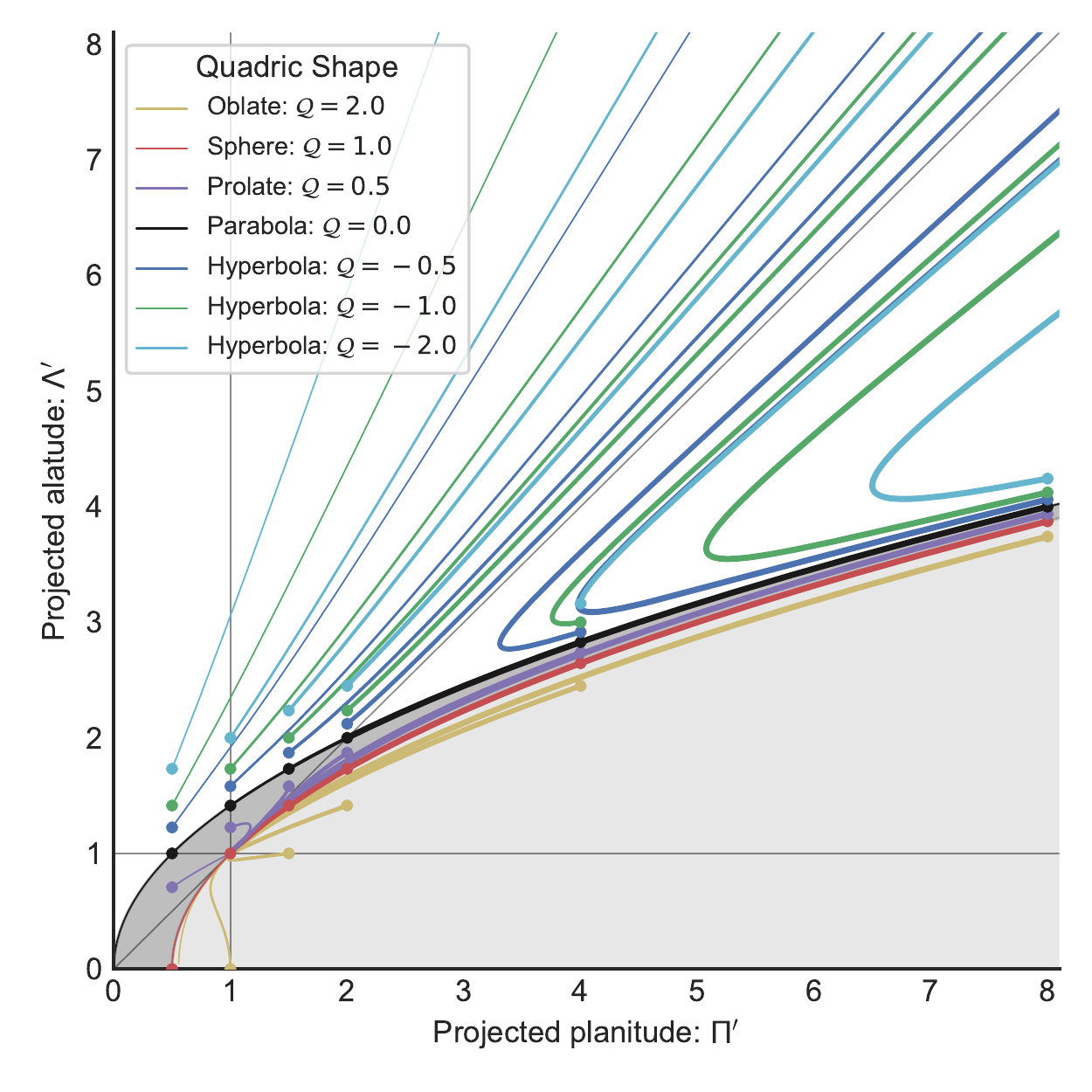}
    & \includegraphics{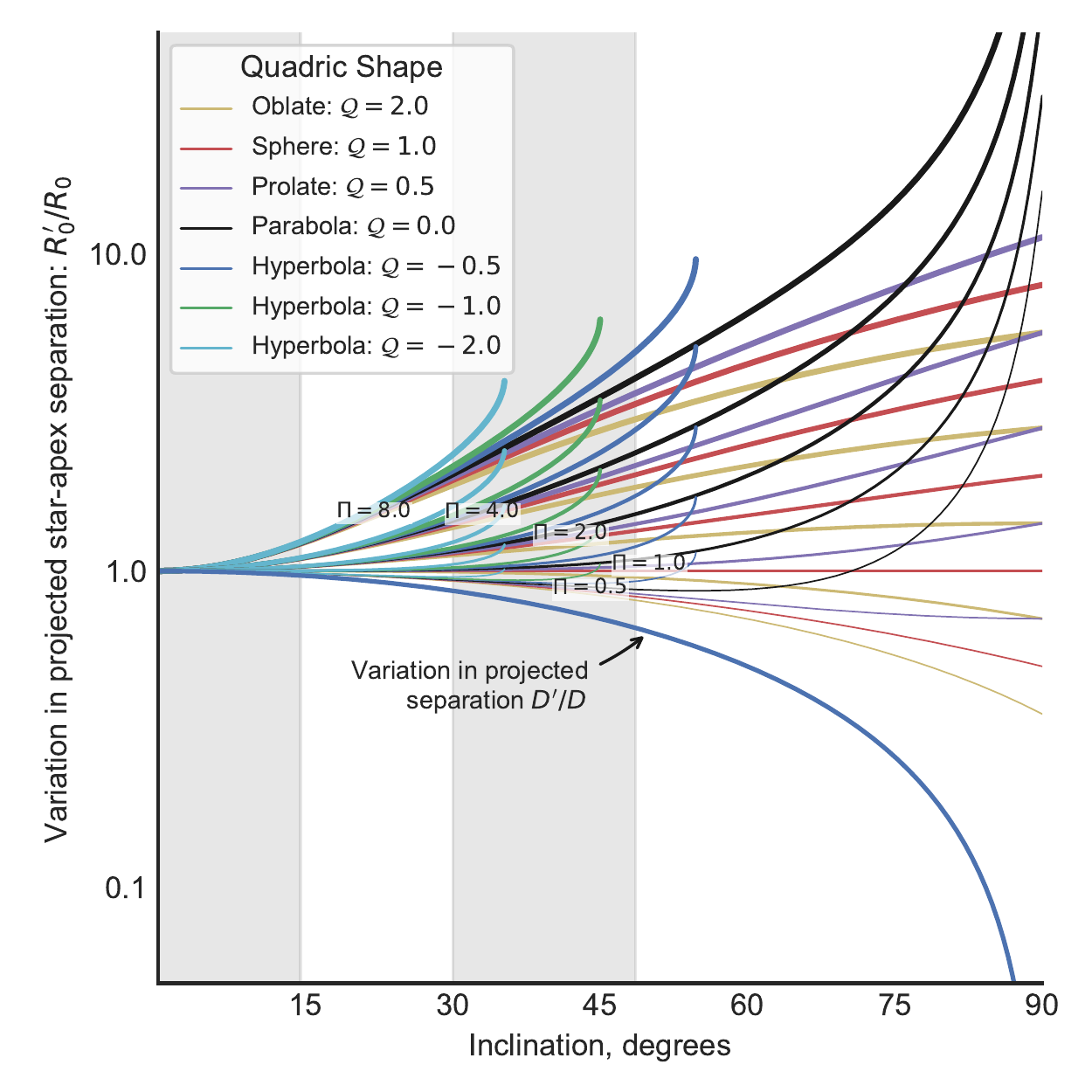}
  \end{tabular}
  \caption[]{ As Figure~\ref{fig:quadric-projection}, but %
    (a)~diagnostic planitude--alatude diagram: \(\Lambda'\) versus
    \(\Pi'\), and %
    (b)~projected/true star-apex distance: \(R_0' / R_0\) versus
    \(\abs{i}\). %
    In (a), shading indicates different classes of quadrics:
    hyperboloids (white), prolate spheroids (dark gray), and oblate
    spheroids (light gray), with the limiting case of paraboloids
    shown by the thin black line.}
  \label{fig:quadric-projection-continued}
\end{figure*}

The set of parameters \((\Q, a, x_0)\) are then sufficient to
characterize the star/bow combination, where \(a\) is the quadric
scale and \(x_0\) is its center displacement from the star.  However,
we can also characterize the star/bow by \((R_0, \Pi, \Lambda)\), where
\(R_0\) is the star-apex distance, and \(\Pi\) and \(\Lambda\) are the
planitude and alatude, see \S~\ref{sec:plan-alat-bow}.  We now derive
the equivalences between these two descriptions.  The apex radius of
curvature for a conic section is
\begin{equation}
  \label{eq:Rc-conic}
  R_c = \frac{b^2}{a} = a \abs{\Q}\ ,
\end{equation}
whereas the perpendicular radius, \(R_{90}\), is the value of \(y\)
when \(x = 0\), which can be found from equations (\ref{eq:conic-x0},
\ref{eq:par-xy}) as
\begin{equation}
  \label{eq:conic-R90}
  R_{90}^2 = \Q \left(a^2 - x_0^2\right) \ .
\end{equation}
Combining equations (\ref{eq:planitude}, \ref{eq:alatude},
\ref{eq:conic-x0}, \ref{eq:Tq}--\ref{eq:conic-R90}) yields
\begin{align}
  \label{eq:R0-from-Q-a-x0}
  R_0 &= x_0 + \sigma a  \\
  \label{eq:Pi-from-Q-a-x0}
  \Pi &= \frac{a \Q}{a + \sigma x_0} \\
  \label{eq:Lambda-from-Q-a-x0}
  \Lambda &= \left( \Q \frac{a - \sigma x_0} {a + \sigma x_0}  \right)^{1/2}
\end{align}
with \(\sigma = \sgn Q\).  It also follows that the quadric parameter in
terms of the planitude and alatude is
\begin{equation}
  \label{eq:Tq-from-Pi-Lambda}
  \Q = 2 \Pi - \Lambda^2 
\end{equation}
Hence, it is the sign of \(2 \Pi - \Lambda^2\) that determines
\(\sigma\) and whether a quadric is a spheroid or a hyperboloid.  For
example, for a constant planitude, \(\Pi\), we can have a family of
different quadric types, with varying alatude, \(\Lambda\), that increases
from oblate, through prolate and paraboloid, to hyperboloid, as
illustrated in Figure~\ref{fig:conics-family}.

\subsection{Parabolic departure function}
\label{sec:parab-depart-funct}

The special case of confocal conic sections
(\(\Lambda = \Pi\)) can be written in polar form as
\begin{equation}
  \label{eq:confocal-polar}
  R(\theta) = R_0 \frac {1 + \ecc} {1 + \ecc \cos \theta}
\end{equation}
where \(\ecc = (1 - \Q)^{1/2}\) is the \textit{conic eccentricity}.
For the confocal parabola (\(\ecc = 1\)), the dimensionless reciprocal
radius is therefore \(R_0/R(\theta) = \tfrac12 (1 + \cos \theta)\), which
suggests the following form for a \textit{departure function} that
measures the difference between a given shape \(R(\theta)\) and the
parabola:
\begin{equation}
  \label{eq:departure-function}
  \Depart\bigl(\cos \theta\bigr) = \frac{R_0} {R(\theta)} - \tfrac12 \left( 1 + \cos \theta \right). 
\end{equation}
From equations~\eqref{eq:confocal-polar} and
\eqref{eq:departure-function} it is clear that \(\Depart\) is a linear
function of \(\cos \theta\) for other confocal conics, being positive for
ellipses (\(\ecc < 1\)) and negative for hyperbolae (\(\ecc > 1\)).
Examples are shown in Figure~\ref{fig:conic-departure} for a grid of 9
conics centered on the confocal parabola, with \((\Pi, \Lambda)\) ranging from
\(3/4\) to \(4/3\) of \((2, 2)\).  The hyperbolae have negative values
of \(\Depart\) in the far wings, with tracks that end at \(\cos \theta_\infty\).

Strictly speaking, the departure function is redundant if one is
interested in only conic sections, since they are fully determined by
\(\Pi\) and \(\Lambda\).  Nonetheless, as we will show in following sections,
it is a useful tool for studying general \(R(\theta)\), being very
sensitive to small variations in the shape.

\subsection{Plane-of-sky projection of quadric surfaces} 

We now apply the machinery of \S~\ref{sec:projection} to find the
projected shape of a quadric bow on the plane of the sky.  The
intrinsic 3D shape of the shell is given by rotating
equations~\eqref{eq:par-xy} about the \(x\)-axis, but it is more
convenient to first transform to a reference frame where the origin is
at the center of the conic section:
\begin{equation}
  \label{eq:xyz-XYZ}
  (X, Y, Z) = (x - x_0, y, z) . 
\end{equation}
In this new frame, the quadric shape is
\begin{gather}
  \label{eq:quadric-XYZ}
  \begin{aligned}
    X &= a\Cos(t) \\ 
    Y &= b \Sin(t)\cos\phi \\
    Z &= b \Sin(t)\sin\phi
  \end{aligned}
\end{gather}
The azimuth of the tangent line as a function of inclination and
parametric variable is then found from equations (\ref{eq:alpha},
\ref{eq:tanphi}) to be
\begin{equation}
  \label{eq:phit-quadric}
  \sin\phi_{\T} = \frac{b \Cos(t)} {a \Sin(t)} \tan i \ .
\end{equation}
Combining equations~(\ref{eq:Trans}, \ref{eq:quadric-XYZ},
\ref{eq:phit-quadric}) gives the observer-frame cartesian
plane-of-sky coordinates of the tangent line:
\begin{gather}
  \label{eq:conic-projected-XY}
  \begin{aligned}
    X_{\T}' & = \frac{\Cos(t)}{a\cos i}
    \left(a^2\cos^2 i + \sigma b^2\sin^2 i\right)
    \\
    Y_{\T}' &= b\Sin(t)
    \left(
      1 - \frac{b^2 \Cos^2(t)}{a^2 \Sin^2(t)}
      \tan^2 i\right)^{1/2}
  \end{aligned}
\end{gather}
We wish to show that this projected shape is a conic section of the
same variety (ellipse or hyperbola) as the one that generated the
original quadric.  If this \emph{were} true, then it would be possible
to write the plane-of-sky coordinates as
\begin{gather}
  \begin{aligned}
    X_{\T}' &= a'\Cos( t')  \\
    Y_{\T}' &= b'\Sin (t')  . 
  \end{aligned}\label{eq:conic-projected-XY-conic}
\end{gather}
Comparing equations (\ref{eq:conic-projected-XY}) and
(\ref{eq:conic-projected-XY-conic}), we find after some algebra that
the two forms for \((X_{\T}', Y_{\T}')\) are indeed consistent, with
the equivalences:
\newcommand\fQi{\ensuremath{f_{\scriptscriptstyle \Q,i}}}
\begin{align}
  \label{eq:a-prime}
  a' &= a \fQi \cos i  \\
  \label{eq:b-prime}
  b' &= b \\
  t' &= \Cos^{-1} \left[ \fQi \Cos(t) \right]  \ ,
\end{align}
where for convenience we define the quadric projection factor:
\begin{equation}
  \label{eq:fQi-factor}
  \fQi = \left( 1 + \Q \tan^2 i \right)^{1/2} \ .
\end{equation}
This demonstrates the original claim that the projected shape is also
a conic section, which means that we can re-use the previous
equations~(\ref{eq:R0-from-Q-a-x0}--\ref{eq:Tq-from-Pi-Lambda}) with
primed quantities substituted for unprimed ones.  From
equations~(\ref{eq:Tq}, \ref{eq:a-prime}, \ref{eq:b-prime}) it follows
that the quadric parameter of the projected shape is
\begin{equation}
  \label{eq:Q-prime}
  \Q' = \frac{ \Q } { \fQi^2 \cos^2 i} \ .
\end{equation}
Finally, we transform the projected reference frame back to be
centered on the star again:
\begin{equation}
  \label{eq:XYZ-xyz-prime}
  (x_{\T}', y_{\T}') = (X_{\T}' + x_0', Y_{\T}') \ , 
\end{equation}
where the projected quadric displacement \(x_0'\) follows from simple
foreshortening:
\begin{equation}
  \label{eq:x0-prime}
  x_0' = x_0 \cos i \ .
\end{equation}


\begin{figure*}
  \centering
  \includegraphics[width=\linewidth]{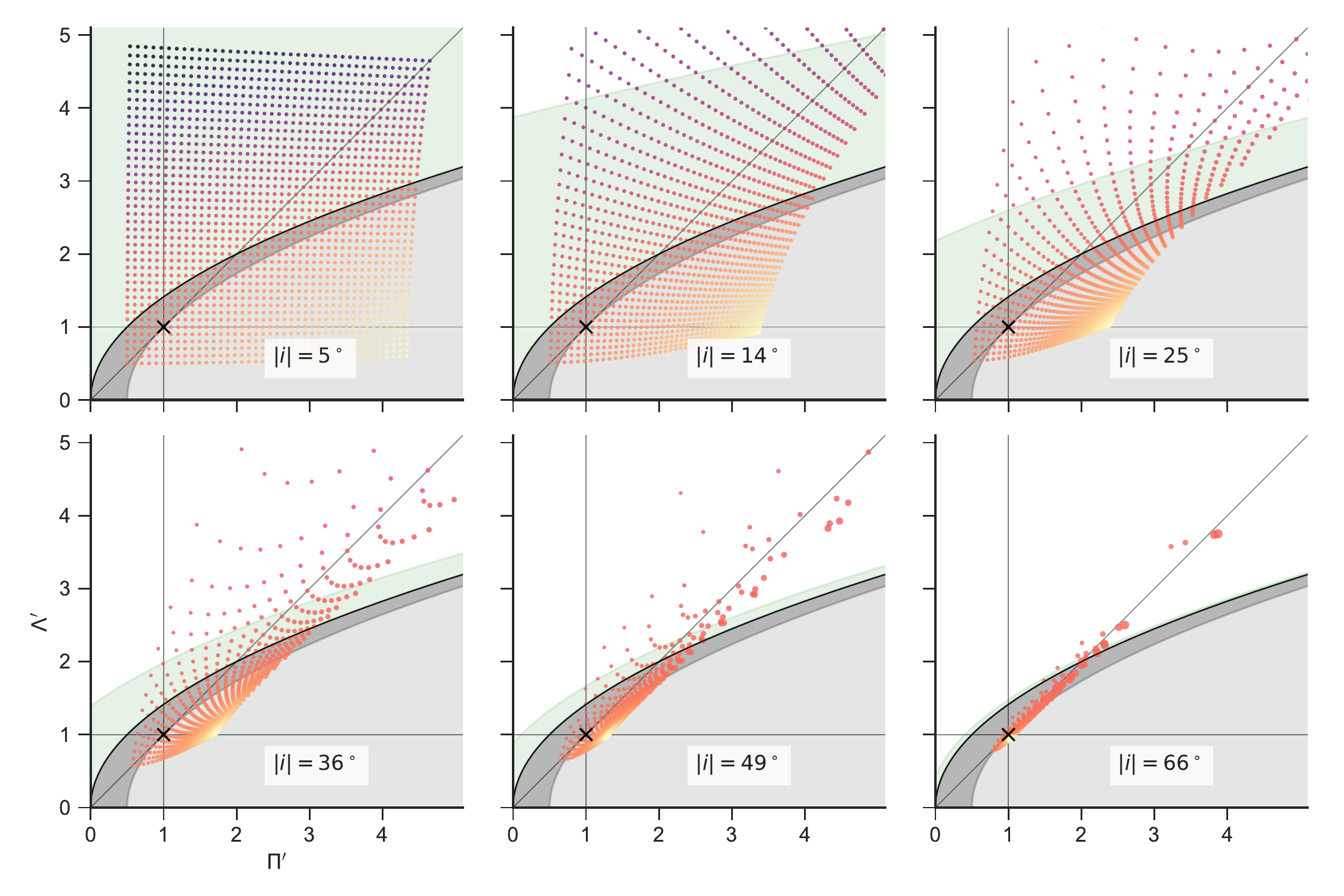}
  \caption[]{Variation with inclination angle of the apparent shape of
    quadric bows with true planitude and alatude that are uniformly
    distributed over the ranges \(\Pi = [0.5, 4.5]\),
    \(\Lambda = [0.5, 4.5]\).  Panels show the apparent
    \((\Pi', \Lambda')\) as the inclination is increased through uniform
    intervals in \(\abs{sin i}\).  Symbol color represents the quadric
    parameter, \(\Q\), increasing from dark blue, through orange, to
    yellow. Symbol size is proportional to the increase in apparent
    star--apex distance, \(R_0'/R_0\).}
  \label{fig:projected-R90-Rc-snapshots}
\end{figure*}

The projection of the apex distance then follows from the primed
version of equation~\eqref{eq:R0-from-Q-a-x0} as
\begin{equation}
  \label{eq:R0-prime}
  \frac{R_0'}{R_0} =
  \cos i \left[ 1 + \frac{\Pi} {\Q} \bigl(\fQi - 1\bigr) \right]
\end{equation}
and the projected planitude and alatude can be calculated from
equations~(\ref{eq:Pi-from-Q-a-x0}, \ref{eq:Tq-from-Pi-Lambda},
\ref{eq:a-prime}, \ref{eq:Q-prime}) as
\begin{align}
  \label{eq:Pi-prime}
  \Pi' &= \frac{\Pi}{(R_0'/R_0)\, \fQi \cos i } \\
  \label{eq:Lambda-prime}
  \Lambda' &= \left( 2 \Pi' - \Q' \right)^{1/2} \ .
\end{align}
These are all shown in Figures~\ref{fig:quadric-projection}
and~\ref{fig:quadric-projection-continued} for a variety of quadric
parameter \(Q\) (line color) and true planitude \(\Pi\) (line
thickness).  The projected planitude and alatude
(Fig.~\ref{fig:quadric-projection}) behave in a qualitatively similar
fashion.  Whatever the true values of \(\Pi\) and \(\Lambda\), all spheroids
(\(\Q > 0\)) tend towards \(\Pi' = 1\) and \(\Lambda' = 1\) as the inclination
increases towards \(\ang{90}\).  This is because when the spheroid is
oriented edge-on, we see its circular cross-section.  Hyperboloids
behave differently: although \(\Pi'\) and \(\Lambda'\) initially decrease with
increasing inclination (for true \(\Pi > 2\)), they turn around and
increase again as \(\abs{i}\) approaches the critical value
\(i_{\mathrm{crit}} = \ang{90} - \abs{\thetaQ}\).  For
\(\abs{i} > i_{\mathrm{crit}}\) the tangent line does not exist (see
\S~\ref{sec:tangent-line}) because the line of sight is ``inside'' the
asymptotic cone of the far wings (with opening half angle
\(\alpha_{\mathrm{min}} = \abs{\thetaQ}\)), and so no limb-brightened shell
would be visible.\footnote{%
  As illustrated in Figure~8 of \citet{Graham:2002a}, the isophotal
  emission contours are elliptical in such a case (assuming
  cylindrical symmetry) and no curved bow shape is apparent.
  Deviations from cylindrical symmetry \emph{can} result in a curved
  emission arc, even for this no-tangent case
  (\citeauthor{Graham:2002a}'s Fig.~9), but that is beyond the scope
  of this paper.} For paraboloids and spheroids,
\(\alpha_{\mathrm{min}} = 0\), which means that the tangent line exists for
all viewing angles.

In Figure~\ref{fig:quadric-projection-continued}a, we show the
inclination-dependent tracks of the quadrics in the diagnostic
\(\Pi'\)--\(\Lambda'\) plane of projected alatude versus projected planitude.
The true planitude and alatude, which are seen for an edge-on viewing
angle \(i = \ang{0}\), are marked by filled circles.  The zones
corresponding to each class of quadric (oblate spheroid, prolate
spheroid, or hyperboloid) are marked by gray shading, and it can be
seen that the tracks never cross from one zone to another. The
convergence of all the spheroid tracks on the point
\((\Pi', \Lambda') = (1, 1)\) is apparent, as is the divergence of the
hyperboloid tracks towards \((\Pi', \Lambda') = (+\infty, +\infty)\), whereas the
paraboloids, by contrast, converge on the point
\((\Pi', \Lambda') = (2, 2)\).  Two special cases are the confocal paraboloid
and the concentric sphere,\footnote{So named because the star is at
  the focus of the parabola, or the center of the sphere.} with true
planitude and alatude \((\Pi, \Lambda) = (2, 2)\) and \((1, 1)\),
respectively, which are the only quadrics whose apparent shape remains
identical for all inclination angles.

Figure~\ref{fig:quadric-projection-continued}b shows how the apparent
star-apex separation varies with inclination.  For moderate
inclinations, \(\abs{i} < \ang{30}\), this depends primarily on the true
planitude \(\Pi\), with very little influence of the quadric parameter
\(\Q\).  For \(\Pi > 1\), the separation increases with \(\abs{i}\), whereas
for \(\Pi < 1\) it decreases slightly.  Note, however, that for cases
where the projected distance to the source of the external flow,
\(D'\), can be measured, then \(R_0'/D'\) is always an increasing
function of \(\abs{i}\).  For larger inclinations, \(\abs{i} > \ang{30}\), the
strands for different \(\Q\) begin to separate, with hyperbolae
showing the strongest increase of \(R_0'\) with~\(\abs{i}\).

A complementary view of the effects of projection is shown in
Figure~\ref{fig:projected-R90-Rc-snapshots}, which shows ``snapshots''
of \((\Pi', \Lambda')\) for a sequence of 6 values of the inclination, equally
spaced in \(\abs{\sin {i}}\), so that each panel is equally likely for
an isotropic distribution of orientations.  The distribution of the
true \(\Pi\) and \(\Lambda\) are each assumed to be uniform on the range
\([0.5, 4.5]\), giving a uniformly filled square of values for
\(\abs{i} = 0\), which becomes increasingly distorted as \(\abs{i}\)
increases.  The color scale represents \(\Q\) and the symbol size is
proportional to \(R_0'/R_0\).  It can be seen that the points tend to
cluster closer and closer to the diagonal, \(\Lambda' = \Pi'\), as the
inclination increases, and that the points just below this line tend
to have the largest values of \(R_0'/R_0\).  The green shaded region
shows the zone of true \(\Lambda, \Pi\) for hyperboloids where the tangent
lines still exists for that value of \(\abs{i}\).  This becomes
smaller and smaller as \(\abs{i}\) increases, which explains why the
hyperboloid zone becomes increasingly depopulated: all quadrics that
lay above this region when \(i = \ang{0}\) will no longer be visible as a
bow for this value of \(\abs{i}\).  Note that this figure is merely
illustrative of the qualitative effects of projection, since in
reality there is no particular reason to expect a uniform distribution
in true \(\Pi\) and \(\Lambda\).






\newcommand\thC{\(\theta^1\)\,Ori~C}
\defcitealias{Canto:1996}{CRW}
\newcommand\CRW{\citetalias{Canto:1996}}

\section{Thin-shell bow shock models}
\label{sec:crw-scenario}

More physically realistic examples of bow shapes are provided by
steady-state hydrodynamic models for the interaction of hypersonic
flows in the thin-shell limit.  The classic examples are the solutions
for the wind--parallel stream and wind--wind problems (see
\S~\ref{sec:intro}) of \citet[][hereafter \CRW{}]{Canto:1996}, where
it is assumed that the two shocks are highly radiative and that the
post-shock flows are perfectly mixed to form a single shell of
negligible thickness. In this approximation, the shape of the shell is
found algebraically by \CRW{} from conservation of linear and angular
momentum, following an approach first outlined in
\citet{Wilkin:1996a}.  For the wind--stream case, the resulting bow
shape was dubbed \textit{wilkinoid} by \citet{Cox:2012a} and has the
form:
\begin{equation}
  \label{eq:wilkinoid-R-theta}
  R(\theta) = R_0\csc\theta\left( 3(1-\theta\cot\theta) \right)^{1/2} \ .
\end{equation}

For the wind--wind case, a family of solutions are found that depend on
the value of \(\beta\), the wind momentum ratio,\footnote{%
  By always placing the weaker of the two winds at the origin, it is
  only necessary to consider \(\beta \le 1\).  } %
see Figure~\ref{fig:crw-schema},
equations~(\ref{eq:stagnation-radius}, \ref{eq:beta-definition}), and
surrounding discussion in \S~\ref{sec:intro}.  We propose that these
shapes be called \textit{cantoids}.  The exact solution for the
cantoid shapes (eqs.~[23, 24] of \CRW{}) is only obtainable in
implicit form, but an approximate explicit solution (eq.~[26] of
\CRW{}) is very accurate for \(\beta \le 0.1\).  The wilkinoid shape
corresponds to the \(\beta \to 0\) limit of the cantoids.  Note that \CRW{}
employ cylindrical polar coordinates, \(z\) and \(r\), see our
Figure~\ref{fig:crw-schema}, and we follow this usage for the
thin-shell models discussed in this section.  \CRW{}'s \(z\) axis
corresponds to the cartesian \(x\) axis used in
sections~\ref{sec:projection} and \ref{sec:conic} of the current
paper, while the \(r\) axis corresponds to \(y\) when \(\phi = 0\).

A generalization of the cantoids to the case of an
anisotropic\footnote{Note that the wind anisotropy axis must be
  aligned with the star--star axis to maintain cylindrical symmetry.}
inner wind is developed next, giving rise
to what we call \textit{ancantoids}, which depend on an anisotropy
index, \(k\), in addition to \(\beta\).  

\subsection{Bow shocks from anisotropic wind--wind interactions}
\label{sec:ancantoid}
\begin{figure}
  \centering
  \includegraphics[width=\linewidth]{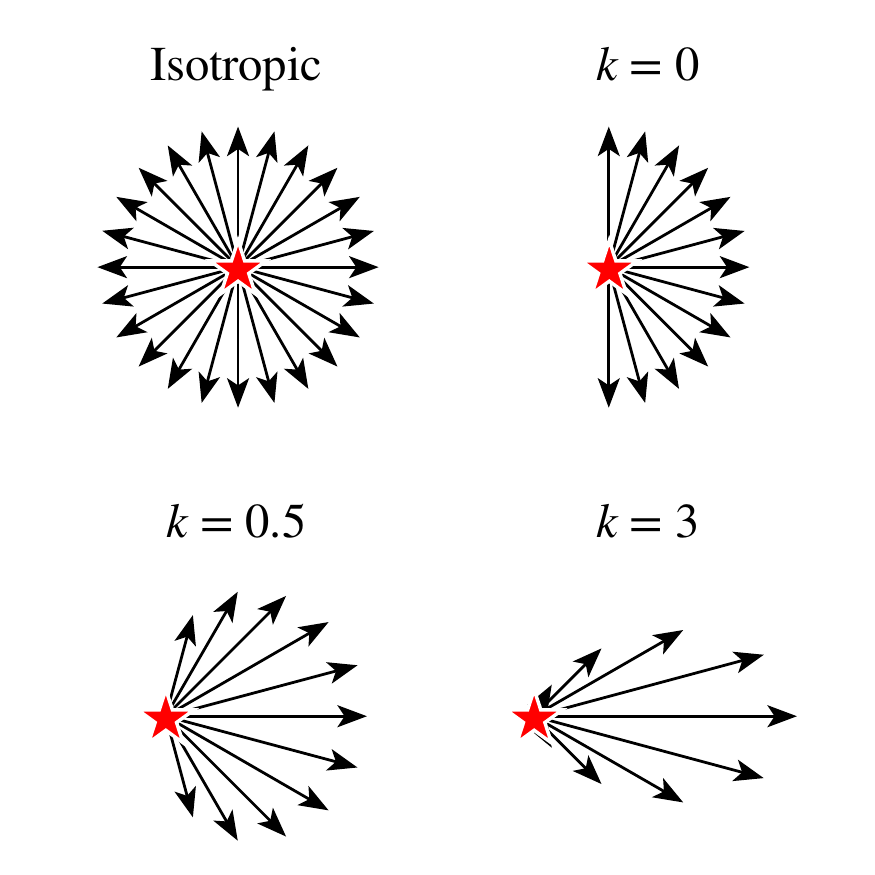}
  \caption[]{Schematic diagram of wind flow patterns in isotropic and
    non-isotropic cases for different values of the anisotropy index,
    \(k\).  Arrow length represents the wind momentum loss rate per
    solid angle.}
  \label{fig:anisotropic-arrows}
\end{figure}

We wish to generalize the results of \citet[\CRW{}]{Canto:1996} to the
case where the inner wind is no longer isotropic, but instead has a
density that falls off with angle, \(\theta\), away from the symmetry axis.
Specifically, at some fiducial spherical radius, \(R_0\), from the
origin, the wind mass density is given by
\begin{equation}
  \label{eq:ancantoid-density}
  \rho(R_0, \theta) =
  \begin{cases}
    \rho_0 \cos^k \theta & \text{for \(\theta \le \ang{90}\)} \\
    0 & \text{for \(\theta > \ang{90}\)}
  \end{cases}
  \ ,
\end{equation}
where \(\rho_0\) is the density on the symmetry axis and \(k \ge 0\) is an
\textit{anisotropy index}.  The wind velocity is still assumed to be
constant and the wind streamlines to be radial, so the radial
variation of density at each angle is
\(\rho(R, \theta) = \rho(R_0, \theta)\, (R/R_0)^{-2}\) and the wind mass loss rate and
momentum loss rate per solid angle both have the same \(\cos^k\theta\)
dependence as the density.  Examples are shown in
Figure~\ref{fig:anisotropic-arrows} for a variety of different values
of \(k\).  As \(k\) increases, the wind becomes increasingly jet-like.

Our primary motivation for considering such an anisotropic wind is the
case of the Orion Nebula proplyds and their interaction with the
stellar wind of the massive star \thC{}
\citep{Garcia-Arredondo:2001a}.  The inner ``wind'' in this case is
the transonic photoevaporation flow away from a roughly hemispherical
ionization front, where photoionization equilibrium, together with
monodirectional illumination of the front, implies that the ionized
hydrogen density, \(n\), satisfies \(n^2 \propto \cos \theta\), which is
equivalent to \(k = 0.5\) in equation~\eqref{eq:ancantoid-density}.
Since the primary source of ionizing photons is the same star that is
the source of the outer wind, it is natural that the inner wind's axis
should be aligned with the star--star axis in this case.  For other
potential causes of wind anisotropy (for instance, bipolar flow from
an accretion disk), there is no particular reason for the axes to be
aligned, so cylindrical symmetry would be broken.  Nevertheless, we
calculate results for general \(k\) with aligned axes, so as to
provide a richer variety of cylindrically symmetric bow shock shapes
than are seen in the cantoids.

\begin{figure}
\includegraphics[width=\linewidth]{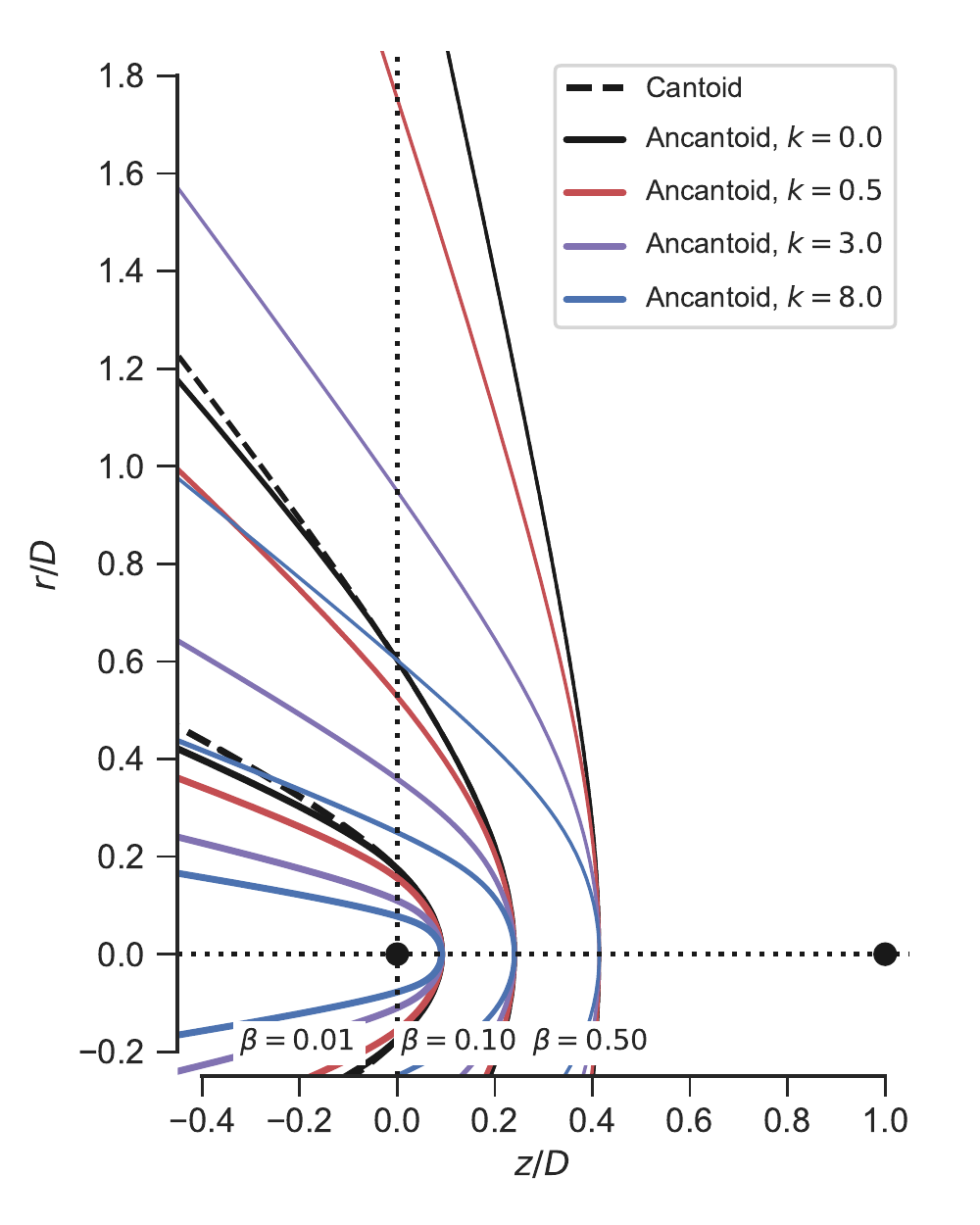}
\caption{Bow shock shapes for interacting winds in the thin-shell
  approximation: cantoids and ancantoids. Coordinates are normalized
  by $D$, the distance between the two wind sources, which are
  indicated by black dots on the axis.  The weaker source is at
  \((0.0, 0.0)\) and the stronger source is at \((1.0, 0.0)\).
  Results are shown for different values of the wind momentum ratio,
  \(\beta\) (different line widths), and for the case where the weaker
  wind is isotropic (black lines) or anisotropic (colored lines).}
\label{fig:r-beta}
\end{figure}

The general solution for the bow shock shape, \(R(\theta)\), in the \CRW{}
formalism is
\begin{equation}
  R(\theta) = \frac {\dot{J}_{\w} + \dot{J}_{\w{}1}}
  {\left(\dot{\Pi}_{\w{}r}+\dot{\Pi}_{\w{}r1}\right)\cos\theta
    - \left(\dot{\Pi}_{\w{}z}+\dot{\Pi}_{\w{}z1}\right)\sin\theta}
  \label{eq:Rmom}
\end{equation}
where \(\dot{\Pi}_{\w{}r}\), \(\dot{\Pi}_{\w{}z}\), \(\dot{J}_{\w}\) are
the accumulated linear radial momentum, linear axial momentum, and
angular momentum, respectively, due to the inner wind emitted between
the axis and \(\theta\). The equivalent quantities for the outer wind have
subscripts appended with ``1''.  The inner wind momenta for our
anisotropic case (replacing \CRW{}'s eqs.~[9, 10]) are:
\begin{gather}
  \label{eq:ancantoid-momenta}
  \begin{aligned}
    \dot{\Pi}_{\w{}z} &= \frac{k + 1}{2(k+2)}\, \dot{M}_{\w}^0 V_{\w}
    \max\left[\bigl(1- \cos^{k+2} \theta\bigr), 1 \right] \\
    \dot{\Pi}_{\w{}r} &= (k + 1)\, \dot{M}^0_{\w} V_{\w}\, I_k (\theta) 
  \end{aligned}
\end{gather}
where
\begin{equation}
  \label{eq:ancantoid-mass-loss}
  \dot{M}^0_{\w} = \frac{2 \pi} {k + 1} r_0^2 \rho_0 V_{\w}
\end{equation}
is the total mass-loss rate of the inner wind. The integral
\begin{equation}
  \label{eq:ancantoid-I-integral}
  I_k(\theta) = \int^{\max(\theta, \pi/2)}_0 \cos^k \theta \sin^2\theta \,d\theta 
\end{equation}
has an analytic solution in terms of the hypergeometric function,
\({}_2 F_1(-\tfrac12; \tfrac{1+k}2; \tfrac{3+k}2; \cos^2 \theta)\), but is
more straightforwardly calculated by numerical quadrature.  The
angular momentum of the inner wind about the origin is
\(\dot{J}_{\w} = 0\) because it is purely radial.  The outer wind
momenta are unchanged from the \CRW{} case, but are given here for
completeness:
\begin{gather}
  \label{eq:ancantoid-momenta-outer}
  \begin{aligned}
    \dot{\Pi}_{\w{}z1} & =
    -\frac{\dot{M}^0_{\w{}1}V_{\w{}1}}{4}\sin^2\theta_1\\
    \dot{\Pi}_{\w{}r1} & =
    \frac{\dot{M}^0_{\w{}1}V_{\w{}1}}{4}\left(\theta_1-\sin\theta_1\cos\theta_1\right)\\
    \dot{J}_{\w{}1} & =
    \frac{\dot{M}^0_{\w{}1}V_{\w{}1}}{4}\left(\theta_1-\sin\theta_1\cos\theta_1\right)D \ .
  \end{aligned} 
\end{gather}

\begin{figure}
  \centering
  \includegraphics[width=\linewidth]{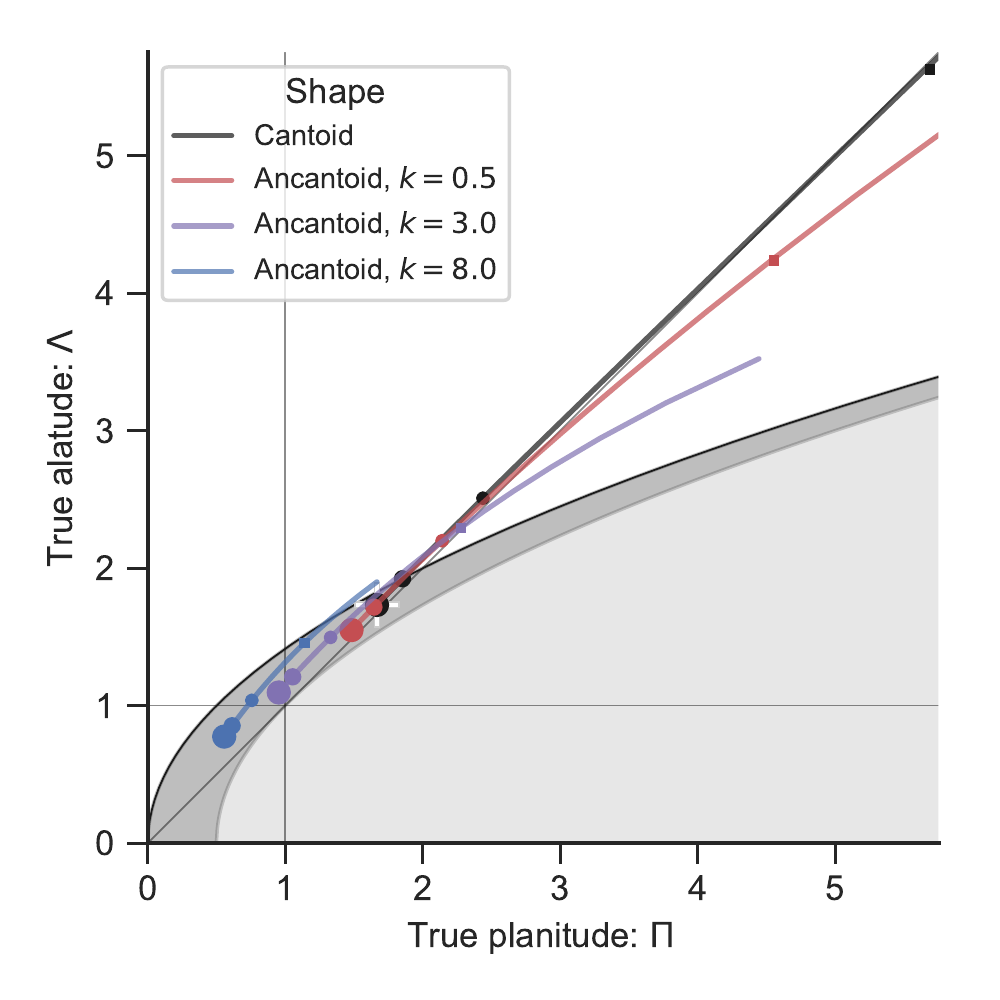}
  \caption[]{True shapes of cantoids and ancantoids in the
    \(\Pi\)--\(\Lambda\) plane, calculated according to results of
    App.~\ref{sec:thin-shell-shapes}.  For each line, \(\beta\) varies
    over the range \([0, 1]\) from lower left to upper right (although
    the black and red lines are truncated on the upper right), and
    line colors correspond to different anisotropy indices, matching
    those used in Fig.~\ref{fig:r-beta}. Circle symbols mark
    particular \(\beta\) values: \(0, 0.01, 0.1\), from largest to
    smallest circle.  Square symbols mark \(\beta = 0.5\), but with
    \(\Lambda\) calculated exactly, instead of using the approximation of
    equation~\eqref{eq:Lambda-approx}.  The white plus symbol marks
    the result for the wilkinoid:
    \((\Pi, \Lambda) = (\frac53, \sqrt{3})\).  Background shading indicates
    the domains of different quadric classes: hyperboloids (white),
    prolate spheroids (dark gray), and oblate spheroids (light gray).}
  \label{fig:ancantoid-Pi-lambda-true}
\end{figure}

We define \(\beta\) in this case as the momentum ratio \emph{on the symmetry axis}, which means that 
\begin{equation}
  \label{eq:ancantoid-momentum-ratio}
  \dot{M}^0_{\w{}1}V_{\w{}1} = 2 (k + 1)\, \beta\, \dot{M}^0_{\w} V_{\w} \ . 
\end{equation}
Substituting
equations~(\ref{eq:ancantoid-momenta}--\ref{eq:ancantoid-momentum-ratio})
into equation~\eqref{eq:Rmom} and making use of the geometric relation
between the interior angles of the triangle shown in
Figure~\ref{fig:crw-schema}:
\begin{equation}
  \label{eq:crw-angles}
  R \sin(\theta + \theta_1) = D \sin \theta_1 \ , 
\end{equation}
yields
\begin{equation}
  \label{eq:ancantoid-theta-theta1-implicit}
  \theta_1 \cot \theta_1 = 1 +
  2 \beta \left(
    I_k(\theta) \cot \theta
    - \frac{1 - \cos^{k+2} \theta} {k + 2} \right)   \ , 
\end{equation}
which is the generalization of \CRW{}'s equation~(24) to the
anisotropic case.  Equation~\eqref{eq:ancantoid-theta-theta1-implicit}
is solved numerically to give \(\theta_1(\theta)\), which is then combined with
equations~(\ref{eq:crw-angles}) and (\ref{eq:stagnation-radius}) to
give the dimensionless bow shape, \(R(\theta; \beta, k)/R_0\), where we now
explicitly indicate the dependence of the solution on two parameters:
axial momentum ratio, \(\beta\), and anisotropy index, \(k\).  We refer to
the resultant bow shapes as \textit{ancantoids}.

\begin{figure}
  \centering
  \includegraphics[width=\linewidth]{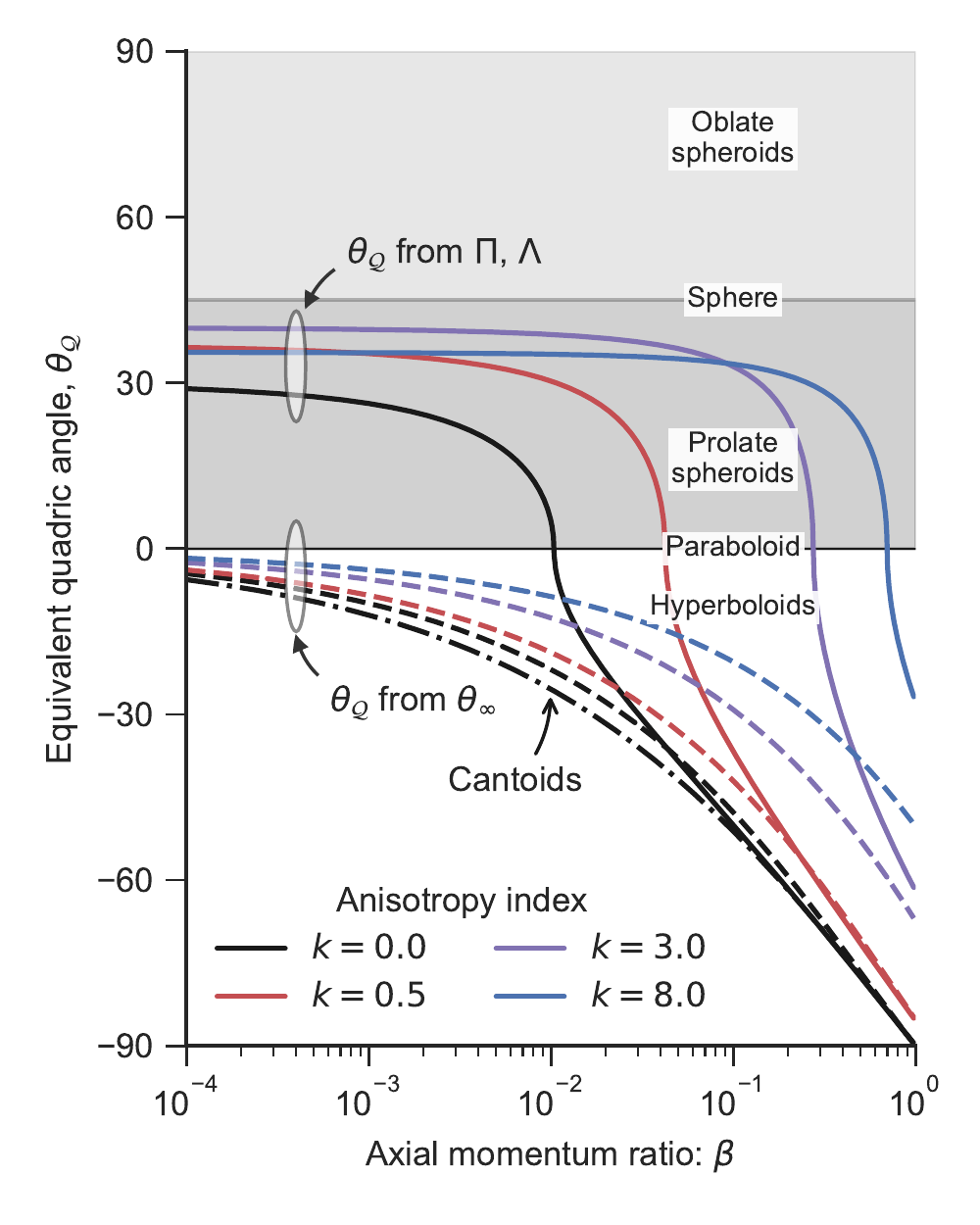}
  \caption[]{Equivalent quadric angles, \(\theta_{\Q}\), for ancantoids and
    cantoids.  Solid lines show values of \(\theta_{\Q}\) calculated from
    \((\Pi, \Lambda)\), which is representative of the shape of the head,
    while dashed lines show \(\theta_{\Q}\) calculated from
    \(\theta_\infty\), which is representative of the tail.  Dot-dashed line
    shows the result for cantoids, which differ from the \(k=0\)
    ancantoids in \(\theta_\infty\), but not in \((\Pi, \Lambda)\). Gray shading and
    line colors have the same meaning as in
    Fig.~\ref{fig:ancantoid-Pi-lambda-true}. }
  \label{fig:ancantoid-angles}
\end{figure}

\begin{figure}
  \centering
  \setkeys{Gin}{width=\linewidth}
  \begin{tabular}{@{}l@{}}
    (a) \\
    \includegraphics{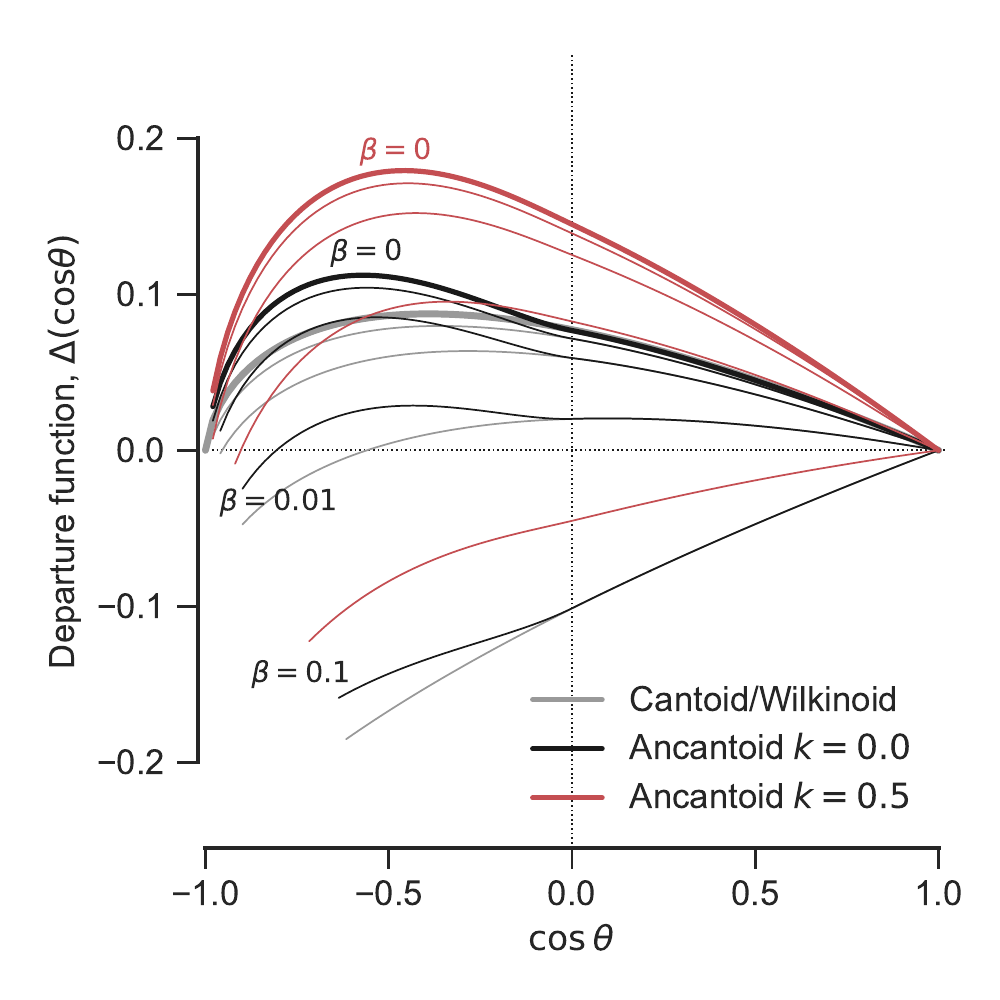} \\
    (b) \\
    \includegraphics{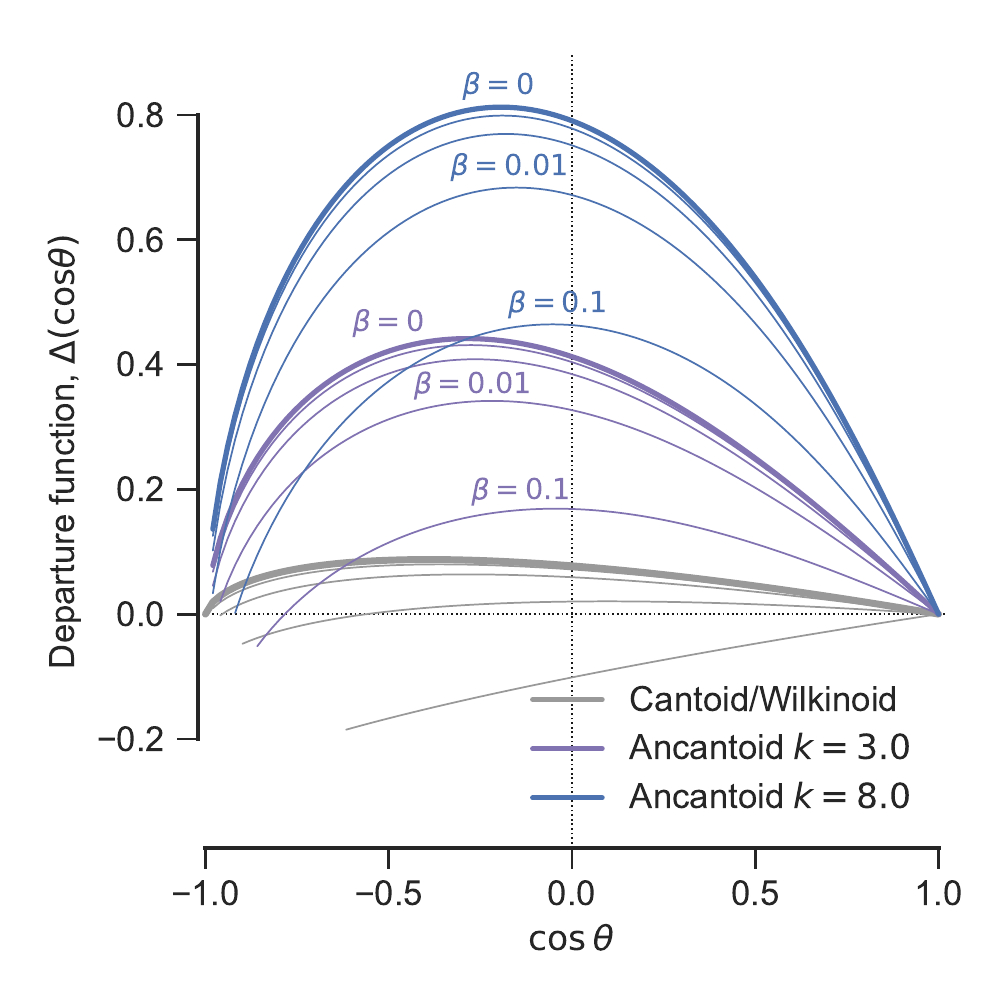}
  \end{tabular}
  \caption[]{Parabolic departure function, \(\Depart(\cos\theta)\), for
    ancantoids and cantoids.  Heavy lines show the \(\beta = 0\) parallel
    stream case (Wilkinoid in the isotropic case).  Light lines show
    increasing values of \(\beta = \num{e-4}\), \num{0.001}, \num{0.01},
    \num{0.1}, as marked.  (a)~Cantoids (gray) and moderately
    anisotropic ancantoids: hemispheric, \(k = 0\) (black), and
    proplyd-like, \(k = 0.5\) (red). (b)~Cantoids (gray) and extremely
    anisotropic, jet-like ancantoids: \(k = 3\) (purple) and \(k = 8\)
    (blue).}
  \label{fig:ancantoid-departure}
\end{figure}

\begin{figure}
  \centering
  \includegraphics[width=\linewidth]{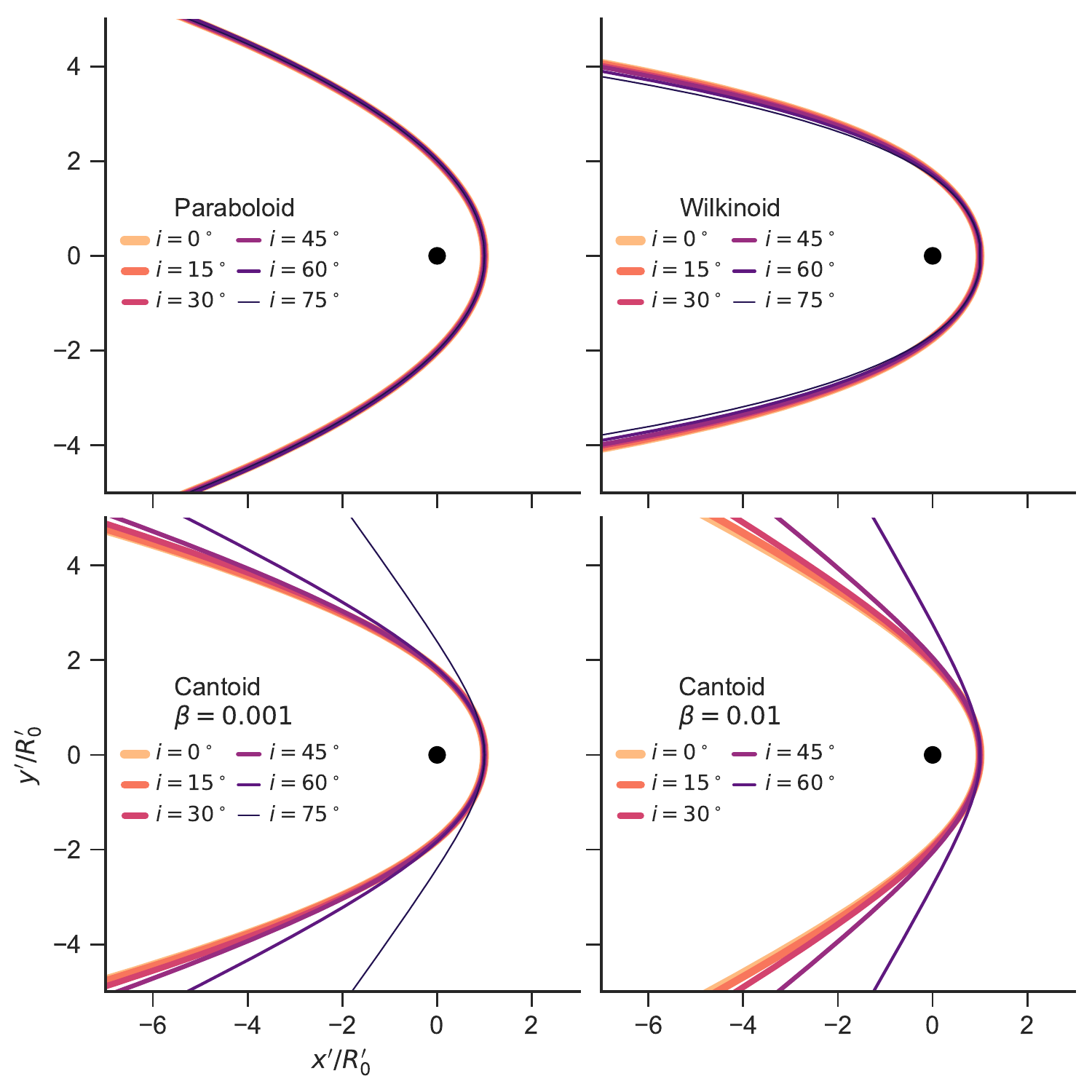}
  \caption{Apparent bow shapes as a function of inclination angle for
    isotropic thin shell models. (a)~Confocal paraboloid for
    comparison (shape independent of inclination).
    (b)~Wilkinoid. (c)~Cantoid, \(\beta = 0.001\). (d)~Cantoid,
    \(\beta = 0.01\). }
  \label{fig:xyprime}
\end{figure}

\begin{figure}
  \centering
  \includegraphics[width=\linewidth]{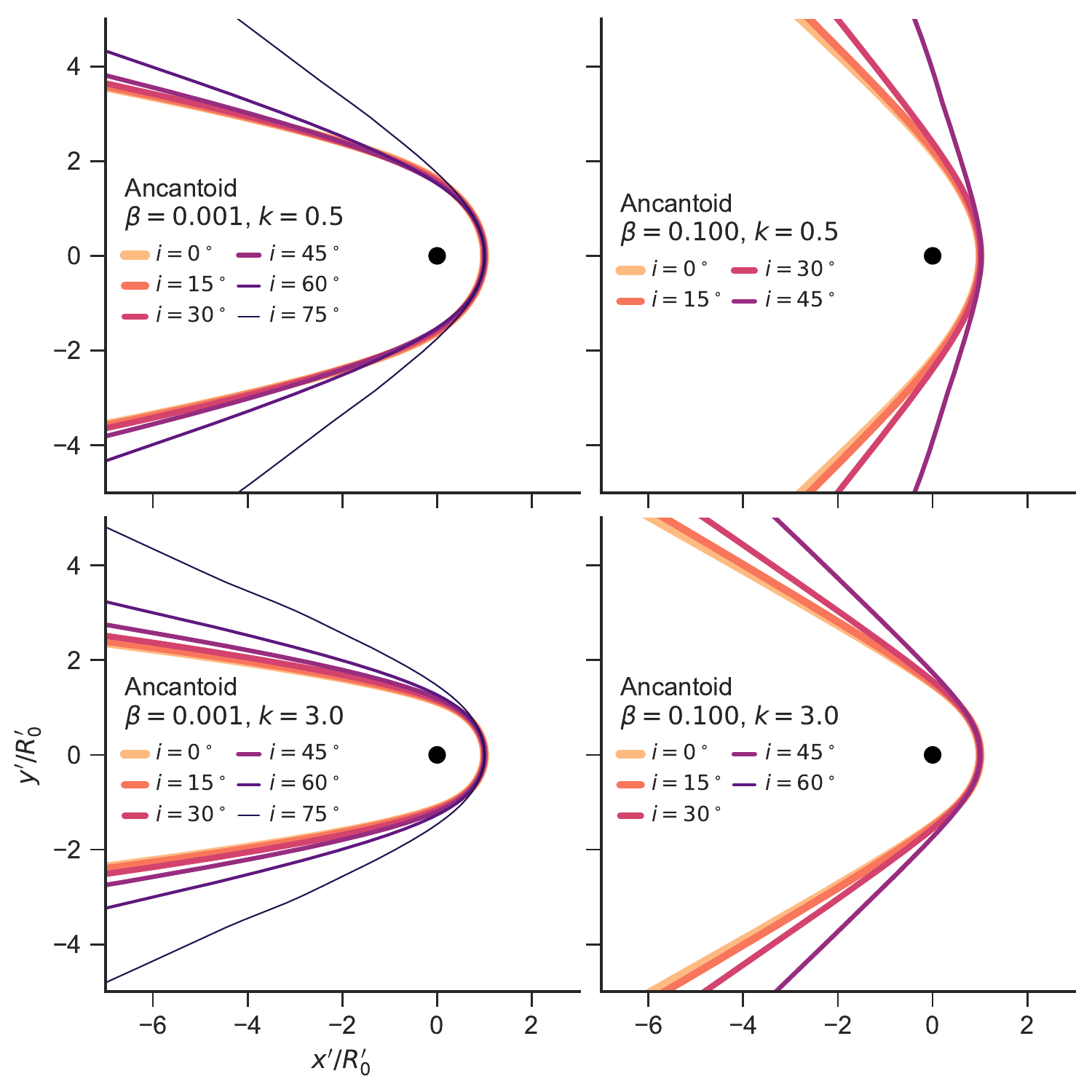}
  \caption{Further apparent bow shapes as a function of inclination
    angle for anisotropic thin shell models (ancantoids).
    (a)~\(\beta = 0.001\), \(k = 0.5\); (b) (a)~\(\beta = 0.1\),
    \(k = 0.5\); (c)~\(\beta = 0.001\) \(k = 3\); (d)
    \(\beta = 0.1\) \(k = 3\).}
  \label{fig:xyprime-ancantoid}
\end{figure}

\subsection{True shapes of cantoids and ancantoids}
\label{sec:true-cantoids-ancantoids}

The shapes of the ancantoid bow shocks are shown in
Figure~\ref{fig:r-beta} for three different values of \(\beta\), and are
compared with the \CRW{} results for cantoids (dashed curves).  The
location of these shapes in the planitude--alatude plane is shown in
Figure~\ref{fig:ancantoid-Pi-lambda-true}, where the gray background
shading indicates the zones of different quadric classes, as in
\S~\ref{sec:conic}, Figures~\ref{fig:quadric-projection-continued}
and~\ref{fig:projected-R90-Rc-snapshots}.  Values of \(\Pi\) and
\(\Lambda\) are calculated via the analytic expressions derived in
Appendix~\ref{sec:ancantoid-planitude} and
\ref{sec:ancantoid-alatude}, respectively, which are only approximate
in the case of \(\Lambda\).  However, the filled square symbols show the
exact results for \(\beta = 0.5\), which can be seen to lie extremely close
to the approximate results, even for the worst case of \(k = 0\). The
leading term in the relative error of
equation~\eqref{eq:Lambda-approx} scales as \((\beta / (k + 2))^2\), so
the approximation is even better for smaller \(\beta\) and larger \(k\).

It is apparent from Figure~\ref{fig:r-beta} that the \(k=0\) ancantoid
is identical to the cantoid for \(\theta \le \ang{90}\) (\(z > 0\), to the right
of vertical dotted line in Fig.~\ref{fig:r-beta}), but is slightly
more swept back in the far wings.\footnote{%
  \label{fn:discontinuity}
  Due to the discontinuity in the inner wind density at
  \(\theta = \ang{90}\) (see Fig.~\ref{fig:anisotropic-arrows}), there is a
  discontinuity in the second derivative of the bow shape.} %
Since the true planitude and alatude depend on \(R(\theta)\) only in the
range \(\theta = [0, \ang{90}]\), the cantoid and the \(k = 0\) ancantoid
behave identically in Figure~\ref{fig:ancantoid-Pi-lambda-true}.
There is a general tendency for the bows to be flatter and more open
with increasing \(\beta\) and decreasing \(k\), with the cantoid being
most open at a given \(\beta\).  All the models cluster close to the
diagonal \(\Lambda \simeq \Pi\) in the planitude--alatude plane, but with a tendency
for \(\Lambda > \Pi\) at higher anisotropy.  There is therefore a degeneracy
between \(\beta\) and \(k\) for higher values of \(\beta\).  The wilkinoid
shape, which corresponds to the \(\beta \to 0\) limit of the cantoids, is
marked by a white plus symbol in
Figure~\ref{fig:ancantoid-Pi-lambda-true}, and lies in the prolate
spheroid region of the plane.  Cantoids lie either in the prolate
spheroid or hyperboloid regions, according to whether \(\beta\) is less
than or greater than about \(0.01\).  For ancantoids of increasing
\(k\), this dividing point moves to higher values of \(\beta\), until
almost the entire range of models with \(k = 8\) are within the
prolate spheroid zone.

However, the true planitude and alatude, which are what would be
observed for a side-on viewing angle (\(i = 0\)), are not at all
sensitive to the behavior of the far wings of the bow shock, which has
a rather different implication as to which variety of quadric best
approximates each shape.  We illustrate this is
Figure~\ref{fig:ancantoid-angles}, which shows two different ways of
estimating the quadric angle, \(\theta_{\Q}\) (see \S~\ref{sec:conic}).
The first is from \((\Pi, \Lambda)\), as in
Figure~\ref{fig:ancantoid-Pi-lambda-true}:
\newcommand\head{^{\text{head}}}
\newcommand\tail{^{\text{tail}}}
\begin{equation}
  \label{eq:thetaQ-head}
  \theta_{\Q}\head =
  \sgn{\bigl(2 \Pi - \Lambda^2\bigr)} \;
  \tan^{-1} \Abs{2\Pi - \Lambda^2}^{1/2} \ ,
\end{equation}
which follows from equations~\eqref{eq:Tq}, \eqref{eq:thetaQ}, and
\eqref{eq:Tq-from-Pi-Lambda}.  The second is from the asymptotic
opening angle of the wings, \(\theta_\infty\) (Fig.~\ref{fig:characteristic-radii}):
\begin{equation}
  \label{eq:thetaQ-tail}
  \theta_{\Q}\tail = \theta_\infty - \ang{180} \ , 
\end{equation}
where \(\theta_\infty\) is calculated from
equation~\eqref{eq:ancantoid-theta-inf} for ancantoids, or
\eqref{eq:cantoid-theta-inf} for cantoids.  If the bow shock shape
were truly a quadric, then these two definitions would agree.
However, as seen in Figure~\ref{fig:ancantoid-angles}, this is not the
case for the cantoids and ancantoids.  While
\(\smash[b]{\theta_{\Q}\head}\) generally corresponds to a prolate spheroid
(except for the largest values of \(\beta\)),
\(\smash[b]{\theta_{\Q}\tail}\) always corresponds to a hyperbola.  This
tension between the shape of the head and the shape of the far wings
has important implications for the projected shapes (as we will see in
the next section), since the far wings influence the projected
planitude and alatude when the inclination is large.

Figure~\ref{fig:ancantoid-departure} shows the parabolic departure
function (see \S~\ref{sec:parab-depart-funct}) for the thin-shell
models. This provides an alternative perspective on the resultant bow
shapes, with two different types of behavior being apparent.  Models
with high \(\beta\) and low anisotropy behave similarly to the
hyperboloids, such as the
\((\Pi, \Lambda) = (\nicefrac32, \nicefrac83)\), \((2, \nicefrac83)\),
\((\nicefrac83, \nicefrac83)\), and \((\nicefrac32, 2)\) cases from
Figure~\ref{fig:conic-departure}.  This is the case for the
\(\beta \ge 0.01\) models in Figure~\ref{fig:ancantoid-departure}a, which
all show departure functions that become negative in the far wings
(more open than parabola) and terminate at a
\(\theta_\infty < \ang{180}\).  The second type of behavior is shown by models with
low \(\beta\) or high anisotropy, which behave like spheroids for positive
and mildly negative values of \(\cos \theta\), but, unlike the spheroids,
all tend towards \(\Depart = 0\) in the far tail as \(\cos\theta \to -1\).

\begin{figure}
  \centering
  \setkeys{Gin}{width=\linewidth}
  \begin{tabular}{@{}l@{}}
    (a) \\
    \includegraphics{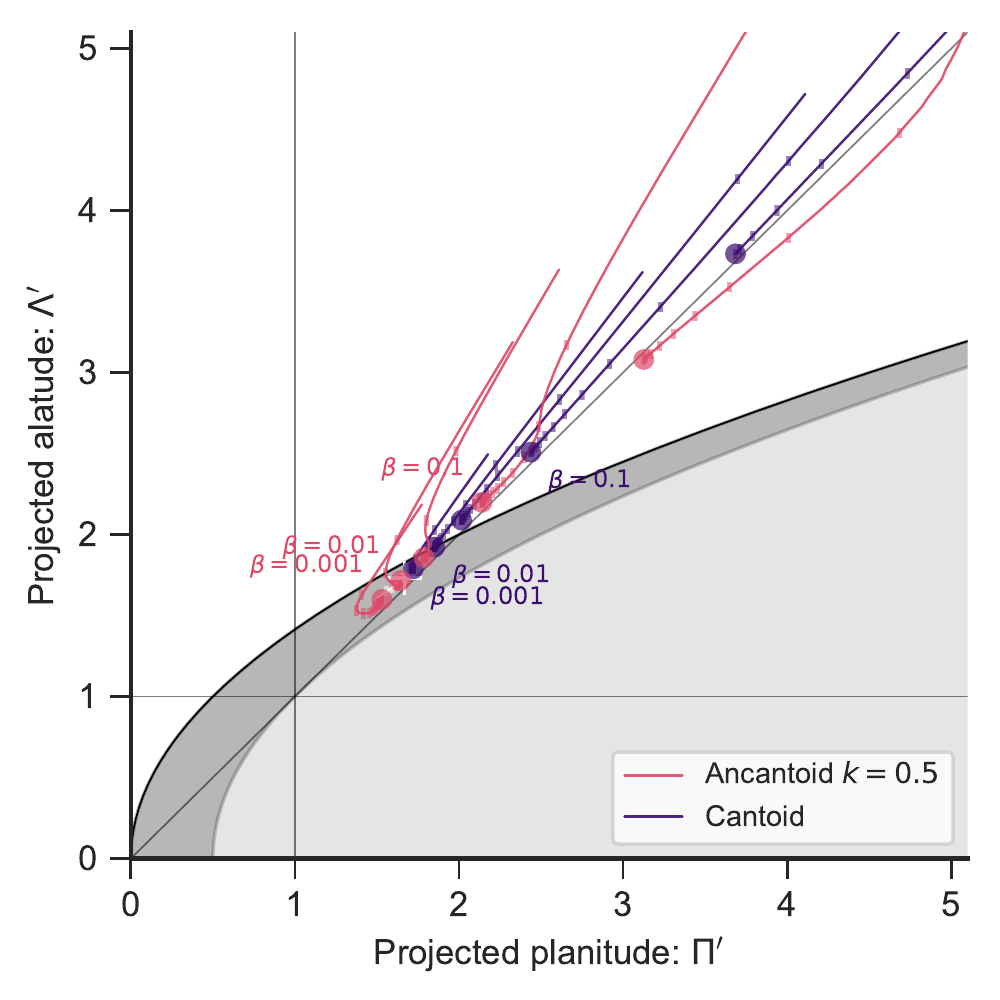} \\
    (b) \\
    \includegraphics{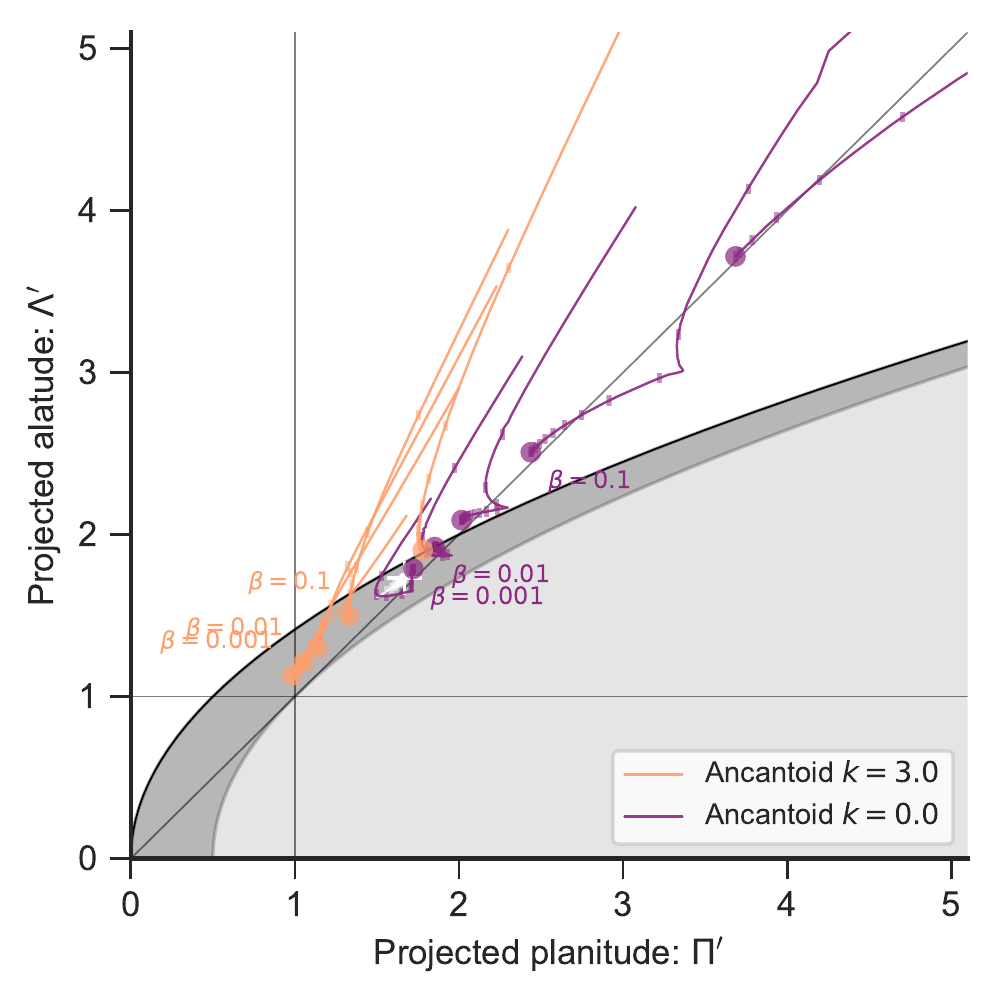}
  \end{tabular}
  \caption[]{Apparent projected shapes of wilkinoid, cantoids and
    ancantoids in the \(\Pi'\)--\(\Lambda'\) plane.  Colored symbols indicate
    the \(\abs{i} = 0\) position for \(\beta = 0.001\), \(0.003\),
    \(0.01\), \(0.03\), \(0.1\), \(0.3\).  Thin lines show the
    inclination-dependent tracks of each model, with tick marks along
    each track for 20 equal-spaced values of \(\abs{\sin i}\).  Gray
    shaded regions are as in
    Fig.~\ref{fig:quadric-projection-continued}a.  The wilkinoid track
    is shown in white. (a)~Isotropic wind model (cantoid) and
    proplyd-like model (ancantoid, \(k = 0.5\)). (b)~Hemispheric wind
    model (ancantoid, \(k = 0\)) and jet-like model (ancantoid,
    \(k = 3\)).}
  \label{fig:thin-shell-R90-Rc}
\end{figure}

\subsection{Apparent shapes of projected cantoids and ancantoids}
\label{sec:proj-shap-cant}

Figures~\ref{fig:xyprime} and~\ref{fig:xyprime-ancantoid} show the
apparent bow shapes of various thin shell models (wilkinoid, cantoids,
ancantoids)\footnote{%
  See also previous studies of the projected shape of the wilkinoid
  \citep{Wilkin:1997a, Cox:2012a, Ng:2017a} and the cantoids
  \citep{Robberto:2005a}.} %
for different inclination angles \(\abs{i}\).  For comparison,
Figure~\ref{fig:xyprime}a shows the confocal paraboloid, whose
apparent shape is independent of inclination (see
Appendix~\ref{app:parabola}).  The wilkinoid (Fig.~\ref{fig:xyprime}b)
shows only subtle changes, with the wings becoming slightly more swept
back as the inclination increases.  The cantoids
(Fig.~\ref{fig:xyprime}c and d) behave in the opposite way, with the
wings becoming markedly more open once \(\abs{i}\) exceeds
\(\ang{60}\) (for \(\beta = 0.001\)), or \(\ang{45}\) (for
\(\beta = 0.01\)).  The ancantoids (Fig.~\ref{fig:xyprime-ancantoid}) can
show more complex behavior.  For instance, in the \(k = 0.5\),
\(\beta = 0.001\) ancantoid (Fig.~\ref{fig:xyprime-ancantoid}a) the near
wings begin to become more closed with increasing inclination up to
\(\abs{i} = \ang{60}\), at which point they open up again, whereas the
opening angle of the far wings increases monotonically with
\(\abs{i}\).

\begin{figure}
  \centering
  \includegraphics[width=\linewidth]{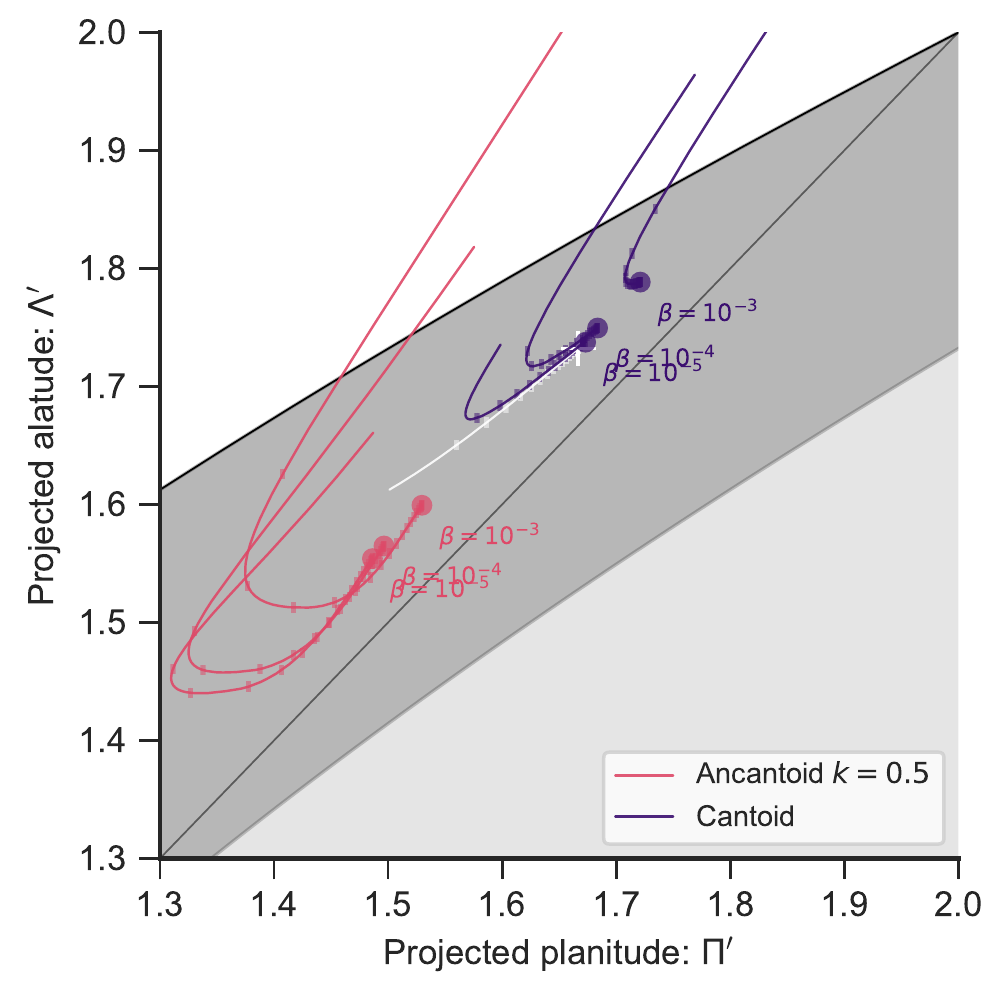}
  \caption[]{As Fig.~\ref{fig:thin-shell-R90-Rc}a but zoomed in to show
    the wilkinoid track (white) and the convergence of the cantoid
    tracks (purple) to the wilkinoid as \(\beta \to 0\).}
  \label{fig:convergence-cantoid-wilkinoid}
\end{figure}

The inclination-dependent tracks that are traced by the thin-shell
models in the projected planitude--alatude plane are shown in
Figure~\ref{fig:thin-shell-R90-Rc}.  The behavior is qualitatively
different from the quadric shapes shown in
Figure~\ref{fig:quadric-projection-continued}a in that the tracks are
no longer confined within the borders of the region of a single type
of quadric (hyperboloid or spheroid). At low inclinations, many of the
models behave like the prolate spheroids, but then transition to a
hyperboloid behavior at higher inclinations, which is due to the
tension between the shape of the head and the shape of the far wings,
as discussed in the previous section. This can be seen most clearly in
the \(\beta = 0.001\), \(k = 0.5\) ancantoid (lowest red line in
Fig.~\ref{fig:thin-shell-R90-Rc}a, see also zoomed version in
Fig.~\ref{fig:convergence-cantoid-wilkinoid}). The track begins
heading towards \((\Pi', \Gamma') = (1, 1)\), as expected for a spheroid, but
then turns around and crosses the paraboloid line to head out on a
hyperboloid-like track.

Ancantoids with different degrees of inner-wind anisotropy are shown
in Figure~\ref{fig:thin-shell-R90-Rc}b.  In all cases, the tracks
follow hyperboloid-like behavior at high inclinations, tending to
populate the region just above the diagonal \(\Lambda' = \Pi'\).  The
\(k = 0\) ancantoids show a kink in their tracks at the point where
the projected apex passes through \(\theta = \ang{90}\), due to the
discontinuity in the second derivative of \(R(\theta)\) there (see
footnote~\ref{fn:discontinuity}).  The wilkinoid has a much less
interesting track, most clearly seen in the zoomed
Figure~\ref{fig:convergence-cantoid-wilkinoid}, simply moving the
short distance from \((\nicefrac53, \sqrt3)\) to
\((\nicefrac32, \smash{\sqrt{\nicefrac83}})\).  Despite its location
in the ellipsoid region of the plane, the fact that it has
\(\theta_\infty = \ang{180}\) means that it behaves more like a parabola at high
inclination, but converges on
\((\nicefrac32, \smash{\sqrt{\nicefrac83}})\) instead of \((2, 2)\)
since the far wings are asymptotically cubic, rather than quadratic.

The local density of tick marks gives an indication of how likely it
would be to observe each portion of the track, assuming an isotropic
distribution of viewing angles.  It can be seen that the ticks tend to
be concentrated towards the beginning of each track, near the
\(\abs{i} = 0\) point, so the hyperboloid-like portions of the tracks
would be observed for only a relatively narrow range of inclinations.
This concentration becomes more marked as \(\beta\) becomes smaller, which
helps to resolve the apparent paradox that the wilkinoid corresponds
to the \(\beta \to 0\) limit of the cantoids, and yet follows a
qualitatively different track.  The detailed behavior of the
small-\(\beta\) cantoid models is shown in
Figure~\ref{fig:convergence-cantoid-wilkinoid}, which zooms in on the
region around the wilkinoid track.  It can be seen that for
\(\beta < 0.001\) the cantoid tracks begin to develop a downward hook,
similar to the \(k = 0.5\) ancantoids discussed above.  For
\(\beta < 10^{-4}\) this begins to approach the wilkinoid track and the
high inclination, upward portion of the track becomes less and less
important as \(\beta\) decreases.



\section{More realistic bow shock models}
\label{sec:more-realistic-bow}

\begin{figure}
  \centering
  \includegraphics[width=\linewidth]{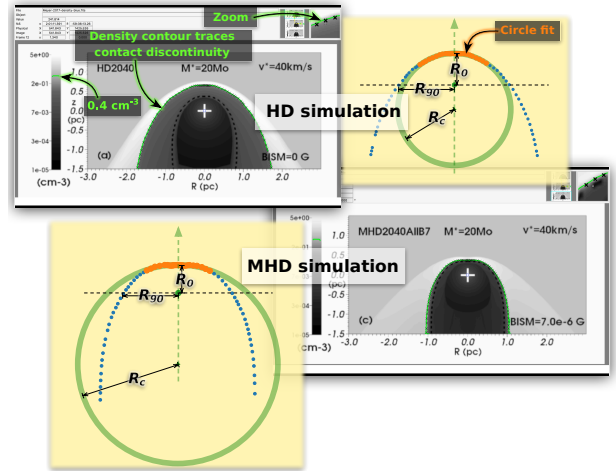}
  \caption[]{Procedure for tracing the contact discontinuity from the
    \citet{Meyer:2017a} simulations.  The density maps from
    \citeauthor{Meyer:2017a}'s Fig.~3 are converted to FITS format and
    displayed using the software SAOImage~DS9 \citep{Joye:2003a}.  The
    density contour at \SI{0.4}{cm^{-3}} is displayed (shown in green
    in the figure) and this is traced by hand by placing ``point
    regions'' on the image (shown by black ``x'' shapes in the
    figure).  The zoom facility of the software allows the points to
    be placed with any required accuracy.  The points are saved to a
    file in the DS9 region file format, which is then read by Python
    programs for further processing.  For example, the yellow boxes
    show circle fits and determination of the parameters \(R_0\),
    \(R_c\), and \(R_{90}\).  In this example, only the points shown
    in orange (within \ang{60} of axis) are used in the fits.}
  \label{fig:meyer-trace}
\end{figure}

\begin{figure}
  (a)\\
  \includegraphics[width=\linewidth]
  {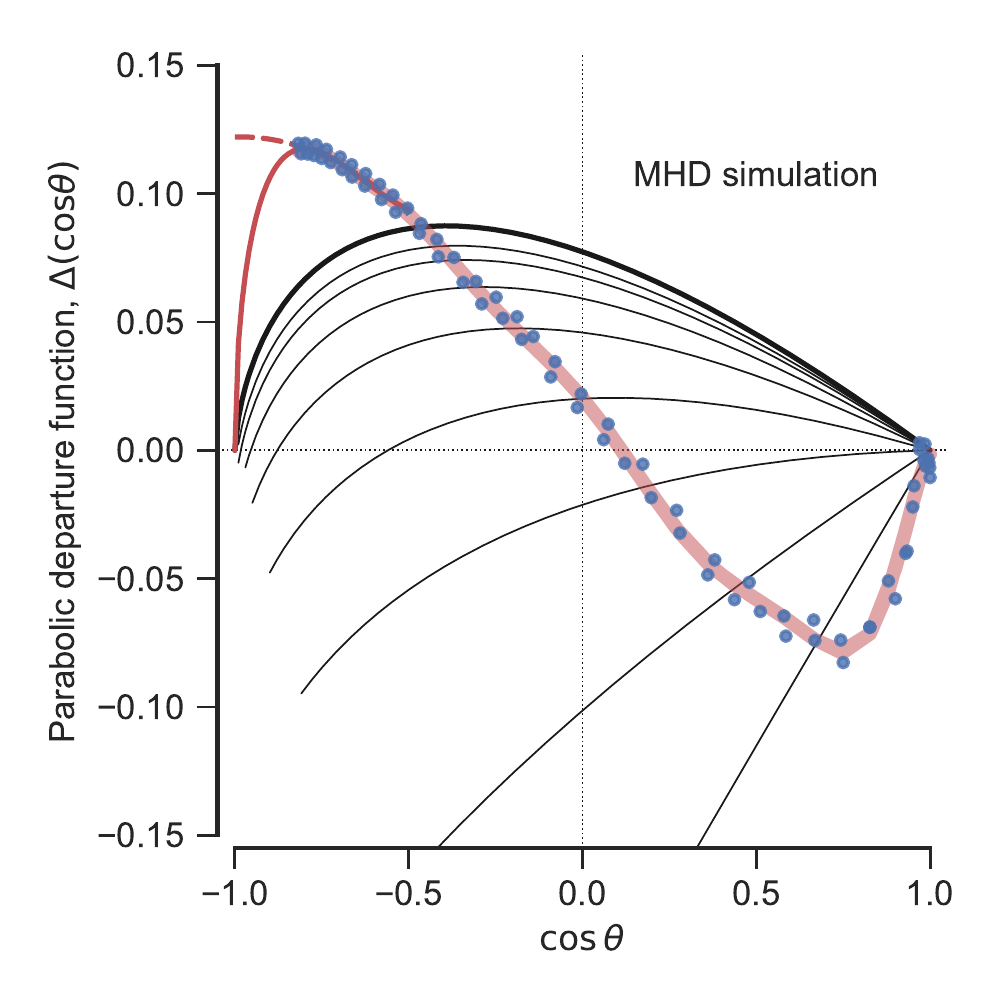}\\[-\baselineskip]
  (b)\\
  \includegraphics[width=\linewidth]
  {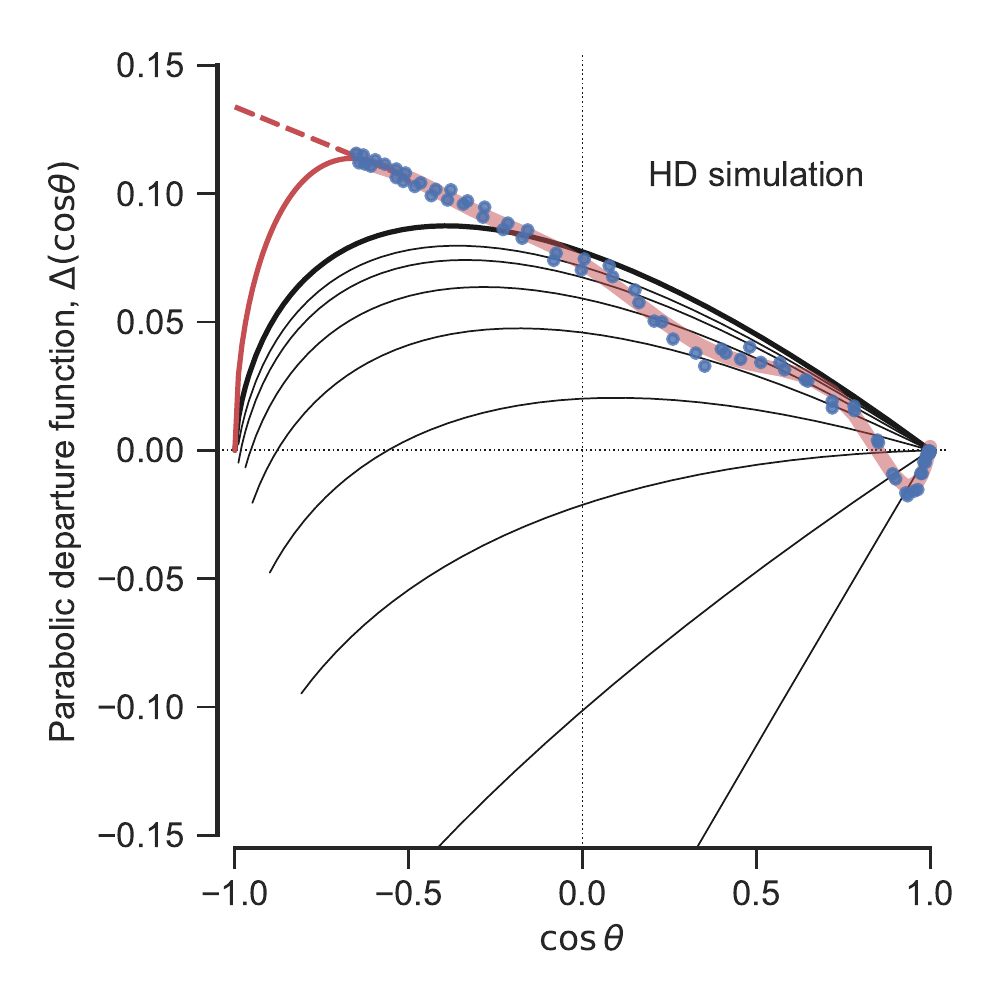}\\[-\baselineskip]
  \caption[]{Departure function for the shape of the contact
    discontinuity, measured from two numerical simulations of a
    \SI{20}{M_\odot} main-sequence star, moving at \SI{40}{km.s^{-1}}
    through a uniform medium of density \SI{0.57}{cm^{-3}}
    \citep{Meyer:2017a}. (a)~Magnetohydrodynamic simulation with
    ambient magnetic field of strength \SI{7}{\micro G}, oriented
    parallel to the stellar velocity. (b)~Hydrodynamic simulation with
    zero magnetic field.  Blue dots show the measured shape, while the
    thick, pale-red line shows a 12th-order Chebyshev polynomial fit.
    The published shapes only extend to
    \(\theta \approx 130\)--\ang{150}, so we extrapolate the shapes out to
    \(\theta = \ang{180}\). Two different extrapolations are shown,
    corresponding to bows that are asymptotically closed (dashed red
    line) or open (solid red line).  For comparison, black lines show
    the departure function for wilkinoid (thick line) and cantoids
    (thin lines).}
  \label{fig:sim-depart}
\end{figure}

The assumptions underlying the models of the previous section may
break down in various ways.  To test whether the planitude--alatude
analysis is still useful in less ``ideal'' situations, we here apply
it to more realistic simulations of stellar bow shocks.  We choose a
pair of hydromagnetic (HD) and magnetohydrodynamic (MHD) moving-star
simulations from \citet{Meyer:2017a}, in which the only difference is
the presence (MHD case) or absence (HD case) of an ambient magnetic
field with strength \(B = \SI{7}{\micro G}\), oriented parallel to the
stellar velocity.  In each case, the inner wind comes from a
\(20\,M_\odot\) main-sequence star, with mass loss rate and terminal
velocity that are roughly constant with time at
\(\dot{M}_{\w} \approx \SI{4e-7}{M_\odot.y^{-1}}\) and
\(V_{\w} \approx \SI{1200}{km.s^{-1}}\), while the outer wind is a
parallel stream due to the star's own motion at \SI{40}{km.s^{-1}}
through a uniform medium of density \SI{0.57}{cm^{-3}}.

For these parameters, the radiative cooling distance for shocked
ambient gas in the bow is a significant fraction (\(\approx 10\%\)) of the
bow shock size, \(R_0\), tending to increase towards the wings, and
the radiative cooling in the shocked stellar wind is even less
efficient.  This represents a significant violation of the assumptions
behind the thin-shell models, since the total shocked shell thickness
is of the same order as \(R_0\).  Nevertheless, the emissivity of
several observationally important emission processes, such as
mid-infrared thermal dust emission and the optical H\(\alpha\) emission
line, is concentrated near the contact discontinuity,\footnote{%
  Note that, in the non-magnetic HD models, efficient thermal
  conduction leads to a thick layer of hot, thermally evaporated
  ambient material that separates the shocked stellar wind from the
  cool, dense shell of shocked ambient gas \citep[see \S~3.3
  of][]{Meyer:2014b}.  In this case, the contact discontinuity is
  taken to be the boundary between hot and cold ambient gas, as
  opposed to the \textit{material discontinuity} between shocked
  ambient gas and shocked wind gas.  In the MHD models, the thermal
  conduction is almost completely suppressed, so that the material and
  contact discontinuities coincide. } %
so it is reasonable to use this surface as a first approximation for
the shape of the bow.

We have traced the contact discontinuity in the two models, using the
procedure outlined in Figure~\ref{fig:meyer-trace}, and show results
for the parabolic departure function (see
\S~\ref{sec:parab-depart-funct}) as blue symbols in
Figure~\ref{fig:sim-depart}.  The MHD simulation shows a strongly
negative dip in the departure function close to the apex
(\(\cos \theta = 1\)), indicating a very flat shape.\footnote{%
  \citet{Meyer:2017a} speculate that this flatness may be the
  signature of the formation of a complex multiple-shock topology at
  the apex \citep{de-Sterck:1999a}.  For our purposes, the reason does
  not matter, merely that the magnetic and non-magnetic models predict
  markedly different shapes. } %
The HD simulation shows only a small negative dip in the departure
function at the apex, but otherwise approximately follows the
wilkinoid curve in the forward hemisphere.  In both cases the
departure function is more positive than the wilkinoid in the far
wings (\(\cos \theta < -0.5\)), but we do not have data for the full range
of \(\theta\), and so two different extrapolations for
\(\theta \to \ang{180}\) are shown.  In the first (dashed red line in
figure), we fit a low-order polynomial of \(\cos \theta\) to the points
with \(\cos \theta < -0.5\) and extend it to \(\cos \theta = -1\), which gives
an asymptotically closed shape.  In the second extrapolation (solid
red line in figure), we fit a polynomial that is multiplied by
\((1 + \cos\theta)^{1/2}\), which forces the departure function to zero at
\(\cos \theta = -1\), giving an asymptotically open shape, as with the
wilkinoid.  In a true steady state, the far wings should be
asymptotically open, but as \(\theta \to \ang{180}\) the flow times become
longer and longer, so that a bow shock with a finite age will be
closed.

\begin{figure}
  \centering
  \includegraphics[width=\linewidth]{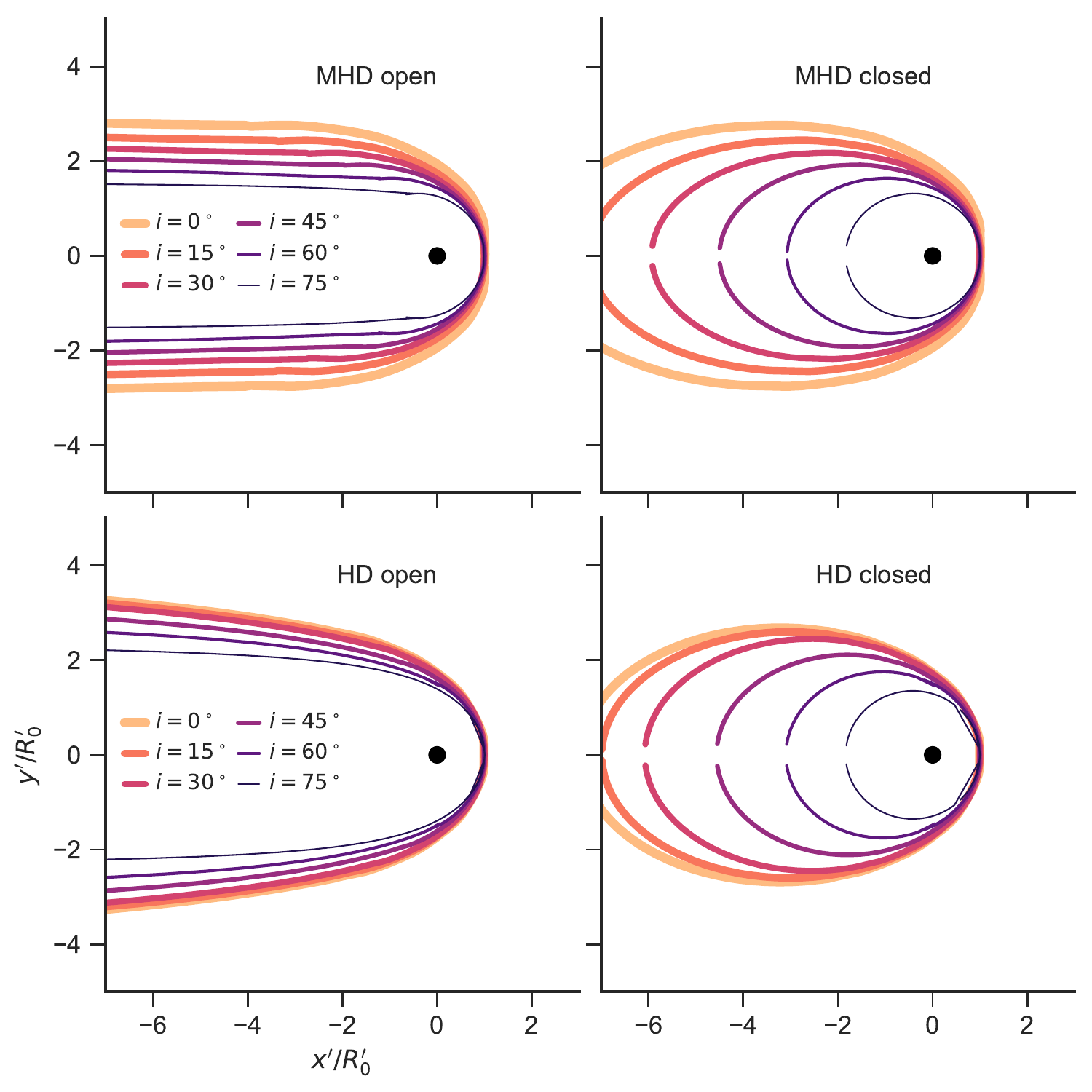}
  \caption[]{Projected shapes of contact discontinuity from
    simulations at different inclinations \(\abs{i}\) (varying line
    color and thickness, see key).  Top row shows magnetized
    simulation of Fig.~\ref{fig:sim-depart}a, bottom row shows
    non-magnetized simulation of Fig.~\ref{fig:sim-depart}b.  Left
    column shows asymptotically open extrapolation, right column shows
    asymptotically closed extrapolation.  All shapes are normalized to
    the projected apex distance, \(R_0'\) }
  \label{fig:sim-xyp}
\end{figure}

Using a 12th-order Chebyshev fit to the traced shapes, we show the
apparent shape of the contact discontinuity at a series of inclination
angles, \(\abs{i}\), in Figure~\ref{fig:sim-xyp}.  The four panels
show the two simulations for each of the two far-wing extrapolations.
Comparison with Figures~\ref{fig:xyprime}
and~\ref{fig:xyprime-ancantoid} shows the general tendency is the same
as with the wilkinoid: that the apex becomes less flat and the wings
less open as the inclination angle is increased.  There is no sign of
the sudden increase in openness at high inclination, as seen in the
cantoids and ancantoids that are asymptotically hyperbolic.  On the
other hand, the projected shapes of both simulations vary much more
strongly with \(\abs{i}\) than the wilkinoid does.  For the HD
simulation, this is mainly apparent for \(\abs{i} > \ang{30}\), but
for the MHD simulation it occurs at all inclinations.

\begin{figure}
  \centering
  \includegraphics[width=\linewidth]{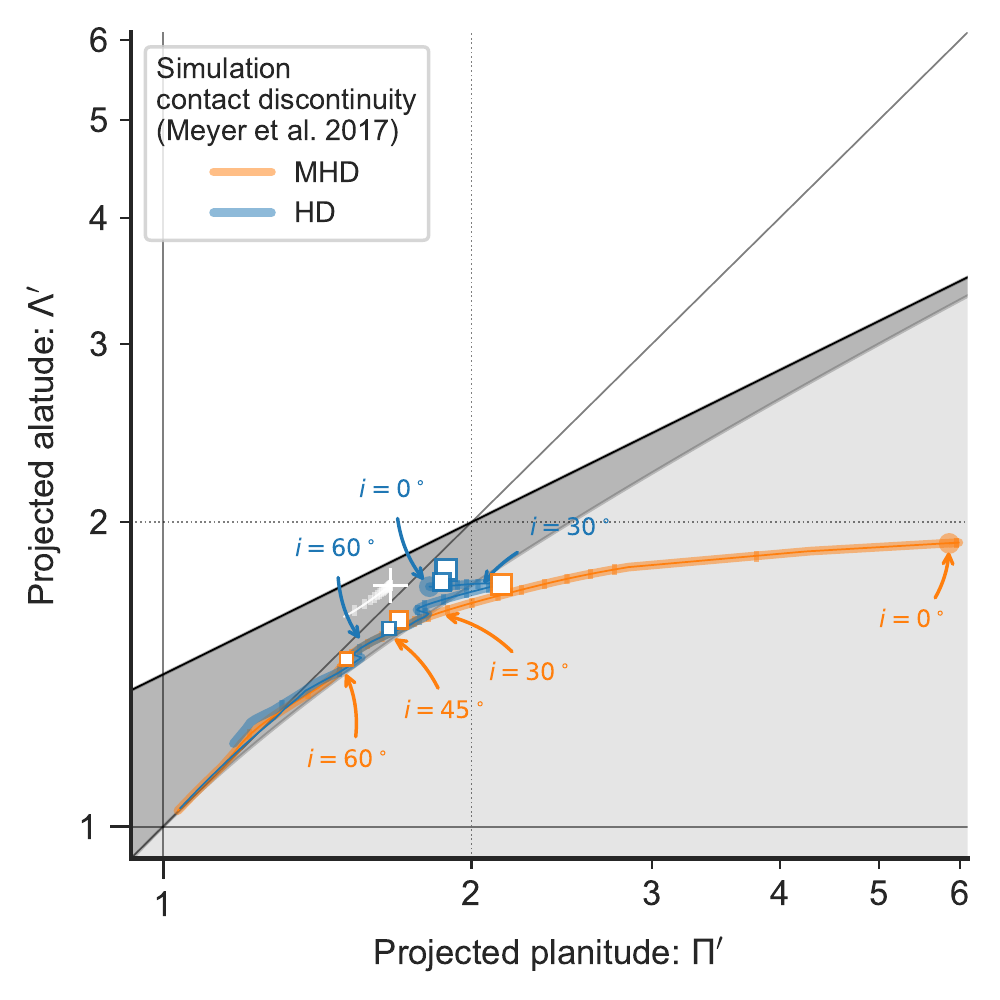}
  \caption[]{Apparent projected shapes of simulations in the
    \(\Pi'\)--\(\Lambda'\) plane.  Thick solid lines show the predicted
    inclination-dependent tracks of the traced contact discontinuity
    shape for the asymptotically open extrapolation, with tick marks
    indicating 20 equal intervals in \(\abs{\sin i}\). Thin solid
    lines show the same for the asymptotically closed extrapolation,
    which only deviates from the open case at the high-\(\abs{i}\) end
    of the HD tracks.  The true planitude and alatude are marked by
    filled circle symbols.  Open square symbols show the shapes traced
    from the dust emission maps at \SI{60}{\um} for inclinations of
    (largest to smallest) \ang{30}, \ang{45}, and \ang{60}. For
    comparison, the wilkinoid track is shown in white. Note that the
    scales of both axes are logarithmic in this case.}
  \label{fig:sim-Pi-Lambda}
\end{figure}

The resultant inclination-dependent tracks in the planitude--alatude
plane are shown in Figure~\ref{fig:sim-Pi-Lambda}.  These are compared
with measurements\footnote{%
  The shape measurements were performed by converting to contours the
  \SI{60}{\um} images in \citet{Meyer:2017a}'s Fig.~10 and then
  tracing the ridge of minimum radius of curvature of the contours.
  Identical results are found from using the \SI{100}{\um} maps
  instead.  For the \SI{25}{\um} maps, although the same results are
  found for low inclinations, in the maps with
  \(\abs{i} \ge \ang{45}\) in the HD case it becomes impossible to trace
  the limb-brightened rim because it becomes fainter than the emission
  from the true apex of the bow.} %
from post-processed infrared dust continuum maps at \SI{60}{\um}
\citep[\S~4.3 of][]{Meyer:2017a}, shown by open square symbols for
\(i = \ang{30}\), \ang{45}, and \ang{60}.  The agreement between the
two is good.  In particular, the \SI{60}{\um}-derived shapes are
always very close to the tracks derived from the contact discontinuity
shape. Also, the ordering of the three inclinations along the tracks
corresponds to what is predicted, although quantitatively there are
some slight deviations.  This close agreement stems from the fact,
emphasized by \citet{Meyer:2014b}, that the long-wavelength dust
emission from hot-star bow shocks tends to be dominated by material
just outside the contact discontinuity.  Note that there is almost no
difference in the planitude--alatude tracks between the closed and open
extrapolations.  This is because \(\Pi'\) and \(\Lambda'\) only depend on the
portion of \(R(\theta)\) between \(\theta_0\) (eq.~[\ref{eq:thetapar}]) and
\(\theta_{90}\) (eq.~[\ref{eq:th90}]), and these are both smaller than the
\(\theta\) range where extrapolation is necessary, except for in the HD
case at the highest inclinations.

\begin{figure}
  \centering
  \includegraphics[width=\linewidth]{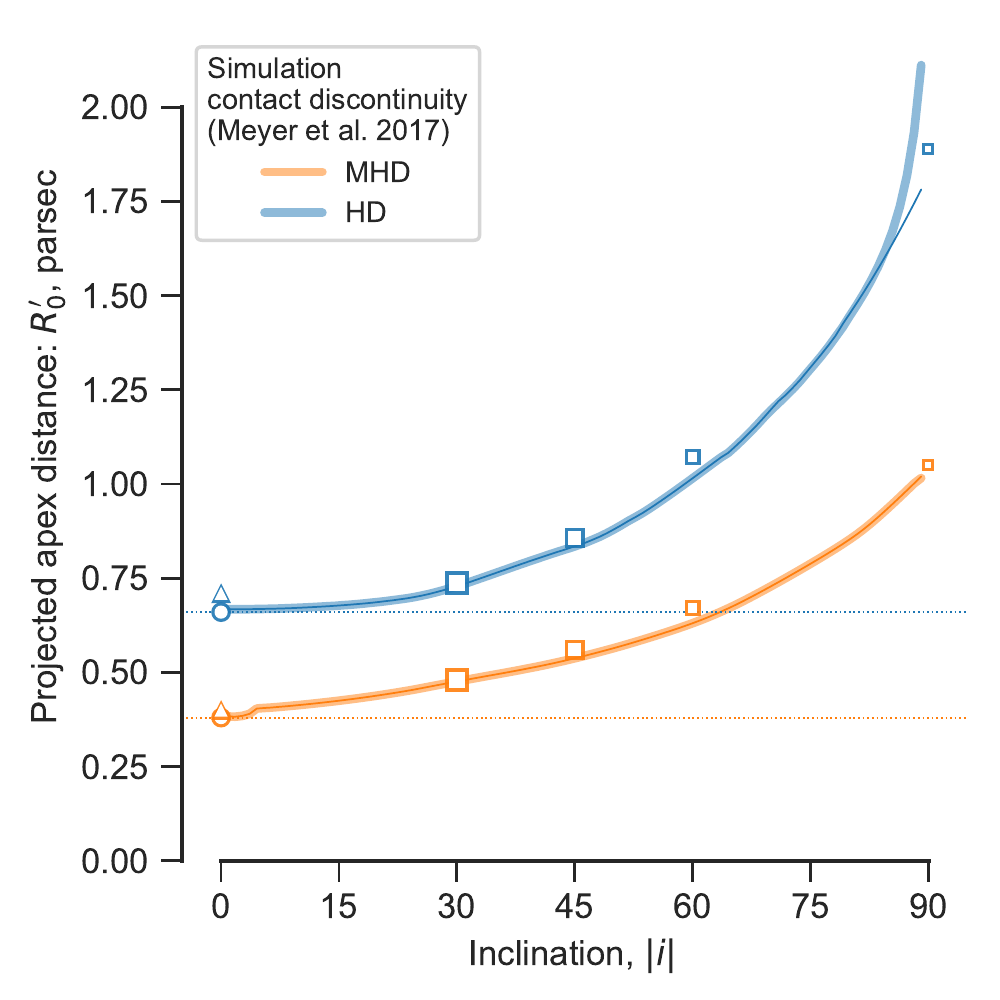}
  \caption[]{Apparent projected apex distance of simulations.  Line
    and symbol meanings are as in Figure~\ref{fig:sim-Pi-Lambda}.  In
    addition, triangle symbols at \(\abs{i} = \ang{0}\) denote radius
    measured on H\(\alpha\) optical emission maps.  Note that the distances
    for the blue square symbols have been adjusted according to the
    correction factor discussed in
    footnote~\ref{fn:meyer-correction}.}
  \label{fig:sim-R0-prime}
\end{figure}

Figure~\ref{fig:sim-R0-prime} shows the inclination
dependence of the projected apex distance, \(R_0'\).  As in the
previous figure, the lines show the prediction based on the shape of
the contact discontinuity, while the square symbols show the results
from the \SI{60}{\um} dust continuum maps.\footnote{%
  \label{fn:meyer-correction}
  There is an apparent error in the spatial scales for the HD
  simulations in Figs.~10 and 11 of \citet{Meyer:2017a}, with the dust
  emission peaks occurring at radii that are clearly too large.  The
  stated apex distance for the contact discontinuity in this
  simulation is \SI{0.69}{pc} from Table~2 of \citet{Meyer:2014b}, and
  the position of the peak in dust column density is \SI{0.70}{pc}
  from Fig.~17a of \citet{Meyer:2014b}.  These are consistent with
  Figs.~3, 4, and 7 of \citet{Meyer:2017a}, but not with Figs.~10 and
  11.  Luckily, the position of the true apex is clearly visible in
  the \SI{25}{\um} maps of Fig.~10 at inclinations of \ang{45} and
  \ang{60}.  The projected separation of the true apex is
  \(R_0 \cos i\), independent of the bow shape, which allows a
  correction factor of \num{0.65} to be found, assuming that the
  on-axis peak in the \SI{25}{\um} emission coincides with the peak in
  dust column density.  This correction has been applied to the blue
  square symbols shown in our Fig.~\ref{fig:sim-R0-prime}.} %
In addition, triangle symbols show results from H\(\alpha\) optical
emission line maps, which are given for \(i = 0\) in Fig.~7 of
\citet{Meyer:2017a}.  Again, the agreement is good between the values
derived from the shape of the contact discontinuity and those derived
from the surface brightness maps. The greatest discrepancy is seen
with the H\(\alpha\) maps and the intermediate inclination dust maps, with
\(R_0'\) being overestimated by a few percent in both cases.  The
differences in behavior between the two simulations are much larger
than this.  The larger true planitude of the MHD simulation means that
the relative increase of \(R_0'\) with \(\abs{i}\) is much stronger
than in the HD simulation for \(\abs{i} < \ang{45}\), as expected from
Figure~\ref{fig:quadric-projection}b.

\begin{figure}
  \centering
  \includegraphics[width=\linewidth]{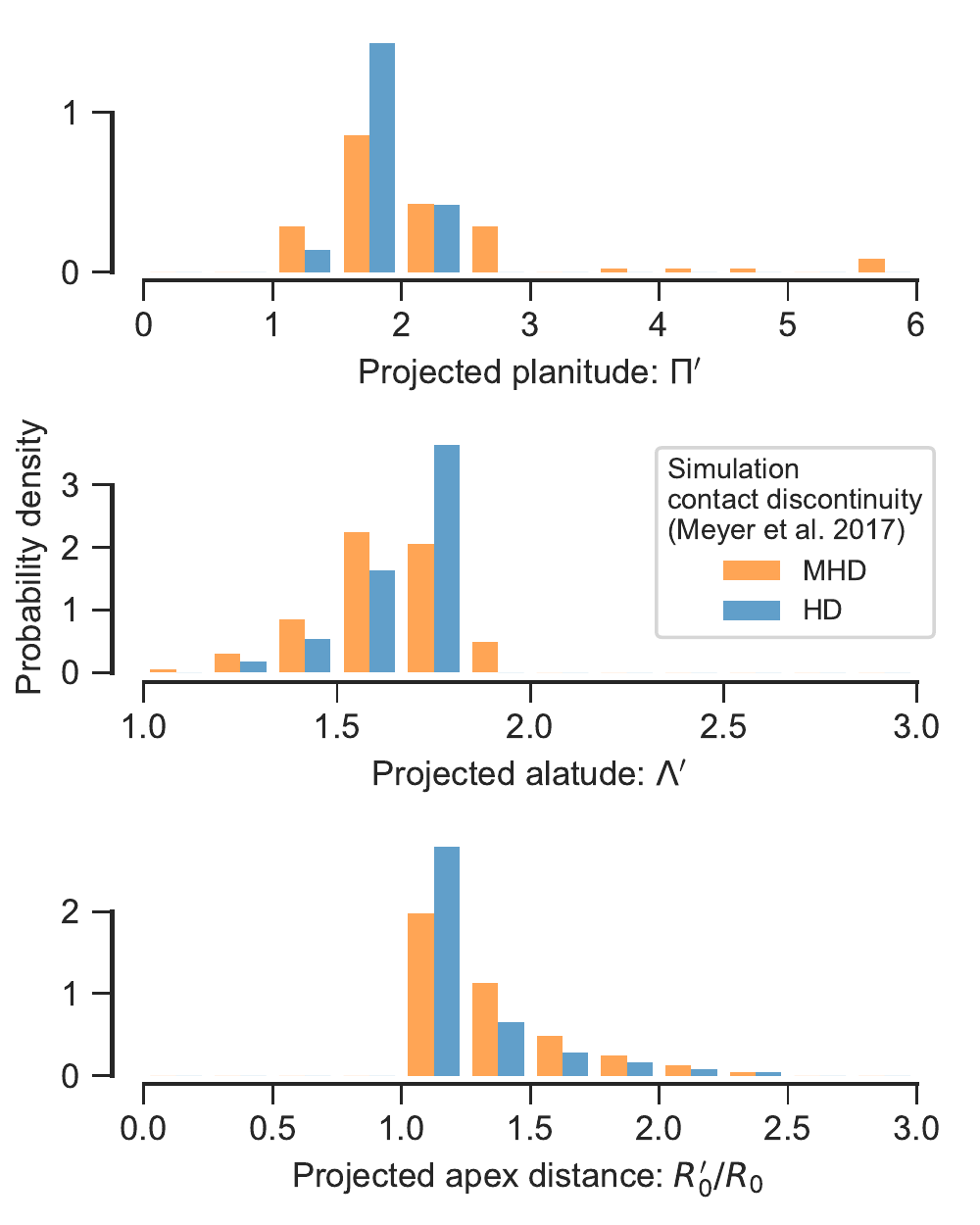}
  \caption[]{Histograms of (top to bottom) projected planitude,
    alatude, and bow shock size for the shape of the contact
    discontinuity in the \citet{Meyer:2017a} simulations. The \(y\)
    axis gives the probability density (per unit \(x\)-axis quantity),
    assuming a uniform distribution of viewing directions.}
  \label{fig:sim-histograms}
\end{figure}

The probability densities\footnote{%
  The probability density is defined so that its integral over the
  full range of the histogrammed variable is unity, making it
  independent of the histogram bin widths.  This means that the
  characteristic width of an approximately unimodal distribution is
  one over the maximum probability density.} %
of the apparent shape and size of the simulation bows (measured at the
contact discontinuity) are shown in Figure~\ref{fig:sim-histograms},
assuming that the viewing direction is uniformly distributed in solid
angle.  The modal value of the projected planitude is similar at
\(\Pi' \approx 1.8\) for both simulations, but the distribution is much
broader in the MHD case, which has a low-level wing extending out to
\(\Pi' \approx 6\).  The projected alatude distributions are both narrower
than the planitude (note the different scale of the histogram axis),
with the MHD case again being the broader of the two and peaking at a
slightly lower value (\(\Lambda' \approx 1.7\) as opposed to
\(\approx 1.8\) for the HD case).  Finally, the distribution of
projected-over-true apex distance is also broader for the MHD case.

\begin{figure*}
  \centering
  \includegraphics[width=\linewidth]{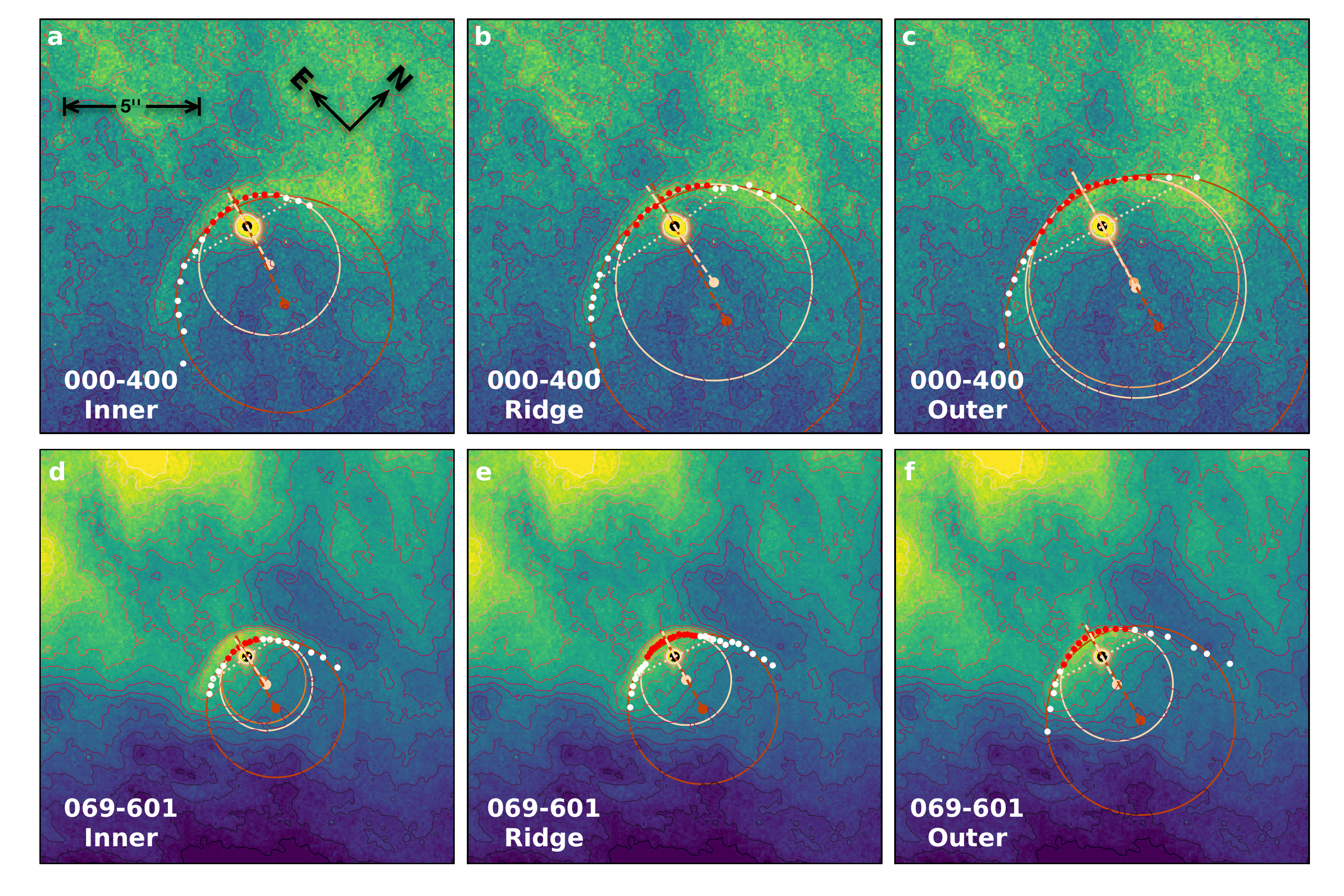}
  \caption[]{Example empirical determination of planitude and alatude
    for observed bow shocks associated with proplyds in the outer
    Orion Nebula (M42).  Color scale and contours show an \textit{HST}
    H\(\alpha\) image (ACS F658N filter, \citealp{Bally:2006a}) of
    M42~000-400 (panels \textit{a--c}) and M42~069-601 (panels
    \textit{d--f}).  The image scale and orientation are indicated on
    panel~\textit{a} and are the same for all panels.  Three different
    bows have been traced by eye on each object (red and white filled
    symbols): (\textit{a, d})~inner edge, (\textit{b, e})~ridge of
    maximum emission, and (\textit{c, f})~outer edge.  For each panel,
    the dark-colored circle shows the initial fit to the full set of
    points (white and red), using the algorithm described in
    Appendix~\ref{app:rcurv-empirical}.  The center of curvature and
    derived axis are shown by a small filled circle and dashed line in
    the same color. Lighter colored circles show three subsequent
    iterations where the fit is restricted to points within
    \(\pm \Delta\theta = \ang{75}\) of the axis.  The subset of points used in
    the final iteration is marked in red.  The perpendicular radii for
    the final iteration are shown by dotted lines. In panels
    \textit{a, b, d--f} the iterations converge immediately, but in
    panel \textit{c} the iterations stably oscillate between two
    slightly different solutions. }
  \label{fig:000-400-fit}
\end{figure*}

\begin{figure*}
  \begin{tabular}{p{0.47\linewidth} p{0.47\linewidth}}
    (a) & (b) \\
    \includegraphics[width=\linewidth]{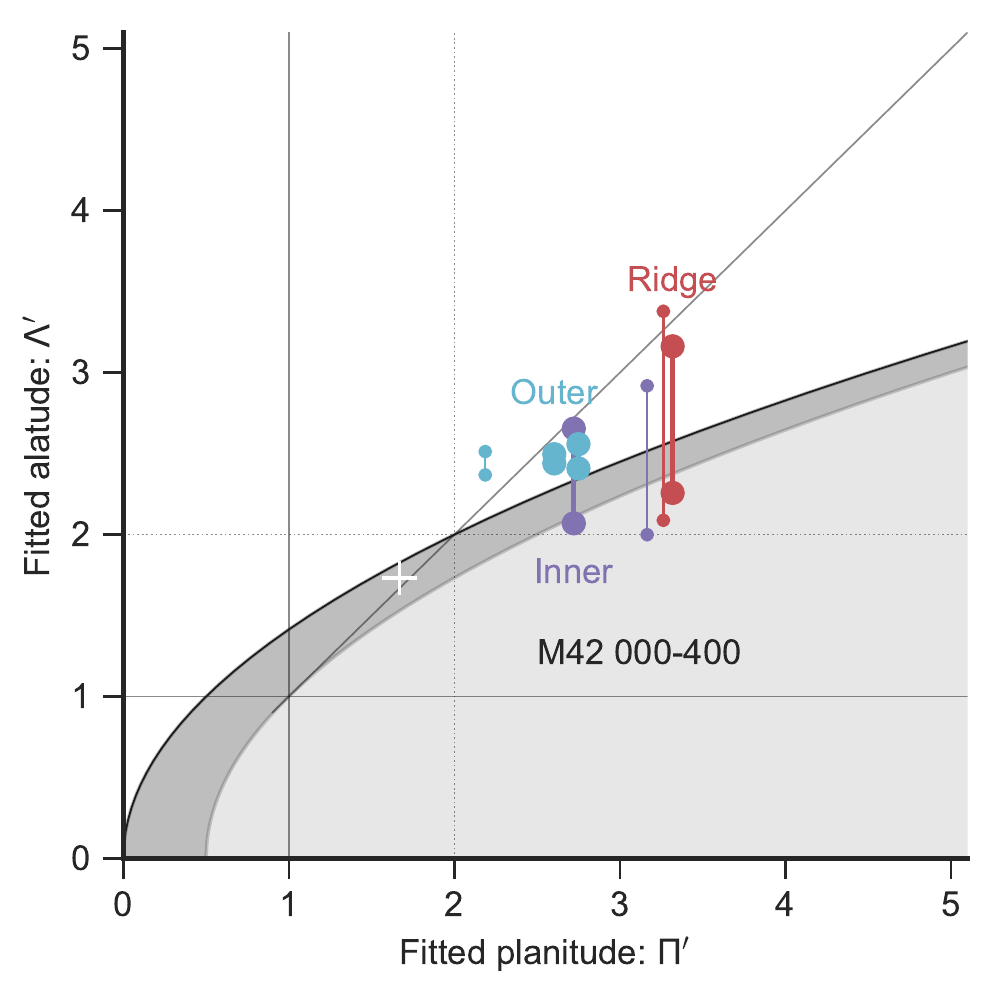}
    & \includegraphics[width=\linewidth]{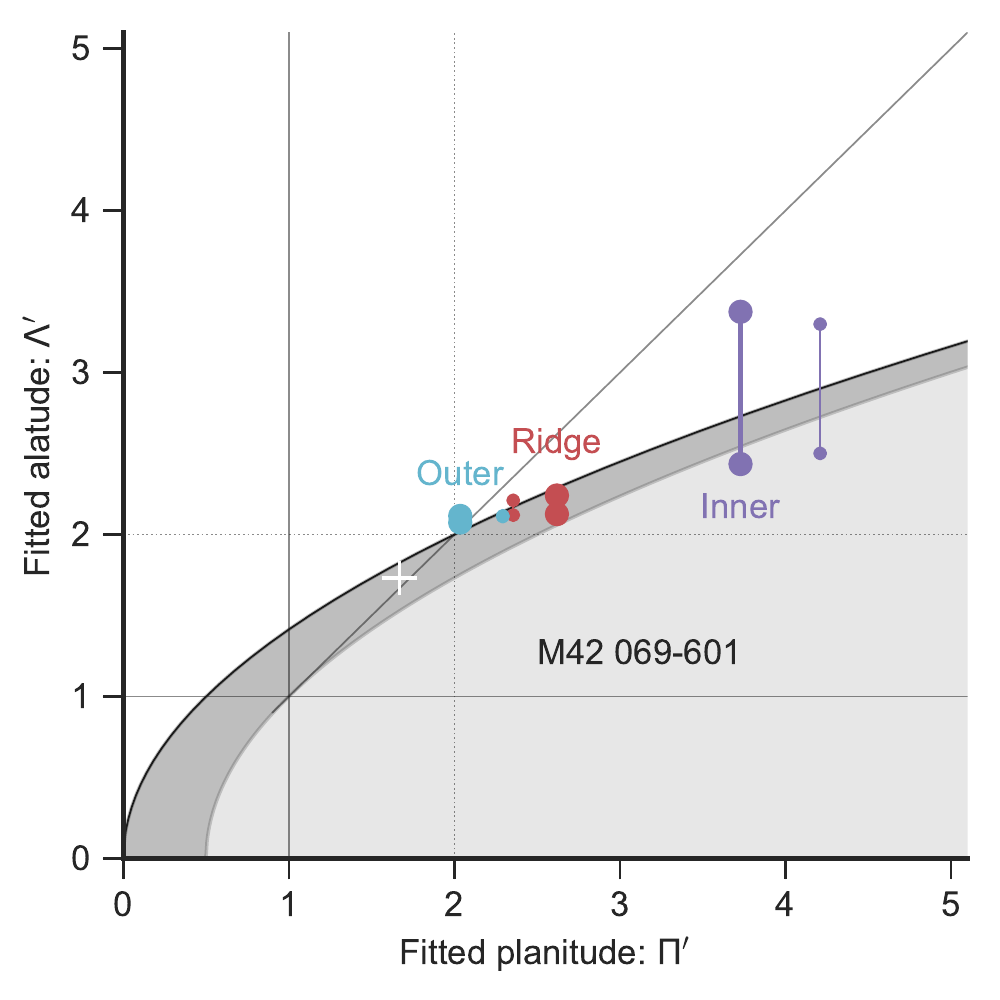}
  \end{tabular}
  \caption[]{Location in the projected planitude--alatude plane of the
    converged circle fits to the M42 bows: (\textit{a})~000-400,
    (\textit{b})~069-601. For each solution, the two values of the
    projected alatude, \(\Lambda_+'\) and \(\Lambda_-'\), corresponding to
    \(R_{90+}\) and \(R_{90-}\), are joined by a vertical line. Large
    symbols show the results from the fits shown in
    Figure~\ref{fig:000-400-fit}, while small symbols show results for
    fits using \(\Delta\theta = \ang{60}\) instead of \ang{75}. In panel
    \textit{a}, two slightly different \(\Pi'\) values are shown for the
    outer bow, since the fit does not converge to a single value (see
    Fig.~\ref{fig:000-400-fit}\textit{c} and
    App.~\ref{app:rcurv-empirical}).}
  \label{fig:000-400-planitude-alatude}
\end{figure*}

\begin{figure*}
  \begin{tabular}{p{0.47\linewidth} p{0.47\linewidth}}
    (a) & (b) \\
    \includegraphics[width=\linewidth]{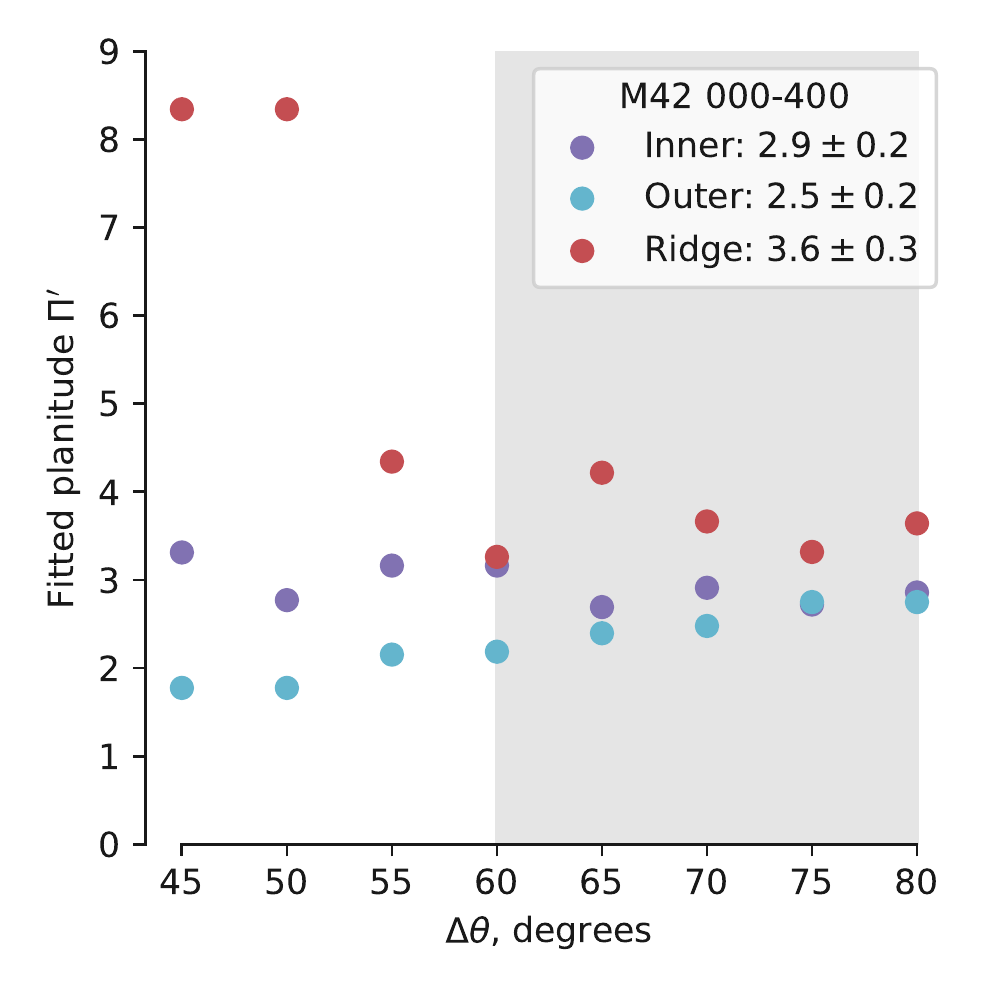}
    & \includegraphics[width=\linewidth]{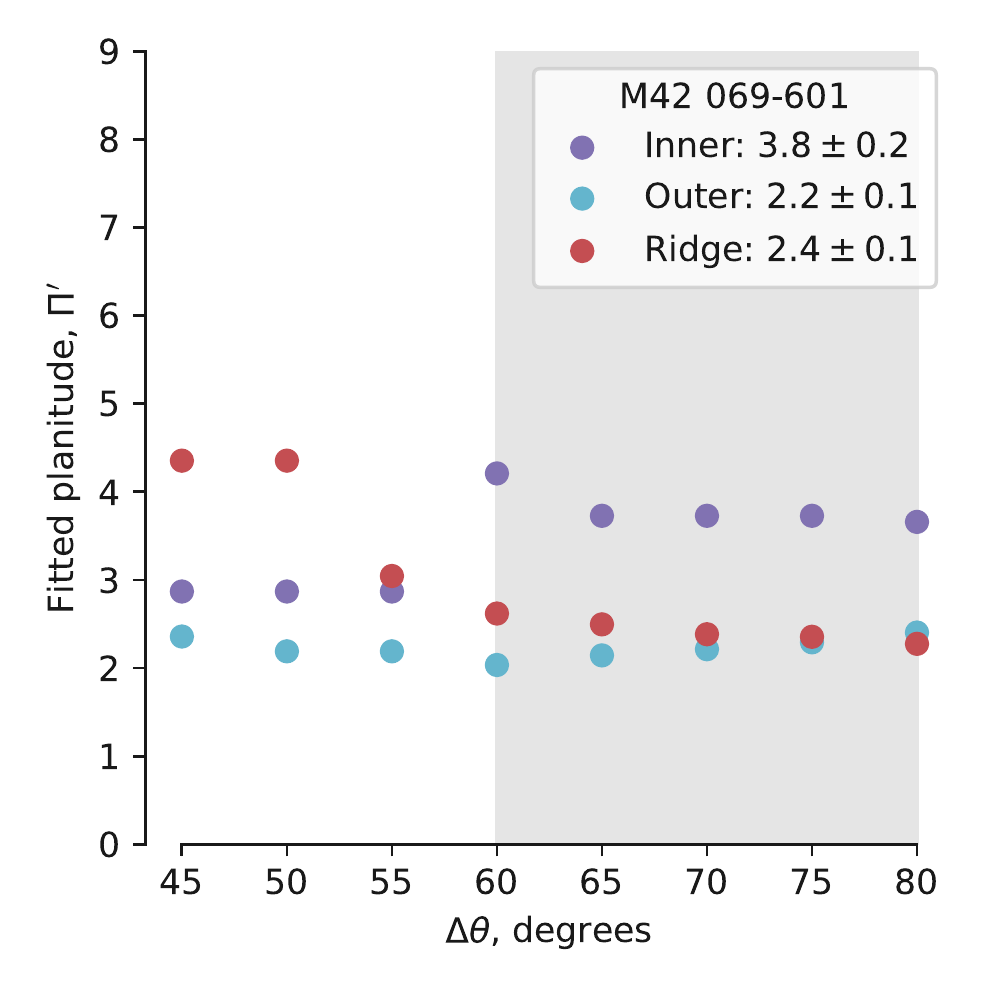}
  \end{tabular}
  \caption[]{Variation of fitted planitude, \(\Pi'\), as a function of
    the parameter \(\Delta\theta\), which controls how close a point must be to
    the axis in order to be included in the circle
    fit. (\textit{a})~M42 000-400, (\textit{b}) M42 069-601.  For the
    three traced bows (inner, outer, and ridge) of each object, the
    symbol key lists the mean and standard deviation of \(\Pi'\),
    calculated over the ``stable'' range \(\Delta\theta = \ang{60}\) to
    \ang{80}, which is indicated by light gray shading.}
  \label{fig:000-400-Delta-theta}
\end{figure*}

\begin{table}
  \caption[]{Fitted planitude and alatude for M42 bows}
  \label{tab:m42-fit}
  \centering
  \begin{tabular}{ll rrr}
    \toprule
    Source & Bow & \(\Pi'\) & \(\langle \Lambda' \rangle\)& \(\Delta\Lambda'\) \\
    \midrule
    000-400 & Inner & \(2.87 \pm 0.19\) & \(2.40 \pm 0.04\) & \(0.35 \pm 0.08\) \\
           &  Ridge & \(3.62 \pm 0.38\) & \(2.73 \pm 0.02\) & \(0.51 \pm 0.10\) \\
           & Outer & \(2.51 \pm 0.24\) & \(2.46 \pm 0.02\) & \(-0.06 \pm 0.02\) \\[\smallskipamount]
           & All & \(3.00 \pm 0.54\) & \(2.53 \pm 0.15\) & \(0.16 \pm 0.16\) \\[\bigskipamount]
    069-601 & Inner & \(3.81 \pm 0.23\) & \(2.90 \pm 0.01\) & \(0.44 \pm 0.05\) \\
           & Ridge & \(2.43 \pm 0.13\) & \(2.17 \pm 0.01\) & \(0.07 \pm 0.02\) \\
           & Outer & \(2.22 \pm 0.15\) & \(2.11 \pm 0.01\) & \(0.01 \pm 0.01\) \\[\smallskipamount]
           & All & \(2.82 \pm 0.75\) & \(2.39 \pm 0.37\) & \(0.17 \pm 0.20\)\\
    \bottomrule
  \end{tabular}
\end{table}

\section{Example application to observations}
\label{sec:obs}

As an example of measuring the projected planitude and alatude of real
bow shocks, we present an analysis of M42~000-400 and M42~069-601,
which are two H\(\alpha\) emission arcs \citep{Bally:2000a,
  Gutierrez-Soto:2015a} associated with proplyds\footnote{%
  The coordinate designation (see \citealp{ODell:1994a} for an
  explanation of the nomenclature) of 000-400 is very imprecise in
  right ascension, but we use it here for consistency with previous
  papers. The associated proplyd is listed with the more correct
  designation 4596-400 in catalogs such as \citet{Ricci:2008a}.  } %
in the west of the Orion Nebula (M42) at a distance of roughly
\SI{0.5}{pc} from the high-mass Trapezium stars that ionize the
nebula.  An image of one of these arcs (M42~069-601) was used in the
illustration of bow shock terminology in
Figure~\ref{fig:bow-terminology}.

\subsection{Empirical determination of bow shock shape}
\label{sec:empir-determ-bow}

We consider three different tracings of the bow shape (see
Fig.~\ref{fig:000-400-fit}): the peak of the emission arc (``ridge''),
and its inner and outer edges.  In all three cases, we placed by eye
the points that define the bow, using SAOImage~DS9 \citep{Joye:2003a}
in a similar fashion to in Figure~\ref{fig:meyer-trace}, and guided by
the image contours.\footnote{%
  In the case of the ``ridge'' method at least, it is possible to
  automate this step, which we will discuss in detail in a following
  paper. } %

We determine the planitude and alatude by fitting a circle to the
traced points within \(\pm \Delta\theta = \ang{75}\) of the bow axis, using the
iterative algorithm described in Appendix~\ref{app:rcurv-empirical}.
The fitted circle, when combined with the position of the central
source, yields the orientation of the bow axis, together with the apex
distance, \(R_0\), radius of curvature, \(R_\C\), and two
perpendicular radii (one for each wing), \(R_{90+}\) and \(R_{90-}\).
These are all indicated on the panels of Figure~\ref{fig:000-400-fit}
by light-colored lines.\footnote{%
  For conciseness, we drop the prime symbol from the radii, both in
  this section and in Appendix~\ref{app:rcurv-empirical}, even though
  they are all projected quantities.} %
The projected planitude and alatude then follow as
\(\Pi' = R_\C/R_0\), \(\Lambda_+' = R_{90+}/R_0\),
\(\Lambda_-' = R_{90-}/R_0\), which are shown in
Figure~\ref{fig:000-400-planitude-alatude}.

\subsection{Analysis of sources of systematic uncertainty}
\label{sec:analys-sourc-syst}

The planitude is found to have a moderate dependence on the choice of
\(\Delta\theta\), as shown in Figure~\ref{fig:000-400-Delta-theta}, where it can
be seen that, although the values of \(\Pi'\) are relatively stable for
\(\Delta\theta \ge \ang{60}\) (light gray shaded region), they can show much
larger variations for \(\Delta\theta < \ang{60}\).  The fact that the radius of
curvature is defined at a point (the projected apex) might seem to
argue for making \(\Delta\theta\) as small as possible, but that would lead to
circle fits that were extremely sensitive to the exact positions of
the few points included in the fit.  A reliable fit requires 4 or more
points, ideally spanning a total separation that is a substantial
fraction of \(R_\C\), which would argue for \(\Delta\theta\) larger than about
\(\Pi'/2\) radians, or \ang{60} to \ang{90}.  On the other hand, if
\(\Delta\theta \ge \ang{90}\) then the planitude and alatude would no longer be
independent since the bow would be forced to lie on the ``sphere''
line, \(\Lambda' = (2 \Pi' - 1)^{1/2}\) (see \S~\ref{sec:conic}).  Balancing
these two concerns suggests an optimal \(\Delta\theta = \ang{75}\), which is
shown in Figure~\ref{fig:000-400-fit}, whereas in
Figure~\ref{fig:000-400-planitude-alatude} we show results for both
\ang{75} (thick lines) and \ang{60} (thin lines).

Unlike all the models considered in \S~\ref{sec:crw-scenario} and
\S~\ref{sec:more-realistic-bow}, the observed bows are not necessarily
symmetrical and so the alatude for the two wings, \(\Lambda_-'\) and
\(\Lambda_+'\), may be different.  We therefore calculate an average
alatude, \(\langle\Lambda'\rangle\), and an alatude asymmetry, \(\Delta\Lambda'\):
\begin{equation}
  \label{eq:alatude-average-and-asym}
  \langle\Lambda'\rangle = \tfrac12 \left( \Lambda_+' + \Lambda_-' \right)
  \quad\quad \Delta\Lambda' = \tfrac12 \left( \Lambda_+' - \Lambda_-' \right)
\end{equation}
The results for these two quantities, together with the planitude, are
shown in Table~\ref{tab:m42-fit}.  For each object and for each tracing
(inner, ridge, outer, see Fig.~\ref{fig:000-400-fit}), the
\(\text{mean} \pm \text{standard\ deviation}\) is listed for circle
fits using \(\Delta\theta = \ang{60}\) to \ang{80} (see
Fig.~\ref{fig:000-400-Delta-theta}).  Additionally, the row ``All''
for each object gives the global mean and standard deviation over all
three tracings.

It can be seen from Table~\ref{tab:m42-fit} that the uncertainty in
the fitted parameters is dominated by the variations between the
different traced shapes.  For example, the one-sigma relative
variation of planitude, \(\Pi'\), is \(< 10\%\) within the individual
tracings, but \(\approx 20\%\) between tracings.  For the mean alatude,
\(\langle\Lambda'\rangle\), the variation within individual tracings is extremely
small\footnote{%
  This is because the only way that variation in the circle fit
  parameters affect the alatudes is through the axis orientation
  vector, and changing the orientation induces roughly opposite
  effects on \(\Lambda_+'\) and \(\Lambda_-'\), which approximately cancel out in
  \(\langle\Lambda'\rangle\).  } %
at \(\approx 1\%\), but is \(\approx 10\%\) between tracings.  The alatude
asymmetry, \(\Delta\Lambda'\), is best interpreted as a difference between the
symmetry axis of the apex region and the symmetry axis of the wings.
In relative terms, this is \(0\)--\(20\%\), but with large systematic
variations between tracings (for instance, in both objects it is very
small for the outer arcs).

It could be argued that much of the between-tracing variations in
\(\Pi'\) and \(\langle\Lambda'\rangle\) are due to real differences between the shapes of
the inner and outer boundaries of the emission arc.  Although this may
be true, in the absence of a robust theory as to exactly what feature
of the observed images constitutes \emph{the} bow shock, such
variations nevertheless serve to limit the precision with which the
bow shock shape can be measured.  We therefore conclude that
conservative estimates of \(20\%\) precision for \(\Pi'\) and \(10\%\)
precision for \(\Lambda'\) are appropriate when analyzing observations of a
similar or better quality\footnote{%
  Since the uncertainties are systematic and due to unavoidably
  subjective decisions, it is unlikely that better quality
  observations would improve the situation, although poorer quality
  observations could make things worse.  } %
to those presented in Figure~\ref{fig:000-400-fit}.  This will be an
important limitation when comparing the statistics of the shapes of
different bow shock populations, as we will present in a following
paper.

\subsection{Derived shape of the M42 arcs}
\label{sec:derived-shape-m42}

From a casual inspection of Figure~\ref{fig:000-400-fit}, it is
apparent that the shapes of the two M42 arcs are closely similar, and
this is confirmed by the numbers in Table~\ref{tab:m42-fit}. Both
000-400 and 069-601 are consistent with
\((\Pi', \Lambda') \approx (3.0, 2.5)\), and if we take the absolute minimum over
all the different tracings and reasonable variations in the
\(\Delta\theta\) fit parameter, we find unassailable lower limits of
\((\Pi', \Lambda') > (2.1, 2.1)\).  In the rest of this section, we consider
only these lower limits, since they are already sufficient for drawing
interesting conclusions.

Comparison with Figure~\ref{fig:sim-Pi-Lambda} shows that both the MHD
and HD simulations of \citet{Meyer:2016a} are inconsistent with the
observations.  Although \(\Pi' > 2.1\) can be satisfied,\footnote{%
  Either by the MHD simulations at low inclinations
  (\(\abs{i} = \ang{0}\)--\ang{20}) or by the HD simulation at
  intermediate inclinations (\(\abs{i} = \ang{30}\)--\ang{40}).} %
the simulations' projected alatude is \(\Lambda' < 1.9\) for all
inclinations, which is significantly less than the observed lower
limit of \(2.1\).  This is not particularly surprising since the
simulations were not tailored to the situation of these proplyd bow
shocks in M42, in which the mildly supersonic photoevaporation flow
from an externally irradiated protoplanetary disk interacts with the
mildly supersonic champagne flow from the core of the Orion Nebula.
The proplyd case has at least four important differences from the
runaway O-star case modeled by \citet{Meyer:2016a}: (1)~The velocity
of the outer wind is \(\le \SI{20}{km.s^{-1}}\) instead of
\SI{40}{km.s^{-1}}; (2)~The outer wind is slightly divergent, rather
than plane-parallel; (3)~Both inner and outer shocks are strongly
radiative, so both shells (see Fig.~\ref{fig:2-winds}) contribute to
the observed emission; (4)~The inner wind is not isotropic, but
instead corresponds to the \(k = 0.5\) case of
equation~\eqref{eq:ancantoid-density}.  The ways in which these
differences may account for the discrepancy with the observations will
be explored in detail in a subsequent paper.

\section{Summary and discussion}
\label{sec:conc}

We have shown that the shapes of stellar bow shocks can be usefully
characterized by two dimensionless numbers: the \textit{planitude},
\(\Pi\), or flatness of the bow's apex, and the \textit{alatude},
\(\Lambda\), or openness of the bow's wings (\S~\ref{sec:plan-alat-bow}).
The planitude and alatude can be estimated from ratios of lengths that
can be straightforwardly measured from observations or theoretical
models.  We develop a general method (\S~\ref{sec:projection}) for
finding the projected shape, \((\Pi', \Lambda')\), of a bow shock's
limb-brightened edge, or \textit{tangent line}, as a function of
inclination angle, \(i\), where the emission shell is idealized as a
cylindrically symmetric surface.

We first apply this method to find inclination-dependent tracks on the
projected planitude--alatude plane for the special case of
\textit{quadric} surfaces (\S~\ref{sec:conic}), such as hyperboloids,
paraboloids, and spheroids, where the tangent line is a conic section.
The spheroids and hyperboloids occupy distinct regions of the plane,
with the paraboloids defining the boundary between the two.  As the
inclination is increased from \(\abs{i} = 0\) (side-on) to
\(\abs{i} = \ang{90}\) (end-on), the tracks first tend to approach the
diagonal \(\Lambda' = \Pi'\), corresponding to confocal conics, always
remaining within their own region.  At the highest inclinations, the
spheroids all converge at \(\abs{i} = \ang{90}\) on the point
\((\Pi', \Lambda') = (1, 1)\) and the paraboloids on the point
\((\Pi', \Lambda') = (2, 2)\).  The hyperboloids, on the other hand diverge as
\((\Pi', \Lambda') \to (\infty, \infty)\) for a finite
\(i_{\mathrm{crit}}\), which depends on the asymptotic opening angle
of the tail.  For \(\abs{i} > i_{\mathrm{crit}}\), the tangent line no
longer exists for the hyperboloid, and it would no longer appear to be
a curved bow shock.  We introduce the parabolic departure function
(\S~\ref{sec:parab-depart-funct}) as tool for visualizing differences
in bow shapes, \(R(\theta)\), over the full range,
\(\theta = [\ang{0}, \ang{180}]\).

We then apply the projection method to a set of thin-shell
hydrodynamic models of bow shocks (\S~\ref{sec:crw-scenario}): the
\textit{wilkinoid} from a wind-parallel stream interaction and the
\textit{cantoids} from wind-wind interactions.  We generalize the
latter to the \textit{ancantoids}, where one of the winds is
anisotropic.  We find that the wilkinoid is confined to a small region
of the \(\Pi'\)--\(\Lambda'\) plane, with projected planitude and alatude
varying with inclination by \(< 15\%\).  The cantoids and ancantoids
with sufficiently small values of \(\beta\), the wind momentum ratio, have
more interesting behavior, with tracks that pass from the spheroid
region at low inclinations to the hyperboloid region at high
inclinations.

In the following section (\S~\ref{sec:more-realistic-bow}), we test the
projected shape analysis methods against the results of computational
fluid dynamic simulations of magnetized and non-magnetized bow shocks
from \citet{Meyer:2017a} of a runaway OB main-sequence star.  We find
that measurements made on maps of infrared dust emission can be
accurate diagnostics of the projected shape of the contact
discontinuity for this type of bow shock
(Fig.~\ref{fig:sim-Pi-Lambda}).  The distributions of projected
planitude and alatude for a population of randomly oriented bow shocks
shows systematic differences between the different simulations.

Finally (\S~\ref{sec:obs}), we give an example of the application of
our methods to observed emission maps of bow shocks, describing a
robust algorithm for empirically determining the projected planitude
and alatude from imperfect real data.  We investigate the sensitivity
of the results to systematic errors due to both observational
uncertainties and subjective choices in the application of the
algorithm.  We find that the projected planitude and alatude can be
determined with precisions of 20\% and 10\%, respectively.  For our
illustrative observations, we show that this is more than sufficient
to rule out certain models.

This paper is the first of a series that will apply our shape analysis
to a wide variety of models and observations of stellar bow shocks.
In a second paper, 
we consider the alternative model of dusty radiation-driven bow wave
\citep{Ochsendorf:2014a}, instead of a hydrodynamic bow shock, and
also calculate the signature in the planitude--alatude plane of
oscillations in the bow shape, which may be due to instabilities or a
time-varying source.  In a third paper, 
we apply our techniques to observational datasets for three different
classes of stellar bow shocks: OB stars \citep{Kobulnicky:2016a}, cool
giants/supergiants \citep{Cox:2012a}, and young stars in the extended
Orion Nebula \citep{Henney:2013a}.  In a fourth paper,
we analyze the proplyd bow shocks in the core of the Orion Nebula
\citep{Garcia-Arredondo:2001a}.



\section*{Acknowledgements}

We are grateful for financial support provided by Dirección General de
Asuntos del Personal Académico, Universidad Nacional Autónoma de
México, through grant Programa de Apoyo a Proyectos de Investigación e
Inovación Tecnológica IN111215.  JATY acknowledges support via a
research studentship from Consejo Nacional de Ciencias y Tecnología,
Mexico.  This work has made extensive use of Python language libraries
from the SciPy \citep{Jones:2001a} and AstroPy
\citep{Astropy-Collaboration:2013a, Astropy-Collaboration:2018a}
projects.  We appreciate the thoughtful comments of the anonymous
referee, which led us to clarify the presentation of our results and
prompted the addition of \S~\ref{sec:obs} and
Appendices~\ref{sec:radius-curvature}
and~\ref{sec:plane-sky-projection}.

\bibliographystyle{mnras}
\bibliography{bowshocks-biblio}
\appendix

\section{Radius of curvature}
\label{sec:radius-curvature}

The radius of curvature of a general curve can be written (e.g.,
eq.~[2-5] of \citealp{Guggenheimer:2012a}):
\begin{equation}
  \label{eq:Rcurv-general}
  R_{\C} \equiv \frac{1}{\abs{\kappa}} = \Abs{\frac{d s}{d \alpha}} \ ,  
\end{equation}
where \(\kappa\) is the \textit{curvature}, \(s\) is the path length along
the curve and \(\alpha\) is the tangent angle (see Fig.~\ref{fig:unitvec}).
In spherical polar coordinates, this becomes \citep{Weisstein:2018a}:
\begin{equation}
  \label{eq:Rcurv-polar}
  R_{\C} = \frac{\left( R^2 + R_\theta^2 \right)^{3/2}}
  {\Abs{R^2 + 2 R_\theta^2 - R R_{\theta\theta}}} \ , 
\end{equation}
where \(R_\theta = d R / d \theta\) and
\(R_{\theta\theta} = d^2 R / d \theta^2\).  At the apex,
\(R_\theta = 0\) by symmetry, which yields
equation~\eqref{eq:radius-curvature} of
\S~\ref{sec:plan-alat-bow}. Note that \(\theta\) is dimensionless and
should be measured in radians \citep{Mohr:2015a, Quincey:2017a}.

\section{Rotation matrices and plane of sky projection}
\label{sec:plane-sky-projection}

The transformation from the body frame (unprimed) to observer-frame
(primed) coordinates is a rotation about the \(y\) axis by an angle
\(i\), which is described by the rotation matrix:
\begin{equation}
  \label{eq:Rotation-matrix-y}
  \mathbfss{A}_y(i) = 
   \begin{pmatrix}
    \cos i & 0 & -\sin i \\
    0 & 1 & 0 \\
    \sin i & 0 &\cos i 
  \end{pmatrix} \ .
\end{equation}
This is used in equation~\eqref{eq:Trans}.  A further application is
to express the observer-frame Cartesian basis vectors in terms of the
body-frame basis:
\begin{align}
  \label{eq:xunit-prime}
  \uvec{x}' &= \mathbfss{A}_y(-i)\, 
              \begin{Vector}
              1 \\ 0 \\ 0  
              \end{Vector}
              =
              \begin{Vector}
              \cos i \\ 0 \\ -\sin i  
              \end{Vector} \ ,
  \\
  \label{eq:yunit-prime}
  \uvec{y}' &= \mathbfss{A}_y(-i)\, 
              \begin{Vector}
              0 \\ 1 \\ 0  
              \end{Vector}
              =
              \begin{Vector}
              0 \\ 1 \\ 0
              \end{Vector} \ , 
  \\
  \label{eq:zunit-prime}
  \uvec{z}' &= \mathbfss{A}_y(-i)\, 
              \begin{Vector}
              0 \\ 0 \\ 1  
              \end{Vector}
              =
              \begin{Vector}
              \sin i \\ 0 \\ \cos i  
              \end{Vector} \ .
\end{align}
Note that in this case the sign of \(i\) is reversed because it is the
inverse operation to that in equation~\eqref{eq:Trans}

Since we are considering cylindrically symmetric bows, all azimuths
\(\phi\) are equivalent, so it is sufficient to work with two-dimensional
curves in the plane \(z = 0\) (which is also \(\phi = 0\)) and then find
the three-dimensional surface by rotating about the \(x\)-axis via the
rotation matrix:
\begin{equation}
  \label{eq:Rotation-matrix-x}
  \mathbfss{A}_x(\phi) = 
   \begin{pmatrix}
    1 & 0 & 0 \\
    0 &\cos\phi & -\sin\phi \\
    0 &\sin\phi & \cos\phi 
  \end{pmatrix} \ ,
\end{equation}
where \(\phi\) takes all values in the interval \([0, 2\pi]\).


\section{Paraboloids and their plane-of-sky projection}
\label{app:parabola}

Equation~\eqref{eq:par-xy} for the \(xy\) coordinates of a quadric in the \(\phi = 0\) plane cannot be used in the case of a paraboloid (\(\Q = 0\)).  Instead, a convenient parametrization is
\begin{gather}
  \label{eq:parabola-xy}
  \begin{aligned}
    x &= R_0 \left(1  - \tfrac{1}{2} \Pi\, t^2\right) \\
    y &= R_0\, \Pi\, t \ ,
  \end{aligned}
\end{gather}
where we have ``baked in'' knowledge of the planitude,
\(\Pi = R_{\C}/R_0\) (see \S~\ref{sec:plan-alat-bow}). The projected
plane-of-sky coordinates of the tangent line follow from
equation~\eqref{eq:Trans} as
\begin{gather}
  \label{eq:parabola-xy-prime-phi}
  \begin{aligned}
    x_{\T}' / R_0 &= \left(1 - \tfrac{1}{2} \Pi\, t^2\right) \cos i
      + \Pi\, t \sin\phi_{\T} \sin i\\
    y_{\T}' / R_0 &= \Pi\, t \cos\phi_{\T}\ ,
  \end{aligned}
\end{gather}
The azimuth of the tangent line is found from
equations~(\ref{eq:alpha}, \ref{eq:tanphi}) as
\(\sin\phi_{\T} = -t^{-1} \tan i \), so that
\begin{gather}
  \label{eq:parabola-xy-prime-final}
  \begin{aligned}
    x_{\T}' / R_0 &= \cos i \left[ 1 + \tfrac{1}{2} \Pi \tan^2 i -
      \tfrac{1}{2} \Pi \left( t^2 - \tan^2 i \right) \right]\\
    y_{\T}' / R_0 &= \Pi \left( t^2 - \tan^2 i \right)^{1/2} \ .
  \end{aligned}
\end{gather}
The projected star--apex distance, \(R_0'\), is the value of
\(x_{\T}'\) when \(y_{\T}' = 0\), yielding
\begin{equation}
  \label{eq:parabola-R0-prime}
  R_0' / R_0 = \cos i \left( 1 + \tfrac{1}{2} \Pi \tan^2 i  \right) \ . 
\end{equation}
Note that this same result can be obtained from a Taylor expansion of
equation~\eqref{eq:fQi-factor} substituted into~\eqref{eq:R0-prime} in
the limit \(\Q \to 0\).

Equation~\eqref{eq:parabola-xy-prime-final} can be rewritten in the
form
\begin{gather}
  \label{eq:parabola-xy-all-primes}
  \begin{aligned}
    x_{\T}' &= R_0' \left(1  - \tfrac{1}{2} \Pi' t'^2\right) \\
    y_{\T}' &= R_0' \Pi' t' \ ,
  \end{aligned}
\end{gather}
where
\begin{align}
  \label{eq:parabola-Pi-prime}
  \Pi' &= \frac{2 \Pi} {2 \cos^2 i + \Pi \sin^2 i} \\
  \label{eq:parabola-t-prime}
  t' &= \cos i \left(t^2 - \tan^2 i\right)^{1/2} \ ,
\end{align}
which demonstrates that the projected shape is also a parabola. It is
apparent from \eqref{eq:parabola-Pi-prime} that the projected
planitude obeys
\begin{equation*}
\lim_{i \to \ang{90}} \Pi' = 2 \ ,
\end{equation*}
for all values of the true planitude \(\Pi\), as is shown by the black
lines in Figure~\ref{fig:quadric-projection}a. The projected alatude can
be found as
\begin{equation}
  \label{eq:parabola-Lambda-prime}
  \Lambda' = \left( 2 \Pi' \right)^{1/2} \ .
\end{equation}
For the special case of the confocal paraboloid,
\(\Pi = \Lambda = 2\), we have \(\Pi' = \Pi\) and
\(\Lambda' = \Lambda\) by equations~\eqref{eq:parabola-Pi-prime}
and~\eqref{eq:parabola-Lambda-prime} for all inclinations, so its
shape is unaffected by projection.

\section{Analytic derivation of thin-shell bow shape parameters}
\label{sec:thin-shell-shapes}

In this appendix, we provide analytic calculations of the planitude,
alatude, and asymptotic opening angle for the wilkinoid, cantoids, and
ancantoids.  We first consider the most general case of the
ancantoids, and then show how results for cantoids and the wilkinoid
follow as special cases.

\subsection{Planitude of ancantoids}
\label{sec:ancantoid-planitude}

From equations~\eqref{eq:radius-curvature} and~\eqref{eq:planitude},
the planitude depends on the apex second derivative,
\(R_{\theta\theta,0}\), as
\begin{equation}
  \label{eq:planitude-from-2nd-derivative}
  \Pi = \left(  1 - R_{\theta\theta,0} / R_0\right)^{-1} \ .
\end{equation}
From equation~\eqref{eq:taylor-R-theta}, the second derivative can be
found from the coefficient of \(\theta^2\) in the Taylor expansion
of \(R(\theta)\).  Since we do not have \(R(\theta)\) in explicit analytic form,
we proceed via a Taylor expansion of the implicit
equations~\eqref{eq:crw-angles}
and~\eqref{eq:ancantoid-theta-theta1-implicit}, retaining terms up to
\(\theta^4\) to obtain from
equation~\eqref{eq:ancantoid-theta-theta1-implicit}:
\begin{equation}
  \label{eq:taylor-expansion-implicit}
  \theta_1^2 = \beta \theta^2 \left( 1 + C_{k\beta} \theta^2\right) + \mathcal{O}(\theta^6)\ , 
\end{equation}
with the coefficient \(C_{k\beta}\) given by
\begin{equation}
  \label{eq:C-k-beta}
  C_{k\beta} = \frac{1}{15} - \frac{3k}{20} - \frac{\beta}{15}  \ .
\end{equation}
Note that it is necessary to include the \(\theta^4\) term in the expansion
of \(\theta_1^2\) so that \(\theta_1/\theta\) is accurate to order
\(\theta^2\).  Then, from equation~\eqref{eq:crw-angles} we find
\begin{align}
  \label{eq:taylor-R-over-D}
  \frac{R}{D} & = \frac{\sin \theta_1} {\sin (\theta + \theta_1)} \nonumber \\
              & = \frac{\beta^{1/2}}{1+\beta^{1/2}}
                \left\lbrace 1 + \theta^2
                \left[ \frac{C_{k\beta}} {2 \left(1+\beta^{1/2}\right)}
                + \frac{1}{6} \left(1+2\beta^{1/2} \right)
                \right]
                \right\rbrace + \mathcal{O}(\theta^4) \ ,
\end{align}
where in the second line we have carried out a Taylor expansion of the
two \(\sin\) terms and substituted
\eqref{eq:taylor-expansion-implicit}.  Comparing coefficients of unity
and \(\theta^2\) between equations~\eqref{eq:taylor-R-theta} and
\eqref{eq:taylor-R-over-D} we find
\begin{align}
  \label{eq:again-R0-over-D}
  \frac{R_0} {D} &= \frac{\beta^{1/2}}{1+\beta^{1/2}} \\
  \label{eq:final-second-derivative}
  \frac{R_{\theta\theta,0}} {R_0} &= \frac{C_{k\beta}}{1+\beta^{1/2}}+\frac{1}{3}\left(1+2\beta^{1/2}\right) \ ,
\end{align}
so that the final result for the planitude, from~\eqref{eq:planitude-from-2nd-derivative}, is
\begin{equation}
  \label{eq:final-planitude}
  \text{ancantoid} \quad
  \Pi = \left[ {1 - \frac{C_{k\beta}}{1+\beta^{1/2}} - \frac{1}{3}\left(1+2\beta^{1/2}\right)}
  \right]^{-1} \ .
\end{equation}

\subsection{Alatude of ancantoids}
\label{sec:ancantoid-alatude}

To find the alatude, \(\Lambda = R_{90} / R_0\), we use
equation~\eqref{eq:crw-angles} at \(\theta = \ang{90}\) to write
\begin{equation}
  \label{eq:Lambda-from-theta-1-90}
  \Lambda = \frac{D} {R_0} \tan \theta_{1,90} \ , 
\end{equation}
where \(\theta_{1,90} = \theta_1(\theta = \ang{90})\), which, following
equation~\eqref{eq:ancantoid-theta-theta1-implicit}, must satisfy
\begin{equation}
  \label{eq:theta-1-90-implicit}
  \theta_{1,90} \cot \theta_{1,90}  = 1 - \frac{2 \beta}{k + 2} \ . 
\end{equation}
Combining \eqref{eq:Lambda-from-theta-1-90} and
\eqref{eq:theta-1-90-implicit} with \eqref{eq:again-R0-over-D} yields
\begin{equation}
  \label{eq:Lambda-beta-xi-theta-1-90}
  \Lambda = \frac{ \left(1 + \beta^{1/2}\right) \,\theta_{1,90}} {\beta^{1/2} \left(1 - \xi_k \beta\right)} \ ,
\end{equation}
where
\begin{equation}
  \label{eq:xi-k}
  \xi_k = \frac{2} {k +2} \ .
\end{equation}
We now take the Taylor expansion of equation~\eqref{eq:theta-1-90-implicit} to find
\begin{equation}
  \label{eq:theta-1-90-Taylor}
  \theta_{1,90}^2 + \tfrac{1}{15}  \theta_{1,90}^4 + \mathcal{O}(\theta_{1,90}^6)
  = 3 \xi_k \beta \ , 
\end{equation}
which, if \(\theta_{1,90}\) is small, has the approximate solution
\begin{equation}
  \label{eq:theta-1-90-approx}
  \theta_{1,90} \approx \left( \frac{3 \xi_k \beta} {1 + \tfrac15 \xi_k \beta} \right)^{1/2} \ .
\end{equation}
Substituting back into equation~\eqref{eq:Lambda-from-theta-1-90}
yields an approximate value for the alatude of
\begin{equation}
  \label{eq:Lambda-approx}
  \text{ancantoid} \quad
  \Lambda \approx \frac {(3 \xi_k)^{1/2} \left( 1 + \beta^{1/2} \right)}
  { \left( 1 + \tfrac15 \xi_k \beta \right)^{1/2} \left( 1 - \xi_k \beta \right)} \ .
\end{equation}
This approximation is surprisingly accurate, with a relative error of
order \(1\%\) even for \(\beta\) as large as \(0.5\) with \(k = 0\).  

\subsection{Planitude and alatude of cantoids and wilkinoid}
\label{sec:cantoid-wilkinoid-shapes}

Since \(\Pi\) and \(\Lambda\) depend on only that portion of the inner wind
emitted in the forward hemisphere, \(\theta \le \ang{90}\), the results for the
cantoids can be found by taking \(k = 0\), in which case
equations~(\ref{eq:C-k-beta}, \ref{eq:final-planitude}, \ref{eq:xi-k},
\ref{eq:Lambda-approx}) yield
\begin{gather}
  \label{eq:cantoid-Pi-Lambda}
  \text{cantoid} \quad
  \begin{cases}
    \quad \Pi &= \dfrac {5} {3 \left( 1 - \beta^{1/2} \right)}\\[12pt]
    \quad \Lambda &= \dfrac { \sqrt 3} {\left( 1 + \tfrac15 \beta \right)^{1/2} \left( 1 - \beta^{1/2} \right)} \ .
  \end{cases}
\end{gather}
The wilkinoid shape is equal to the \(\beta \to 0\) limit of the cantoid, so
its planitude and alatude are given by:
\begin{gather}
  \label{eq:wilkinoid-Pi-Lambda}
  \text{wilkinoid} \quad
  \begin{cases}
    \quad \Pi &= \dfrac {5} {3}\\[10pt]
    \quad \Lambda &= \sqrt 3 \ .
  \end{cases}
\end{gather}
The wilkinoid results can also be obtained directly from
equation~\eqref{eq:wilkinoid-R-theta}, and in the case of \(\Lambda\) this
has already been noted by several authors \citep{Cox:2012a,
  Meyer:2016a}.

\subsection{Asymptotic opening angle}
\label{sec:asympt-open-angle}

The asymptotic opening angle
of the far wings, \(\theta_\infty\), can be found from
equation~\eqref{eq:ancantoid-theta-theta1-implicit} for the
ancantoids, together with the condition that
\(\theta_\infty + \theta_{1\infty} = \pi\).  These yield the implicit equation
\begin{equation}
  \label{eq:ancantoid-theta-inf}
  \theta_\infty - \left( \frac {k + 2 (1 - \beta)} {k + 2} \right) \tan \theta_\infty
  = \pi + 2 \beta I_k(\pi/2) \ ,
\end{equation}
where
\begin{equation}
  I_k(\pi/2) = \frac{\sqrt \pi}{4} \,
      \frac{  \GammaFunc\left( \frac{k+1}{2} \right)} {\GammaFunc\left(\frac{k+4}{2}\right)}
\end{equation}
and \(\GammaFunc\) is the usual Gamma function.  This can be compared with the equivalent result obtained by \CRW{} for the cantoids:
\begin{equation}
  \label{eq:cantoid-theta-inf}
  \theta_\infty - \tan \theta_\infty = \frac{\pi}{1 - \beta} \ .
\end{equation}
Note that, unlike in the cases of \(\Pi\) and \(\Lambda\),
equation~\eqref{eq:ancantoid-theta-inf} does \emph{not} reduce to
equation~\eqref{eq:cantoid-theta-inf} in the limit \(k \to 0\).  This is
because, for \(\theta > \ang{90}\), the \(k = 0\) ancantoid differs from the
cantoid since the former has no wind in the backward hemisphere (see
Figure~\ref{fig:anisotropic-arrows}).  Therefore there is less inner
support for the far wings of the bow, and so \(\theta_\infty\) is smaller than
in the cantoid case.  The wilkinoid result again follows from
\(\beta \to 0\), implying that \(\theta_\infty = \pi\), or, in other words, that the far
wings are asymptotically parallel to the symmetry axis, as is the case
for the paraboloid (App.~\ref{app:parabola}).  In the case of the
wilkinoid, however, the behavior is cubic in the wings,
\(z \sim r^3\), as opposed to quadratic as in the paraboloid.

\section{Empirical determination of radius of curvature for a bow shock of unknown orientation}
\label{app:rcurv-empirical}

Consider a set of \(N\) points on the plane of the sky,\footnote{%
  In the main body of the paper, the prime symbol (\('\)) is used to
  distinguish projected from ``true'' quantities.  In this Appendix,
  for simplicity, we omit the primes since \emph{all} quantities are
  projected.} %
with Cartesian coordinates \(\bm{r}_k = (x_k, y_k)\) for
\(k = 1 \dots N\).  We wish to estimate the radius of curvature of the
smooth curve that the set of points is presumed to be sampled from.
To do this, we fit a circle to the points as follows.  The circle is
defined by its center, \(\bm{r}_{\C} = (x_{\C}, y_{\C})\), and radius,
\(R_{\C}\).  For a given circle, we define a mean radius of the set of
points from the circle center:
\begin{equation}
  \label{eq:rcurv-mean-radius}
  \bar{R}_\C(x_{\C}, y_{\C}) = \frac{1}{N}  \sum_{k = 1}^{N} \Abs{\bm{r}_k - \bm{r}_{\C}}  \ .
\end{equation}
We then optimize
to find best-fit values \((x_{\C}^*, y_{\C}^*)\), which minimize the
objective function
\begin{equation}
  \label{eq:rcurv-objective-function}
  f(x_{\C}, y_{\C}) = \sum_{k = 1}^{N} \left(
    \abs{\bm{r}_k - \bm{r}_{\C}}  - \bar{R}_\C (x_{\C}, y_{\C}) \right)^2 \ .
\end{equation}
The best-fit radius of curvature is then given by
\(R_{\C}^* = \bar{R}_\C(x_{\C}^*, y_{\C}^*)\).

If we also know the position, \(\bm{r}_0 = (x_0, y_0)\), of the bow's
central source, then we can find the unit vector in the direction of
the bow's projected axis as
\begin{equation}
  \label{eq:rcurv-axis-vector}
  \uvec{\xi} = \frac {\bm{r}_0 - \bm{r}_{\C}^*} { \abs{\bm{r}_0 - \bm{r}_{\C}^*}} \ , 
\end{equation}
and the apex distance from the source as\footnote{%
  This is only valid if the resultant \(R_0 < R_{\C}^*\), otherwise
  the opposite sign of \(\uvec{\xi}\) must be taken.}
\begin{equation}
  \label{eq:rcurv-apex-distance}
  R_0 = \Abs{\bm{r}_{\C}^* + R_{\C}^* \uvec{\xi} - \bm{r}_0} \ .
\end{equation}
A refinement of the method is then to iteratively repeat the circle
fit after restricting the set of points to those lying within a
certain angle \(\Delta\theta\) of the bow axis, where we find that best results
are obtained with \(\Delta\theta \approx \ang{60}\) to \ang{75}.  That is, 
\begin{equation}
  \label{eq:rcurv-delta-theta}
  \Abs{\theta_k} < \Delta\theta \ , 
\end{equation}
where the signed angle \(\theta_k\) of each point from the
axis,\footnote{%
  Although the sign of \(\theta_k\) is not relevant to
  equation~\eqref{eq:rcurv-delta-theta}, it is used below in
  calculating the perpendicular radii.  } %
measured at the source position \(r_0\), can be calculated as
\begin{equation}
  \label{eq:rcurv-theta-k}
  \theta_k = \arctan \left[  
    \frac{\left( \bm{r}_k - \bm{r}_0 \right) \cdot \uvec{\xi}^\perp}
    {\left( \bm{r}_k - \bm{r}_0 \right) \cdot \uvec{\xi}}
  \right] \ .
\end{equation}
In the preceding equation, the ``perpendicular'' operator (\(\perp\))
rotates its vector argument counter-clockwise by \ang{90}, so that
\((x, y)^\perp = (-y, x)\).
  
Two or three iterations are sufficient for convergence in most cases,
although in some cases it is possible that the process will converge
to a stable flip-flop oscillation between two different solutions.
This is due to the dependence of the \(\theta_k\), via \(\uvec{\xi}\), on
the \(\bm{r}_{\C}^*\) of the previous iteration, which can lead to
points entering and leaving the fitted set.  We have not found this to
be a serious problem in practice, since the two solutions tend to be
very close to one another.  It could be mitigated by averaging
\(\bm{r}_{\C}^*\) over two previous iterations.  The alternative of
measuring the angle with respect to the center of curvature,
\(\bm{r}_{\C}^*\), instead of the source, \(\bm{r}_0\), is found to be
much less stable.

If quantitative estimates exist for the uncertainties, \(\epsilon_k\), in the
measurements of \(\bm{r}_k\), then it is appropriate to incorporate
weights of \(\epsilon_k^{-2}\) in the objective function.  However, it is
rare for the \(\epsilon_k\) to be objectively quantifiable, since the
uncertainties are often systematic and/or subjective.  In cases
where the bow shape is traced by eye, based on real or synthetic
observations, a more practical approach is to maintain uniform
weighting but to place a greater density of points \(\bm{r}_k\) in
regions where the shape is well-determined and to place them more
sparsely in regions where the shape is less certain.

Since there is no guarantee of symmetry about the axis \(\uvec{\xi}\),
the perpendicular radius will in general be different in the two wings
of the bow, with values \(R_{90+}\) and \(R_{90-}\).  These can be
estimated by defining
\begin{equation}
  \label{eq:rcurv-R-k}
  R_k = \Abs{\bm{r}_k - \bm{r}_0}
\end{equation}
and linearly interpolating between the points \((\theta_k, R_k)\) at
\(\theta = \pm \ang{90}\).

Our Python language implementation of this algorithm is freely
available at \url{https://github.com/div-B-equals-0/circle-fit}.  An
example application to real data is given in \S~\ref{sec:obs} and
Figure~\ref{fig:000-400-fit}.  Note that this method is not necessary
if the orientation of the bow axis is known a~priori, in which case
the Taylor series method described in \S~\ref{sec:plan-alat-bow} is
more efficient and accurate.


\bsp	
\label{lastpage}
\end{document}